\documentclass[11pt] {article}
\usepackage[a4paper,margin=1in]{geometry}

\usepackage[T1]{fontenc} %
\usepackage[normalem]{ulem} %

\usepackage[usenames,dvipsnames,svgnames,x11names]{xcolor}

\newcommand{\modc}[1]{{\textcolor{blue}{#1}}}
\newcommand{\addc}[1]{\textcolor{teal}{#1}}
\newcommand{\delc}[1]{ {\textcolor{gray} {\sout{#1}} }}
\newcommand{\repc}[2]{ {\textcolor{gray} {\sout{#1}} }{\textcolor{teal} {#2}}}

\newcommand{\modcminor}[1]{{\textcolor{blue}{#1}}}
\newcommand{\addcminor}[1]{\textcolor{teal}{#1}}

\newcommand{\delcminor}[1]{ {\textcolor{gray} {\sout{#1}} }}

\renewcommand{\modc}[1]{#1} %
\renewcommand{\addc}[1]{#1} %
\renewcommand{\delc}[1]{} %
\renewcommand{\repc}[2]{#2} %

\renewcommand{\modcminor}[1]{#1} %
\renewcommand{\addcminor}[1]{#1} %
\renewcommand{\delcminor}[1]{} %

\usepackage{balance}

\usepackage{listings}

\lstdefinelanguage{XML}
{
basicstyle=\ttfamily\footnotesize,
  morestring=[b]",
  moredelim=[s][\bfseries\color{Maroon}]{<}{\ },
  moredelim=[s][\bfseries\color{Maroon}]{</}{>},
  moredelim=[l][\bfseries\color{Maroon}]{/>},
  moredelim=[l][\bfseries\color{Maroon}]{>},
  morecomment=[s]{<?}{?>},
  morecomment=[s]{<!--}{-->},
  commentstyle=\color{gray},
  stringstyle=\color{blue},
  identifierstyle=\color{red}
}

\usepackage{moreverb}

\usepackage[nounderscore]{syntax}

\usepackage[leqno,centertags,reqno,,leqno,centertags,reqno,,cmex10]{amsmath}
\usepackage{amssymb}
\usepackage{amsthm}
\usepackage{amsfonts}
\usepackage{textcomp}
\usepackage{gensymb}
\usepackage{siunitx}
\newcommand{\mud}{\mu\$}

\usepackage[caption=false]{subfig} %

\usepackage{multirow} %
\usepackage{rotating} %
\usepackage{booktabs} %
\usepackage{colortbl} %
\usepackage{tablefootnote} %
\usepackage{tabularx}

\usepackage[colorlinks,bookmarksopen,bookmarksnumbered,citecolor=blue,urlcolor=blue,linkcolor=blue]{hyperref}

\usepackage{xspace}

\usepackage{enumitem}

\newcommand{\hide}[1] {}

\usepackage{fontawesome}
\newcommand{\xmark}{\textcolor{Mahogany!50}{\faTimesCircle}}%
\newcommand{\cmark}{\textcolor{LimeGreen}{\faCheckCircle}}%
\newcommand{\pcmark}{\textcolor{ForestGreen}{\faCheck}}%
\newcommand{\namark}{\textcolor{gray!50}{\faCircle}}%
\usepackage{makecell}

\usepackage{blindtext}

\usepackage{titlesec}

\newcounter{takeaway}

\titleclass{\takeaway}{straight}[\part]
\titleformat{\takeaway}[hang]{\normalfont\em}{}{0ex}{\colortakeaway}
\titlespacing*{\takeaway}{0pt}{1ex}{1ex}
\renewcommand{\thetakeaway}{\textbf{Observation~\arabic{takeaway}}}
\newcommand{\colortakeaway}[1]{%
  \colorbox{blue!15}{\parbox{\dimexpr\columnwidth-2\fboxsep}{\thetakeaway.~#1}}}

\begin{document}

\title{Characterizing FaaS Workflows on Public Clouds:\\The Good, the Bad and the Ugly\footnote{This is a preprint version of an article accepted for publication in \textit{IEEE Transactions on Parallel and Distributed Systems,2026} (TPDS). DOI: 10.1109/TPDS.2026.3678606.}}

\author{Varad Kulkarni~$^1$, Nikhil Reddy~$^1$, Tuhin Khare~$^2$,\\
Abhinandan S. Prasad~$^3$, Chitra Babu~$^4$ and Yogesh Simmhan~$^1$\\
\small
$^1$~Indian Institute of Science, Bangalore, India\\
\small
$^2$Georgia Institute of Technology, USA\\
\small
$^3$~Indian Institute of Technology (IIT), Palakkad, India\\
\small
$^4$~IIT, Madras, India\\
\small
Email: \{varadk, simmhan\}@iisc.ac.in
}

\date{}
\maketitle
\begin{abstract}
Function-as-a-service (FaaS) is a popular serverless computing paradigm for event-driven functions that elastically scale on public clouds. FaaS workflows (e.g., \textit{AWS Step Functions} and \textit{Azure Durable Functions}), are composed from FaaS functions (e.g., AWS Lambda and Azure Functions) to build practical applications. But, the complex interactions between functions in the workflow and limited visibility into the internals of proprietary FaaS platforms are major impediments to \repc{gaining a deeper understanding of}{analyzing} a FaaS workflow's \repc{platform}{performance}. While several works characterize FaaS platforms to derive such insights, \addc{or offer FaaS Workflow benchmarks}, there is a lack of a principled \repc{and rigorous study for}{of} FaaS workflow platforms, which have unique scaling, performance and costing behavior influenced by the platform design, dataflow and workloads. In this article, 
we perform extensive evaluations of three popular FaaS workflow platforms from AWS and Azure, running $25$ micro-benchmark and application workflows over $139k$ invocations.
Our detailed analysis confirms some conventional wisdom but also uncovers unique insights on the function execution, workflow orchestration, inter-function interactions, cold-start scaling and monetary costs. Our observations help developers better configure and program these platforms, set performance and scalability expectations, and identify research gaps on enhancing the platforms.
\end{abstract}

\section{Introduction}%

\textit{Serverless computing} is a cloud computing paradigm for the development, deployment, and execution of cloud applications without explicit resource management by a user. 
\textit{Function-as-a-service (FaaS)} is a serverless variant where functions, triggered by events or requests, are the unit of application definition, deployment, and scaling. FaaS is popular in enterprises~\cite{sscfaas}, science~\cite{sesw}, and machine learning~\cite{ammesfaas} due to its automated elasticity, ease of management, and billing per-invocation. 
\textit{FaaS \uline{Workflows}} \addc{usually} compose functions into a Directed Acyclic Graph (DAG) to form non-trivial applications \modc{with defined dataflow between functions and execution order}~\cite{scsaiew}. \addc{While FaaS workflows may include iterative constructs, we focus on DAG-based workflow that are common in literature.} 
\modc{Shahrad~\textit{et al.} report that $54\%$ of FaaS applications have one function and $95\%$ have at most $10$ functions, i.e., a sizable fraction have multiple functions that interact~\cite{swcosw} --}
\addc{we refer to this growing class as \textit{FaaS workflows}, which can be designed and executed using AWS Step Functions, Azure Durable Functions, Google Workflows, etc., on public clouds.}

\subsection{Motivation and Gaps}
A deeper understanding of the design, performance, and scalability of FaaS workflow platforms enables better design of workflow applications and configuration of FaaS platforms~\cite{swcosw}.
\modc{The low-latency and rapid scaling of FaaS functions may not directly translate to workflows due to dependencies and orchestration overheads.} 
However, understanding the design trade-offs of these platforms is non-trivial. 
\textit{First,} FaaS platforms from public (commercial) Cloud Service Providers (CSPs) are \emph{proprietary}. 
Developer documents have scarce details on their internal design, motivating the need for empirical studies for reverse engineering. 
\textit{Second,} FaaS workflows have fine-grained and complex interactions between the functions through parameter passing, whose performance side-effects are not obvious~\cite{bmspec}. The interplay between the platform architecture and workflow behavior is non-trivial.

Recent efforts have conducted \emph{black-box} profiling of FaaS platforms~\cite{swcosw,peak-behind}. \modc{But these are limited to individual functions, and the learnings do not fully map to workflow platforms due to message passing and orchestration overheads, resource sharing, and asymmetric scaling.}
\modc{Benchmarks for FaaS~\cite{faasdom,fbench,copik2021sebs,crossfit,befaas} and, recently, for FaaS Workflows~\cite{sebsflow} exist.
These help compare platforms for the same application but do not characterize the workflow's behavior under sustained and dynamic workloads, with the goal of understanding the performance stability, scaling, and cost impact of the platforms.}
\modc{There is also a large body of research into \textit{optimizing FaaS platforms}~\cite{baosfs,sequoia}. Similar opportunities can arise from our systematic study of FaaS workflow platforms.}

In particular, there is a lack of a holistic study of diverse FaaS workflow characteristics to understand their performance, scalability, and costs on public clouds. We identify 
an exhaustive set of dimensions essential for developers and systems researchers to assess the workflow's performance, 
 understand their scaling, overheads due to container coldstarts, time spent on function execution vs. inter-function communication and coordination, 
and the monetary costs. These 
are closely tied to the underlying platform, requiring their characterization and analysis. 
\modc{Existing works limit their study to just functions and/or only to a subset of these features (Table~\ref{tbl:cmp})}.

\subsection{Contributions}
We make several key contributions to address this gap.
\begin{enumerate} [leftmargin=*]
    \item We define key dimensions to be analyzed to understand the performance of FaaS workflows, \modc{and position existing literature against these}.
    \item \modc{These guide our principled benchmarking of the performance and cost of three popular FaaS workflow platforms:}
\textit{AWS Step Functions} and two flavors of \textit{Azure Durable Functions} -- Default Storage (AzS) and Netherite (AzN).
    \item \modc{We analyze these benchmark outcomes to synthesize}
    key \textit{observations}. These complement, and sometimes contradict, the existing wisdom on FaaS functions and expose ``the good, the bad, and the ugly'' of FaaS workflow platforms. \addc{These are finally summarized as design insights.}
\end{enumerate}

\addc{In the course of this study,} we have deployed 
\modc{$6$~FaaS workflows with $38$~functions  and $160$ workload configurations on AWS and Azure for this study; cumulatively, the workflows are invoked $\approx 139k$ times and functions $\approx 5.8M$ times.}
We use our \textit{XFaaS} multi-cloud FaaS workflow deployment tool~\cite{xfaas} to run realistic workflows and workloads from our \textit{XFBench} workflow benchmark~\cite{xfbench}. Provenance logs from XFaaS and cloud traces provide metrics for our analysis, and micro-benchmarks help analyze specific patterns.
We target the public FaaS platforms 
on Amazon AWS and Microsoft Azure, which account for $56\%$ of the cloud market.\footnote{\href{https://www.srgresearch.com/articles/cloud-market-gets-its-mojo-back-q4-increase-in-cloud-spending-reaches-new-highs}{Cloud Market Gets its Mojo Back; AI Helps Push Q4 Increase in Cloud Spending to New Highs}, Synergy Research Group, 2024}

\addc{These result in} some key insights, including:
\begin{itemize}[leftmargin=*]
\item \emph{Cold-start and Scaling:} 
Our analysis shows the amplified effects of function cold-starts on workflows 
due to the workflow structure, e.g., the cold-start execution times of longer workflows are $\approx 74\%$ higher than warm-starts, while workflows with more task-parallelism see a $\approx 50\%$ higher cold-start overhead. AWS is able to scale faster than Azure.
\item \emph{Inter-function vs. Function execution latencies:}  
The time required to transfer messages between functions can dominate workflow execution, accounting for $83$--$96\%$ of overall time. AzN has the least transfer latency for payloads up to $256KB$, while AzS has the highest. AWS has the least variability.

\item \emph{Cost Analysis}: FaaS workflow billing is complex with different components.
Our models identify orchestration and data transfers as the dominant cost ($\approx$ 99\%) rather than function execution.
While AWS performs best, it is also costlier.
\end{itemize}

The rest of this article is organized as follows: \S~\ref{sec:relwork} provides an overview of current research and gaps; \S~\ref{sec:bg} reviews the design of FaaS workflow platforms from AWS and Azure; 
\S~\ref{sec:xfw} defines the workflow and workflow benchmarks used while \S~\ref{sec:exp} describes our experiment harness; \S~\ref{sec:results} provides detailed results, observations and analysis from the benchmarks; 
\addc{\S~\ref{sec:insights} summarizes these into practical design insights;}
and lastly,
\S~\ref{sec:conclusion} presents our conclusions and future directions.

\section{Related Work}\label{sec:relwork}

\begin{table*}[h]
  \centering
  \scriptsize
  \caption{Comparison of Literature on FaaS and FaaS Workflow Characterizations Studies}
  \label{tbl:cmp}
  \setlength{\tabcolsep}{0.8pt}
  \begin{tabularx}{\textwidth}{>{\columncolor{gray!18}}l || *{8}{c} p{3.22cm}}
    \hline
    \bf Lit.~$\downarrow$~~|~~Char.~$\rightarrow$ & \bf Functions & \bf Workflows & \bf Scaling & \bf Coldstarts & \bf Func. Exec. & \bf Inter-func. & \bf E2E & \bf Costs & \bf CSPs\\
    \hline\hline
    
    \bf Eismann et al.~\cite{pcswicpe} & \cmark & \cmark & \xmark & \xmark & \xmark & \xmark & \cmark & \pcmark & GCP \\\hline
    
    \bf Li et al.~\cite{aossiseke} & \cmark & \xmark & \cmark & \xmark & \cmark & \namark & \namark & \xmark & Knative, OpenFaas, Nuclio, Kubeless \\\hline
    
    \bf \addc{Li et al.}~\cite{faasflow2022} & \cmark & \cmark & \xmark & \xmark & \cmark & \cmark & \cmark & \xmark & Private Implementation\\\hline
    
    \bf Lin et al.~\cite{moptpds} & \cmark & \cmark & \xmark & \xmark & \xmark & \xmark & \cmark & \cmark & AWS \\\hline
    
    \bf Lopez et al.~\cite{ucccfo} & \cmark & \cmark & \cmark & \xmark & \xmark & \xmark & \xmark & \xmark & IBM, AWS, Azure \\\hline
    
    \bf Mahgoub et al.~\cite{wisefusesigm} & \cmark & \cmark & \xmark & \xmark & \cmark & \cmark & \cmark & \pcmark & AWS, GCP, Azure\\\hline
    
    \bf \addc{Mahgoub et al.}~\cite{mahgoub2022orion} & \cmark & \cmark & \cmark & \cmark & \cmark & \cmark & \cmark & \cmark & AWS\\\hline
    
    \bf Ristov et al.~\cite{ctwsapplied} & \cmark & \xmark & \cmark & \cmark & \cmark & \namark & \namark & \xmark & AWS, GCP, IBM \\\hline
    
    \bf \addc{Schmid et al.}~\cite{sebsflow} & \cmark & \cmark & \cmark & \cmark & \cmark & \cmark & \cmark & \pcmark & AWS, GCP, Azure\\\hline
    
    \bf Shahidi et al.~\cite{cppeiisw} & \cmark & \cmark & \xmark & \cmark & \cmark & \cmark & \cmark & \pcmark & AWS, Azure \\\hline
    
    \bf Shahrad et al.~\cite{swcosw} & \cmark & \xmark & \xmark & \cmark & \cmark & \namark & \namark & \xmark & Azure, OpenWhisk \\ \hline
    
    \bf Ustiugov et al.~\cite{baosfs} & \cmark & \xmark & \xmark & \cmark & \xmark & \namark & \namark & \xmark & Knative \\\hline
    
    \bf Wang et al.~\cite{epmicsoc} & \cmark & \xmark & \xmark & \xmark & \cmark & \namark & \namark & \xmark & Azure \\ \hline
    
    \bf Wang et al.~\cite{peak-behind} & \cmark & \xmark & \cmark & \cmark & \xmark & \namark & \namark & \xmark & AWS, Azure, GCP \\\hline
    
    \bf Wen et al.~\cite{ccsepjsep} & \cmark & \xmark & \cmark & \cmark & \cmark & \namark & \namark & \xmark & AWS, Azure, Alibaba, GCP \\\hline
    
    \bf Wen et al.~\cite{aessarx} & \cmark & \cmark & \xmark & \xmark & \xmark & \xmark & \cmark & \xmark & AWS, Azure, Alibaba, GCP \\\hline
    
    \bf Wen et al.~\cite{msswicws} & \cmark & \cmark & \cmark & \xmark & \cmark & \cmark & \cmark & \xmark & AWS, Azure, Alibaba, GCP\\\hline
    
    \bf Yu et al.~\cite{cspssocc} & \cmark & \cmark & \xmark & \cmark & \cmark & \cmark & \cmark & \pcmark & OpenFaas, AWS, FN Project\\\hline
    
    \rowcolor{CornflowerBlue!25}\bf \textit{This Article} & \cmark & \cmark & \cmark & \cmark & \cmark & \cmark & \cmark & \cmark & AWS, Azure\\\hline
    \multicolumn{10}{l}{\it \cmark~Examined, \pcmark~Partly examined, \xmark~Not examined, \namark~Not applicable since only functions are studied}
  \end{tabularx}
\end{table*}

\subsection{Serverless and FaaS Characterization Studies}
The proprietary nature of the commercial FaaS workflow platforms poses a significant impediment to understanding their performance trade-offs~\cite{bmspec}. While they provide developer documentation and performance expectations, their detailed design and operational performance insights ``in the wild'' is rather limited~\cite{swcosw}. 
Contemporary FaaS performance studies
fail to offer a principled and rigorous evaluation of FaaS \textit{workflow} platforms along key dimensions \addc{we identify} (Table \ref{tbl:cmp}).

\subsubsection{Support for FaaS Functions vs. Workflows}
This is a key distinction of our study, which goes beyond individual functions and holistically explores the behavior of workflows composed from them. Workflows have overheads of data transfers, orchestration, cascaded cold-start and scaling. 
Also, workflow costs vary for CSPs, and include non-trivial data transfer and orchestration costs.

Scheuner and Leitner~\cite{faaspemlr} review articles between $2016$ and $2019$ and highlight research on the performance of FaaS functions on AWS Lambda. Our work extends to detailed first-hand performance and cost analysis of FaaS workflows on AWS and Azure.
Microsoft Azure~\cite{swcosw} characterize production workloads on Azure Durable Functions using the default storage provider (AzS). They analyze the function's trigger types, invocation frequencies, patterns and resource needs. 
\modc{$40\%$ of FaaS applications have $2$--$10$ functions with $75\%$ of these executing within $10$s.} The benchmark workflows in our study fall within this spectrum. \modc{As a CSP, their paper optimizes their platform; we take a developer's view in understanding the platforms and optimizing workflows.}

\subsubsection{Scaling Properties}
Rapid scaling is a unique benefit of FaaS.
So, it is important to understand the responsiveness of FaaS workflow platforms to dynamic workloads.
Wang et al.~\cite{peak-behind} and 
Ristov et al.~\cite{ctwsapplied} analyze the scaling of functions \repc{in managing}{to} sudden bursts of traffic.
This is extend to FaaS workflows~\cite{msswicws}, reporting AWS as the best; this matches our observations.
\modc{Our study performs a more rigorous evaluation of the workflow scaling
and concurrency levels on AWS, AzS and AzN, and the consequent response times. We also analyze billed costs with a cost model.}

\subsubsection{Cold-start Overheads}
Cold-start
is the overhead paid for container initialization when scaling out in response to increased request rates to a function.
These effects can cascade in workflows, and form an important metric for our analysis~\cite{ctwsapplied}.
Shahrad et al.~\cite{swcosw} and Wang et al.~\cite{peak-behind} show that cold-starts in AWS and Azure cause considerable delays in function execution. 
\modc{They also report Azure to have a high cold-start latency, which
we too observe.}
Our study analyzes the impact of these cascading overheads on workflows, 
and disaggregate function cold-starts and the inter-function data transfer overheads for a deeper insight on the variability of workflow executions.

\subsubsection{Workflow Latencies}
The end-to-end (E2E) execution time for workflows is an important user-facing metric.
Understanding the effect of workflow patterns -- number of functions, workflow length and number of fan-outs/fan-ins -- on latency is critical~\cite{wisefusesigm}.
The E2E time also includes overheads such as data transfers between adjacent functions and workflow coordination. 
Li et al.~\cite{aossiseke} examine function performance for Knative and OpenFaaS but lack an analysis of these additional latencies. 
Shahidi et al.~\cite{cppeiisw}, 
Yu et al.~\cite{cspssocc} and Mahgoub et al.~\cite{wisefusesigm} identify orchestration overheads and inter-function latencies as key factors of E2E latencies in AWS, Azure and OpenFaaS but
do not analyze the impact of workflow structure, unlike us.
Data transfer latencies can affect the E2E latency, depending on the payload sizes.
Mahgoub et al.~\cite{wisefusesigm} and Shahidi et al.~\cite{cppeiisw} suggest that inter-function overheads is high and AzS is particularly bad. We extend this comparison to Azure's recent Netherite (AzN) and see better performance.
We also establish the reasons for inter-function overheads
based on the underlying system design, advancing prior research.

\subsubsection{Monetary Cost Analysis}
Lastly, 
the cost to execute the workflow is a vital metric for users but understudied. 
While this is simple for functions that are billed based on the memory usage and function invocation time (GB-s),  workflows have complex execution and also opaque costing.
Data transfer and coordination overheads are billed separately from function executions, and in fact dominate. 
Wen et al.~\cite{ccsepjsep} offers a cost breakdown for AWS, Azure, and Alibaba functions, but falls short on workflow insights. 
Eismann et al.~\cite{pcswicpe} and Mahgoub et al.~\cite{wisefusesigm} both explore cost models for FaaS workflows for AWS and GCP, but fail to include 
data transfer costs that form the bulk of the cost.
Lin et al.~\cite{moptpds} provides a cost analysis for AWS workflows, which is relatively simple and matches our model. 
Cost modeling for Azure is challenging since many granular moving parts that are billed separately.

\subsection{Serverless and FaaS Benchmarks}
Several FaaS and serverless benchmark \modc{suites have} been proposed.
However, \modc{many focus on specific aspects of performance and do not undertake a detailed cross-platform study.}
SeBS~\cite{copik2021sebs} FaaSdom~\cite{faasdom}, \modc{and CrossFit~\cite{crossfit} provide microbenchmarks to measure system parameters such as CPU utilization, memory, disk I/O and network characteristics. FunctionBench~\cite{fbench} and SeBS~\cite{copik2021sebs} also include real-world functions and application workloads, with SeBS also providing a workflow-focused extension, SeBS-Flow~\cite{sebsflow}.}
 BeFaaS~\cite{befaas} and Deathstar~\cite{deathstar} are application-centric benchmarks with complex interactions. 
BeFaaS, however, offers only a single eCommerce application and performs a limited empirical study with it, while Deathstar~\cite{deathstar} is more for long-running serverless applications rather than FaaS workflows.

\begin{figure*}[t!]
\centering%
  \subfloat[AWS Lambda Step Functions]{
    \includegraphics[width=0.23\textwidth]{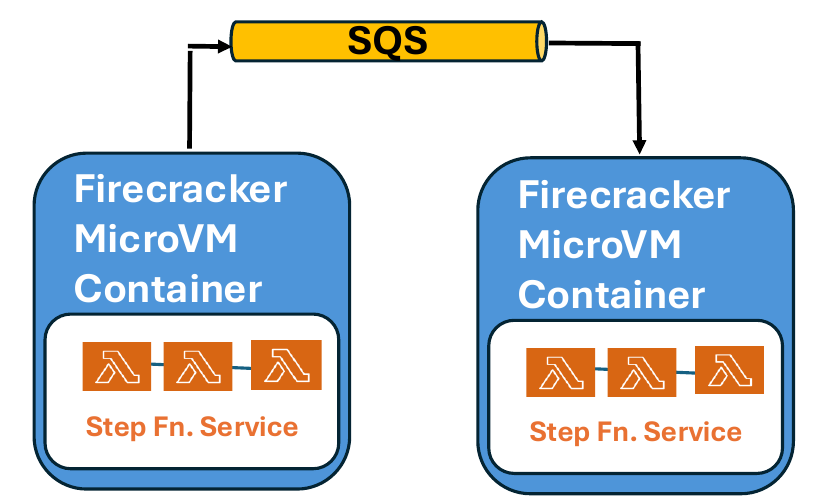}%
    \label{fig:faas-arch:aws}
  }~~
    \subfloat[Azure Durable Functions--Storage (AzS)]{
   \includegraphics[width=0.35\textwidth]{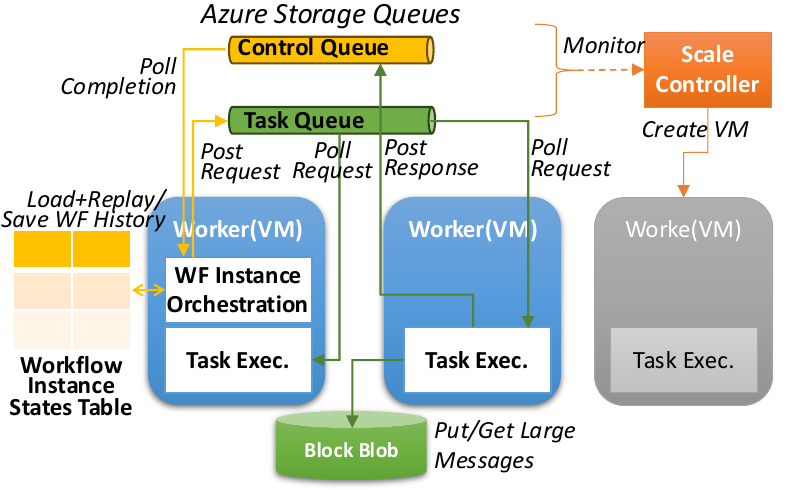}%
    \label{fig:faas-arch:azs}
  }~~
  \subfloat[Azure Durable Functions--Netherite (AzN)]{\includegraphics[width=0.35\textwidth]{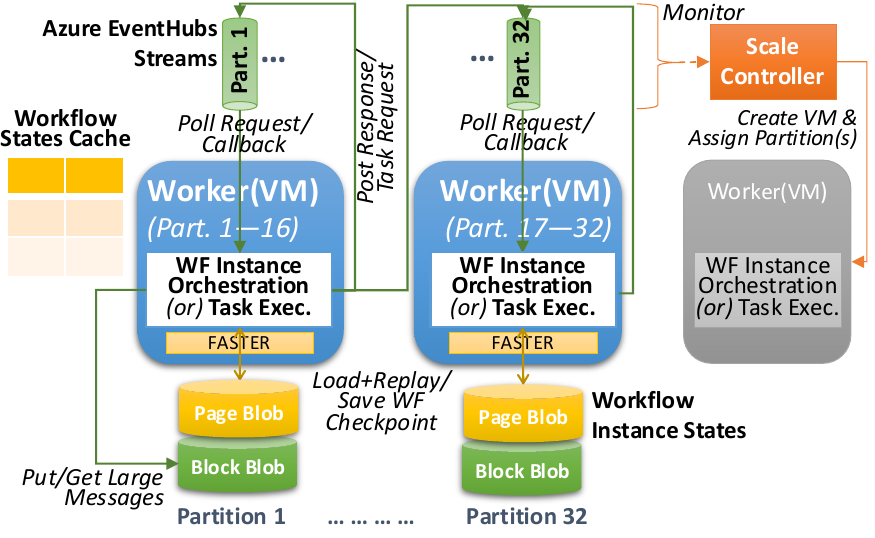}%
    \label{fig:faas-arch:azn}
  }%
\caption{Architecture of AWS (inferred) and Azure AzS and AzN FaaS Workflow Platforms.}
\label{fig:faas-arch}
\end{figure*}

Our recent work, XFBench~\cite{xfbench}, proposes a novel FaaS workflow benchmark that is deployed using our XFaaS multi-cloud FaaS framework~\cite{xfaas}. But it only validates the workflow suite rather than perform a detailed analysis. \modc{We use these prior works to perform an extensive characterization.}
\addc{Schmid et al.~\cite{sebsflow} propose SeBS-Flow, another recent work that extends from SeBS~\cite{copik2021sebs} to benchmark serverless workflows. They propose six real-world applications and four micro-benchmarks capturing diverse computational and coordination patterns. 
This is used to systematically evaluate performance, cost, scalability and runtime deviations across AWS, AzS and GCP. While they offer realistic application workflows and their performance analysis, our work complements it by focusing on detailed empirical characterization of hybrid-cloud deployments, analyzing fine-grained latency--cost tradeoffs, and exposing CSP-specific variability not captured in SeBS-Flow, coupled with insights on designing better FaaS workflows. We leverage their Trip Booking workflow as one of the realistic applications in our benchmark.}

\subsection{Research on Serverless and FaaS}
\addc{While optimizing FaaS or FaaS workflow platforms is not a direct goal for our article, our characterization study can be leveraged for future optimizations.} There is already a large body of work on 
optimizing FaaS platforms, but often for private clouds~\cite{lass,faastlane}; public FaaS platforms can only be optimized by the CSP, such as reported in Netherite~\cite{netherite}.

\addc{Complementary studies have focused on 
workflow-level bottlenecks. 
Mahgoub et al.~\cite{mahgoub2022orion} identify key performance factors in serverless DAGs and propose right-sizing, bundling, and prewarming techniques to mitigate end-to-end latency variability. 
Yu et al.~\cite{yu2023following} advocate a data-centric approach to workflow orchestration, 
and trigger primitives to coordinate executions efficiently.
FaaSFlow~\cite{faasflow2022} improves workflow efficiency via a dataflow model and adaptive storage to minimize data movement overheads. This targets architectural efficiency within a controlled system.}
\addc{A recent survey~\cite{serverless_survey_csur} provides a comprehensive taxonomy of serverless system layers and highlights open challenges in orchestration and coordination.
}

\addc{Our empirical findings complement  these system-level enhancements by exposing platform behavior insights that offer opportunities for adaptive scaling, scheduling, and deployment strategies across heterogeneous and wide-used public CSP environments.}
We limit our study to public CSPs because they are widely adopted compared to FaaS in private clouds~\cite {swcosw}.

\section{AWS and Azure FaaS Workflow Platforms}\label{sec:bg}%

\modc{Amazon’s \textit{AWS Lambda} and Microsoft’s \textit{Azure Functions} are two widely used FaaS offerings. Their workflow counterparts are \textit{AWS Step Functions} and \textit{Azure Durable Functions}. Azure provides two storage backends: the default \textit{Azure Storage (AzS)} and the newer, higher-performance \textit{Netherite (AzN)}~\cite{mnetherite}.}
\addc{This section summarizes their essential orchestration design that influences latency, cost, and scalability.}
\subsection{Development and Deployment of FaaS Workflows}
\modc{Developers define individual functions and compose them into a workflow DAG. AWS uses a YAML state-machine file, whereas Azure requires user-written orchestration code.}
\modc{At deployment, users specify the cloud region and the trigger type (HTTP, REST, or message).}
\modc{Functions remain \textit{cold} until the first invocation creates an execution container -- Firecracker microVMs on AWS or worker VMs on Azure. This initialization is the \textit{cold-start} overhead; later invocations may reuse existing containers.}
\modc{Both platforms use proprietary orchestration logic to determine downstream function execution, handle parameter passing, and manage scale-in/out actions.}

\subsection{AWS Step Functions Orchestration}

\modc{AWS Step Functions model workflows as state machines whose states correspond to Lambda functions. 
Fig.~\ref{fig:faas-arch:aws} reflects design behavior inferred from \addc{limited }documentation and prior works~\cite{peak-behind,ctwsapplied,aws-containers-on-demand,awssqs,aws-step-fns}.}
\modc{Each function executes inside a Firecracker container under a selected runtime (e.g., Python, Java, Node.js). The Step Functions service coordinates transitions and data exchange among functions.}
\modc{In our experiments, workflows are triggered asynchronously through REST APIs.}
\modc{Data exchange between functions likely relies on Amazon SQS, consistent with the 256KB payload limit.}
\modc{AWS Lambda spawns a new container when all warm instances are busy, loading the image from S3.\footnote{\href{https://docs.aws.amazon.com/lambda/latest/dg/lambda-concurrency.html}{Understanding Lambda function scaling}, AWS, 2024} CPU allocation scales with configured memory.}
\modc{Lambda can handle up to $15,000$ concurrent cold starts per second with sub-50\,ms latency. Each container can serve up to 10 executions per second, with an account-level concurrency cap of $1000$~\cite{aws-containers-on-demand}.}
\addc{Prior work indicates Lambda uses bin-packing for container placement~\cite{peak-behind}.}
\modc{Idle containers are reclaimed after $\approx5$~mins to control resource usage~\cite{aws_timeout}.}

\subsection{Azure Durable Functions Orchestration}
\modc{A workflow in Azure Durable Functions is composed through a user-defined \textit{orchestrator} code that coordinates multiple \textit{activity} functions. These are implemented in Python, .NET, or JavaScript and run as processes within worker VMs (2 vCPUs, 1.5GB RAM).}
\modc{Azure provides two orchestration runtimes: with default Storage Provider (AzS) and Netherite Storage provider (AzN)~\cite{netherite}.}

\subsubsection{Azure Default Storage Provider (AzS)}
\label{sec:azs}
\modc{AzS (Fig.~\ref{fig:faas-arch:azs}) orchestrates executions via a \textit{task hub} comprising of four storage entities: (i) Azure Tables (history), (ii) Control Queues (orchestration state), (iii) Activity Queues (inputs $\leq$ 64 KB), and (iv) Blob Storage (large payloads).}
\modc{The orchestrator enqueues tasks, receives results through the control queue, and updates the workflow state in the history table.}
\modc{A Scale Controller monitors queue lengths and adjusts the number of VMs (max 100 per-workflow), removing idle VMs after 5~mins.\footnote{\href{https://learn.microsoft.com/en-us/azure/azure-functions/event-driven-scaling?tabs=azure-cli}{Event-driven scaling in Azure Functions}, Microsoft 2024}}
\addc{Scaling is at the VM level, shared by multiple functions, unlike AWS where functions scale independently.}

\subsubsection{Netherite Storage Provider (AzN)}
\label{sec:bg-neth}

\label{sec:bg-neth}

\modc{AzN (Fig.~\ref{fig:faas-arch:azn}) replaces queues and tables with optimized services -- \textit{Azure Event Hubs} for messaging and \textit{Page and Block Blobs} for durable state -- overcoming AzS's performance limitations~\cite{netherite}.}
\modc{Orchestration state is partitioned across multiple logical partitions, each managed by one or more worker VMs. Event Hubs mediate inter-partition communication, while Page and Block Blobs persist orchestration and activity data using FASTER-based indexing for efficient I/O.}
\modc{A Scale Controller dynamically adjusts partitions and workers (max 32).\footnote{\href{https://github.com/microsoft/durabletask-netherite}{Durable Task Netherite}, Microsoft, 2024}}
\modc{While finer details are omitted, AzN has lower orchestration overheads and better parallelism than AzS~\cite{netherite}.}

\section{FaaS Workflow and Workload Suite}\label{sec:xfw}%

\begin{table}[t!]
    \caption{Workflow Functions and Input Sizes: \underline{S}mall, \underline{M}edium, \underline{L}arge}
    \label{tbl:workflows:desc}
    \includegraphics[width=\columnwidth]{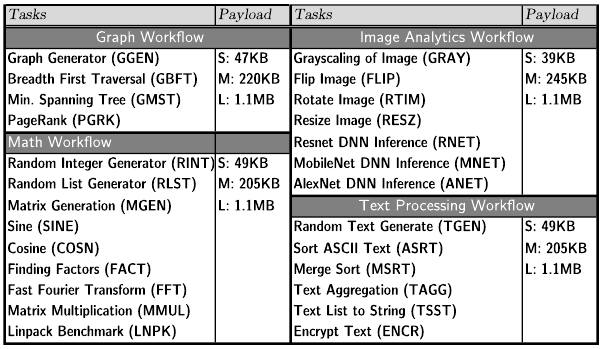}
\end{table}

\subsection{Functions and Workflows}

\modc{
Recent works such as XFBench~\cite{xfbench} and SeBS-Flow~\cite{sebsflow} propose benchmarking of FaaS workflows.
We build upon XFBench to perform a detailed characterization of public-cloud workflow platforms, with emphasis on sustained execution behavior and architectural causes of performance differences.}
XFBench~\cite{xfbench} assembles popular functions from literature~\cite{faasdom,fbench,copik2021sebs,crossfit},
along with our own, for a comprehensive suite of \textit{$26$ FaaS functions} (Table~\ref{tbl:workflows:desc}). These cover compute-intensive numerical operations, graph algorithms with irregular access patterns, text processing of Enterprise data, and DNN/ML models for image analytics and inferencing.

\begin{figure}[t]
\centering%
    \includegraphics[width=1\columnwidth]{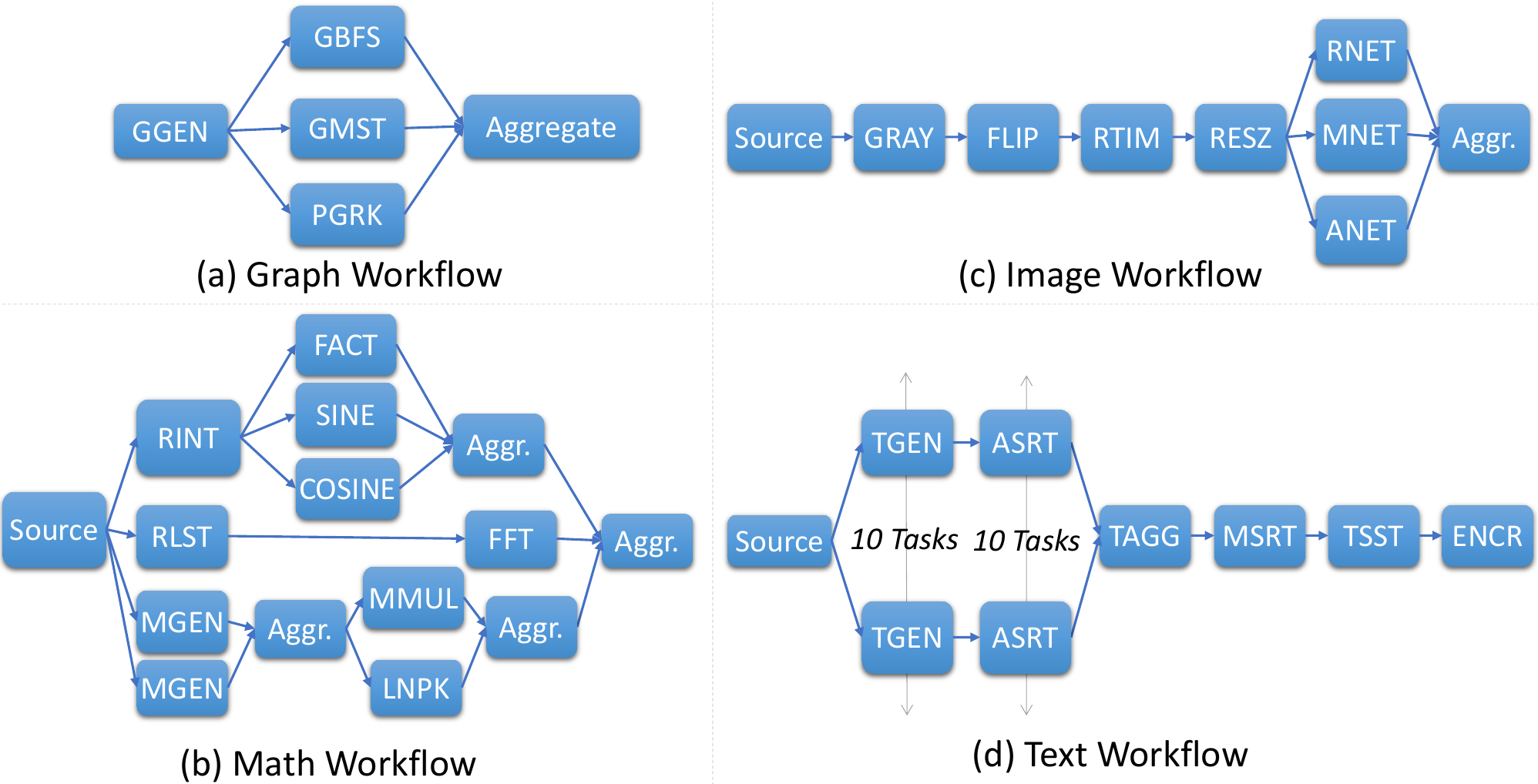}
    \caption{Workflow DAG Compositions from XFBench used in our study.}
    \label{fig:workflows:dag}
\end{figure}

\modc{These functions are composed into two synthetic (\textit{Graph processing} and \textit{Math computation}) and two realistic (\textit{Image analytics} and \textit{Text processing})} workflows with diverse interaction patterns (Fig.~\ref{fig:workflows:dag}).
The workflows also differ in their complexity: number of tasks ($5$--$25$), length of critical path ($3$--$7$), fan-ins and fan-outs ($2$--$20$), nested task parallelism ($0$--$2$), and take between $1$--$900$s to run, depending on the workload. These align with the workflow sizes in Azure deployments~\cite{swcosw}. \modc{These workflow patterns 
affect the data flowing between functions, concurrent tasks executed, control logic/state transitions, etc., all of which affect the performance, scalability, overheads, and costs of the FaaS workflow platforms.}

\begin{figure}[t]
\centering%
    \includegraphics[width=0.85\columnwidth]{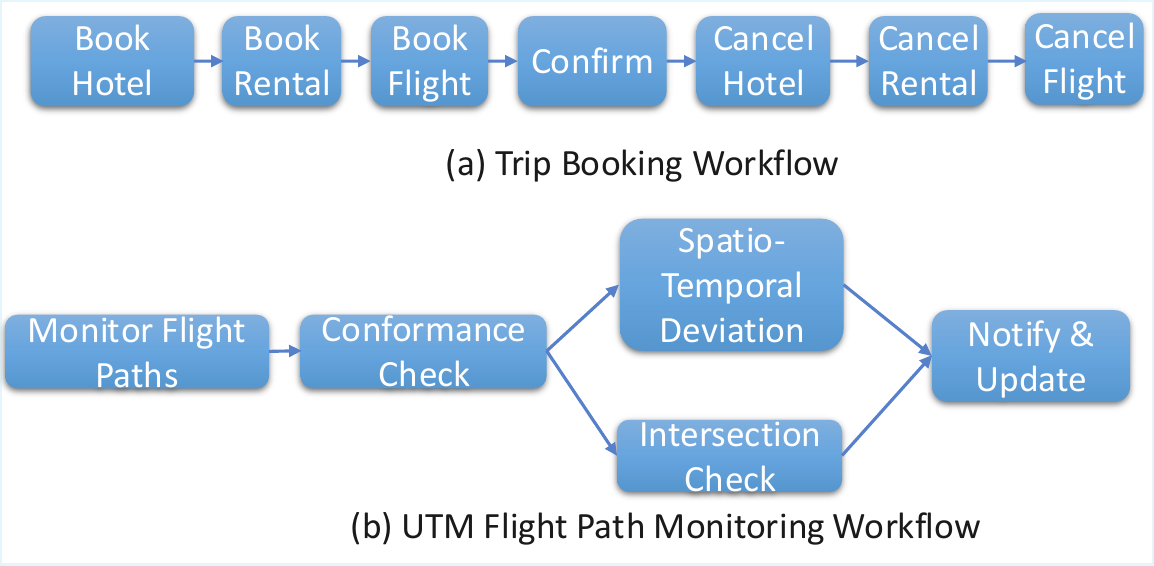}
\caption{Trip Booking and Drone UTM Application Workflows.}
\label{fig:new-wfs}

\end{figure}

\addc{We complement these with two more application-oriented workflows (Fig.~\ref{fig:new-wfs}). The \textit{Trip Booking Workflow} from SeBS-Flow~\cite{sebsflow} is a Saga-style transactional workflow that models the booking and cancellation of hotel, flight, and rental services with compensation logic. It uses native NoSQL stores to store the trip details: DynamoDB for AWS and CosmosDB for Azure.
Our proposed \textit{UTM Flight Path Monitoring Workflow} is inspired by the FAA's Unmanned Aircraft System Traffic Management (UTM)~\cite{UTM} application.
It models real-time drone flight management by registering drone flight plans, checking their flight-path conformance, and analyzing real-time in-flight spatio-temporal telemetry for deviations before issuing notifications. It uses a VM-hosted \textit{PostgreSQL} database with tables that store geo-restrictions, planned trajectories, and flight paths.
These real-world workflows help confirm the performance characteristics observed in the synthetic ones and include novel functionality, such as database calls.
}

\subsection{Workloads}

Workloads determine the clients' workflow invocation patterns. We consider three factors: the \textit{size of the input payload} to the workflow, which has cascading effects on the size of downstream messages; the \textit{Requests per Second (RPS)} sent to a workflow, and \textit{Dynamism} in the RPS across time. We combine these to imitate real-world execution loads~\cite{swcosw}. 

We use \textit{three payload sizes} for each workflow: Small, Medium, and Large (Table~\ref{tbl:workflows:desc}). We use three \textit{static request rates} for coverage: $1$~RPS, $4$~RPS, and $8$~RPS. While these are the rates to the initial function, fan-outs and fan-ins based on the workflow structure can increase the RPS on downstream functions. E.g., the text workflow executes $80$ concurrent \texttt{TGEN} downstream functions for an $8$~RPS workflow input. Lastly, we use three \textit{dynamic input rates}: Step, Sawtooth, and ``Alibaba''. \textit{Step} sends the following RPS for $2mins$ each: $[1, 2, 4, 8, 4, 2, 1]$, slowly scaling up/down. \textit{Sawtooth} has a sharper rise with these RPS for $10s$ each: $[1, 2, 3, 4, 5, 6, 7, 8]$, and then drops to $0$~RPS for $60s$; this repeats thrice. Lastly, we scale and use the real-world \textit{Alibaba} micro-services trace~\cite{simmhan2024tcc} (Fig.~\ref{fig:graph-alibab-scaling}, red line), which has a peak of $17$~RPS. 
\addc{Given its specialized nature, we construct a novel workload for the UTM workflow by combining real-world 24-hour flight arrival and departure data for Bangalore (ICAO: VOBL) from OpenSky Network, 
and pair each request with a unique synthetic UAV trajectory (Fig.~\ref{fig:utm-timeline}).}
All workloads run for $5mins$, except Step that runs for $14mins$, \addc{and the UTM that runs for $10mins$.}

These are representative of enterprise, IoT, and inferencing workloads. e.g., in IoT scenarios~\cite{riotbench} the workloads may be bursty (sawtooth) when sensors periodically send events with a similar frequency while eCommerce and user-facing workloads may show a gradual diurnal pattern (step).

\section{Experiment Setup}\label{sec:exp}%

\subsection{Benchmarking Harness}
We use our open-source XFaaS framework~\cite{xfaas} to implement the above functions and workflow DAGs in Python in a CSP-agnostic manner. XFaaS automatically generates each CSP's native Python function wrappers with additional code to monitor function and workflow execution times, and translates the workflow DAG JSONs to the CSP's workflow definition language. 
XFaaS also deploys these to specific CSP regions.
XFBench invokes the workloads on the deployed workflows using the Apache JMeter v5.6.2 load testing tool. All the workflows are HTTP-triggered. 

In addition to the workflows and workload above, we also run micro-benchmark functions and workflows to evaluate specific characteristics of FaaS workflow platforms.
\addc{Artifacts to reproduce the benchmarking study are hosted at \url{https://github.com/dream-lab/FaaS-WF-Char}.}

\subsection{Cloud FaaS Platform Setup}

We refer to AWS Step Functions, Azure Durable Functions using the default Azure Storage Provider, and the more recent Azure Durable Functions using Netherite
as \textbf{AWS}, \textbf{AzS} and \textbf{AzN}.
\addc{While AWS and Azure also provide premium options, e.g., AWS Lambda's Provisioned Concurrency and Azure Functions' Premium Plan, 
we evaluate only the default plans to understand their standard execution behavior and costs.}

We use regions: \texttt{ap-south-1} for AWS, and \texttt{centralindia} for Azure; selected experiments are run in US-East and Central Europe. 
JMeter clients run on an overprovisioned VM in the same region as the workflows to avoid wide-area network latency and performance overhead. We use Standard D8s v3 VM with $8$ vCPUs and $32$GB RAM for Azure and EC2 C5.4X Large instance with $16$ vCPUs and $32$GB RAM for AWS.

We use the system defaults for all FaaS workflow platforms unless noted otherwise. E.g., the \texttt{maximum partition count} (PC) in AzN is $12$, the \texttt{max\-Concurrent\-Activity\-Functions} (MA) in AzS is $10$, and the container timeout for both AWS and Azure are $5mins$.
As discussed in~\ref{ta:scaling:complex} we change some defaults of AzS and AzN to ensure that most of the workflows and workloads can run. For AzS we change the \texttt{max\-Concurrent\-Orchestrator\-Functions} (MO) to $8$ and MA to $1$, and for AzN we change the PC to $32$. 
AWS Step Functions have no such knobs to tune. 
The container memory in AWS is set to $128MB$ for most functions; we use $256MB$ for Alexnet and $512MB$ for Resnet to ensure the models fit.
Azure does have a setting at a function level.
We use the default \textit{Consumption Plan}, which assigns $1.5GB$ memory and $1$ vCPU per worker VM. \addcminor{Most experiments were conducted between September and November 2024. Additional experiments corresponding to Figures~\ref{fig:summary-aws-knobs},~\ref{fig:utm-timeline},~\ref{fig:saga-wfs}, and~\ref{fig:cost-analysis} were performed in November and December 2025 using the publicly available versions of the respective cloud platforms at that time.}

\subsection{Metrics}

We measure, report, and analyze key performance metrics from various logs in this characterization study.
For each workflow invocation, we give the \textit{End-to-End (E2E) latency} or \textit{makespan}, which is the time between the start of the first function execution to the end of the last function execution along the critical path of that workflow. We also split this into the \textit{function execution times} along the critical path and the \textit{inter-function times}. These are measured by XFaaS using telemetry code in the function logic and using workflow ID/function IDs in the inter-function message headers.
The time between the upstream function completing and the downstream function initiating is the \textit{inter-function time}.

We also collect custom logs for our analysis, e.g., tracking the container IDs, CPU architectures, AWS Cloudwatch logs, and costing from the AWS and Azure portals.
These help track \textit{cold-start overheads} 
and the scale-in/-out of containers,
in relevant analyses. 
For AWS, we get a count of the concurrent containers indirectly through the number of concurrent executions since only one executes per container, reported every $1min$.\footnote{\href{https://docs.aws.amazon.com/lambda/latest/dg/monitoring-concurrency.html}{Lambda Monitoring concurrency}, AWS, 2024} \addc{
These are validated using direct measurements based on unique invocation identifiers and cold-start flags embedded in the function state~\cite{copik2021sebs,sebsflow}. 
We use a similar approach for Azure.} 
\modc{Our custom logger records the container ID from the function's state and timestamps per execution to identify a count of unique workers and executions at any time for a workflow.}
For AWS, invocations that exceed the $95^{th}$ percentile of E2E execution time are classified as a cold-starts; all AWS runs have stable and reproducible E2E times and this way of detecting cold-starts suffices. For AzS and AzN, we detect the new containers on seeing a new container ID for an invocation.

\section{\modc{Empirical Results and Analysis}} \label{sec:results}

\subsection{Scaling with Input Rates }
\label{sec:results:scaling}

A key defining feature of FaaS is its ability to scale the number of FaaS instances when the input load is variable, while maintaining stable latency.
We examine if these properties extend to FaaS workflows and provide a stable E2E time.

\takeaway{Azure Durable Functions (AzS and AzN) are sensitive to configurations that affect concurrency and scaling, requiring tuning to achieve stable performance.
\label{ta:scaling:complex}}

\begin{figure}[t!]
\centering%
\subfloat[AzS Default \textit{(MO=5, MA=10)}]{%
    \includegraphics[width=0.5\columnwidth]{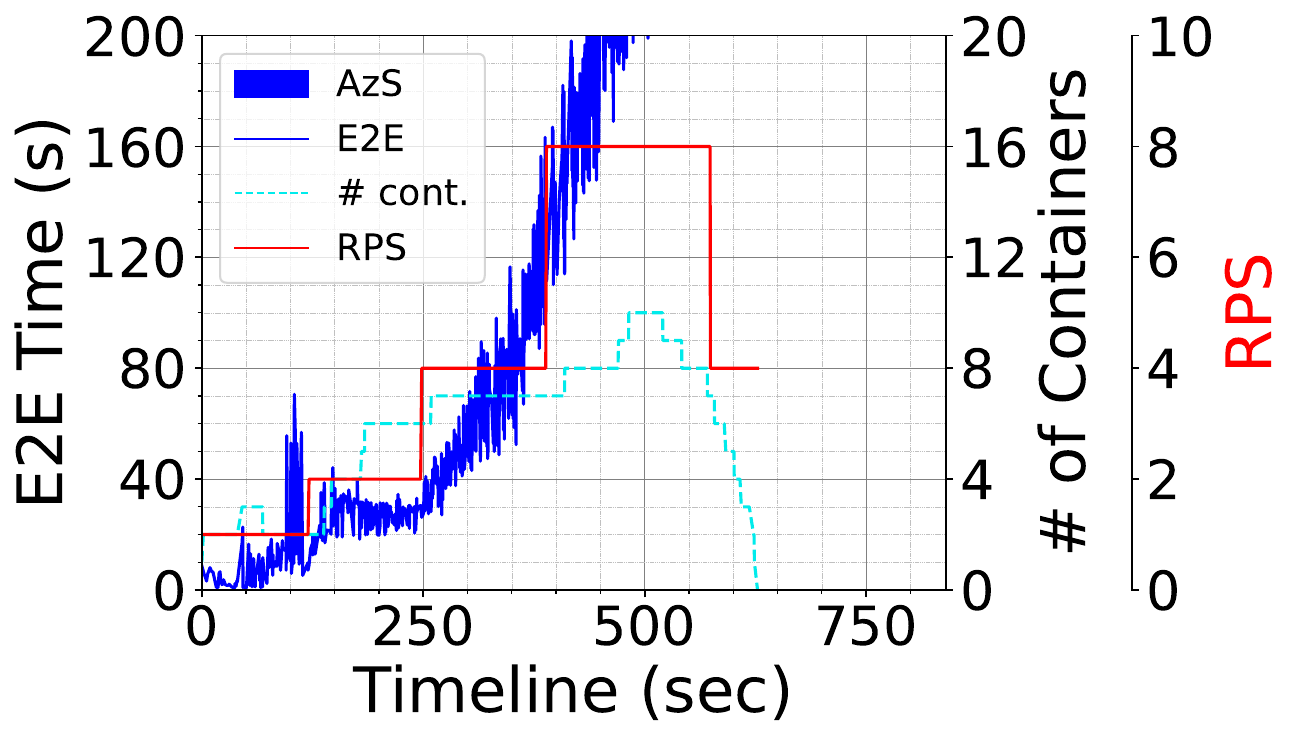}%
    \label{fig:graph-az1-step-knobs:mo5-ma10}%
  }%
  \subfloat[AzS Custom \textit{(MO=8, MA=1)}]{%
    \includegraphics[width=0.5\columnwidth]{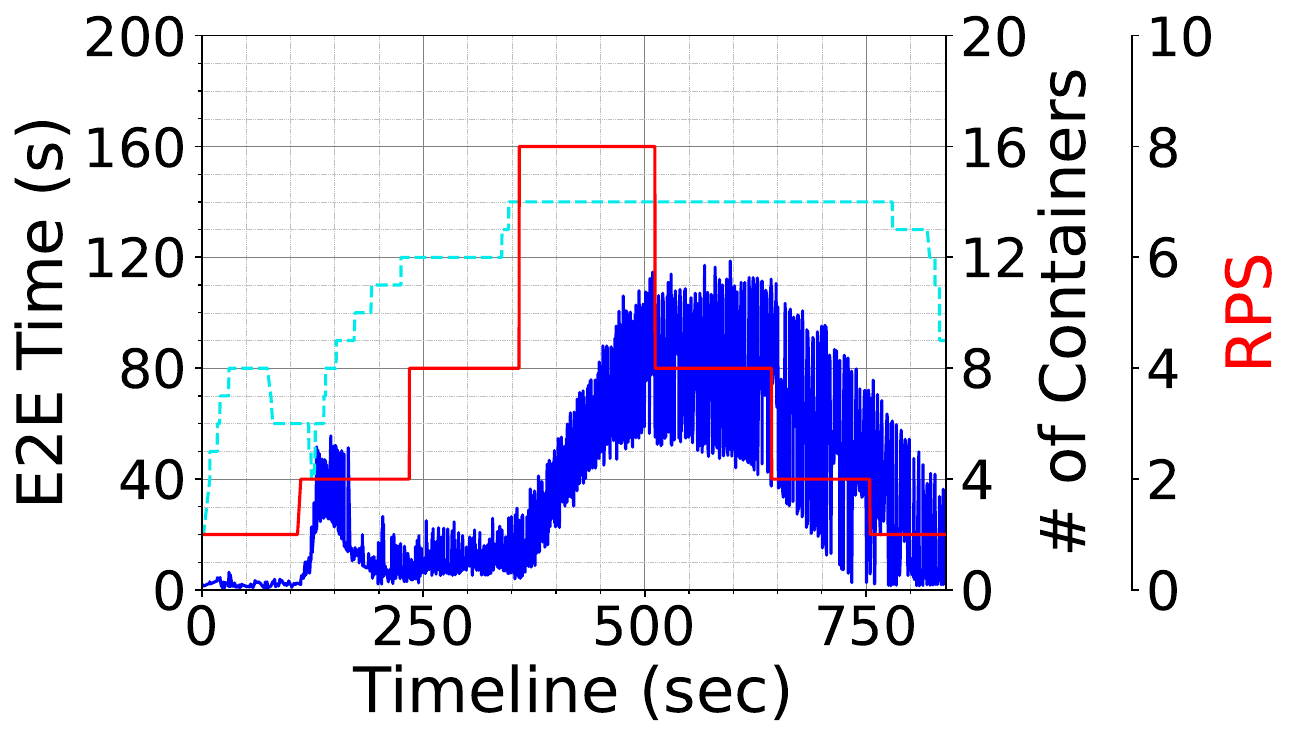}%
    \label{fig:graph-az1-step-knobs:mo8-ma1}%
  }%
\caption{AzS for \textit{Graph} Workflow with \textit{Medium} payload size and \textit{Step} Workload, using different $MO$ and $MA$ configurations.}
\label{az1-step-knobs}
\end{figure}

\begin{figure}[t!]
\centering%
  \subfloat[Default, \textit{PC=12}]{%
     \includegraphics[width=0.245\textwidth]{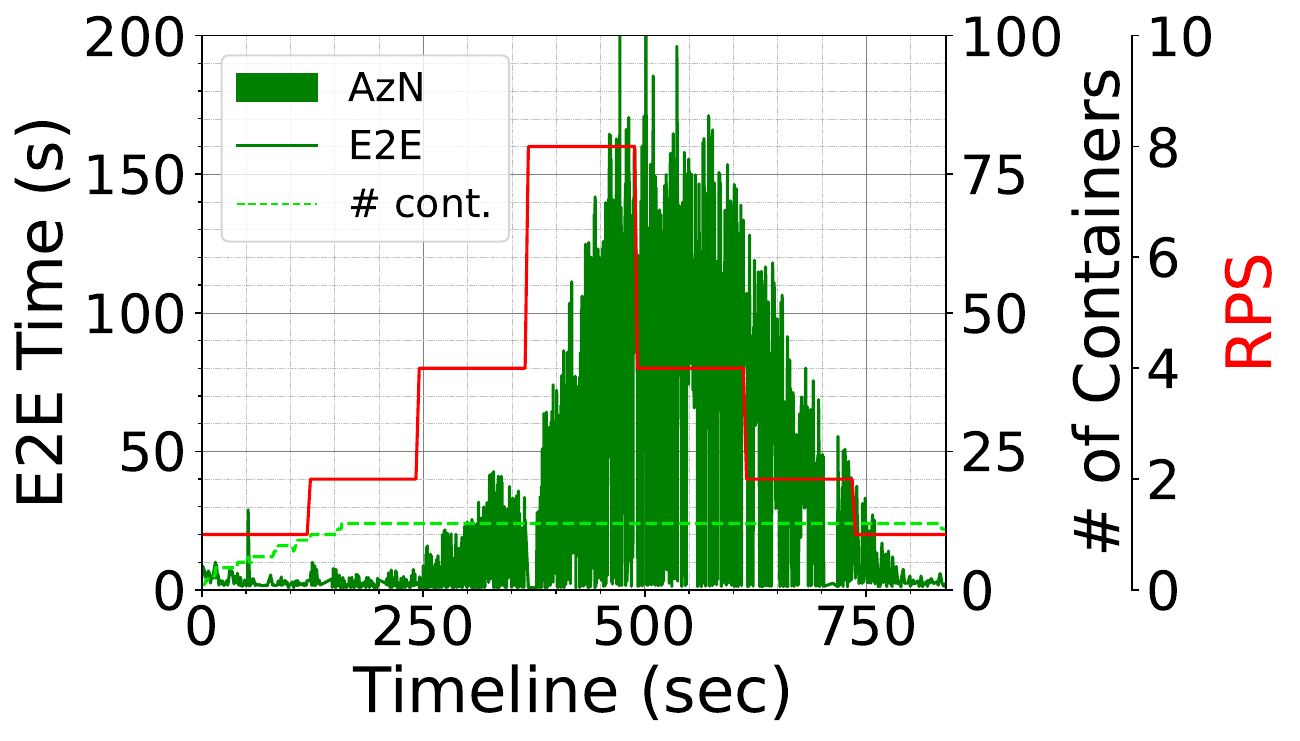}
    \label{fig:graph-az2-step-knobs:mo8-ma1-pc12}%
  }
  
  \subfloat[Article, \textit{PC=32}]{%
    \includegraphics[width=0.245\textwidth]{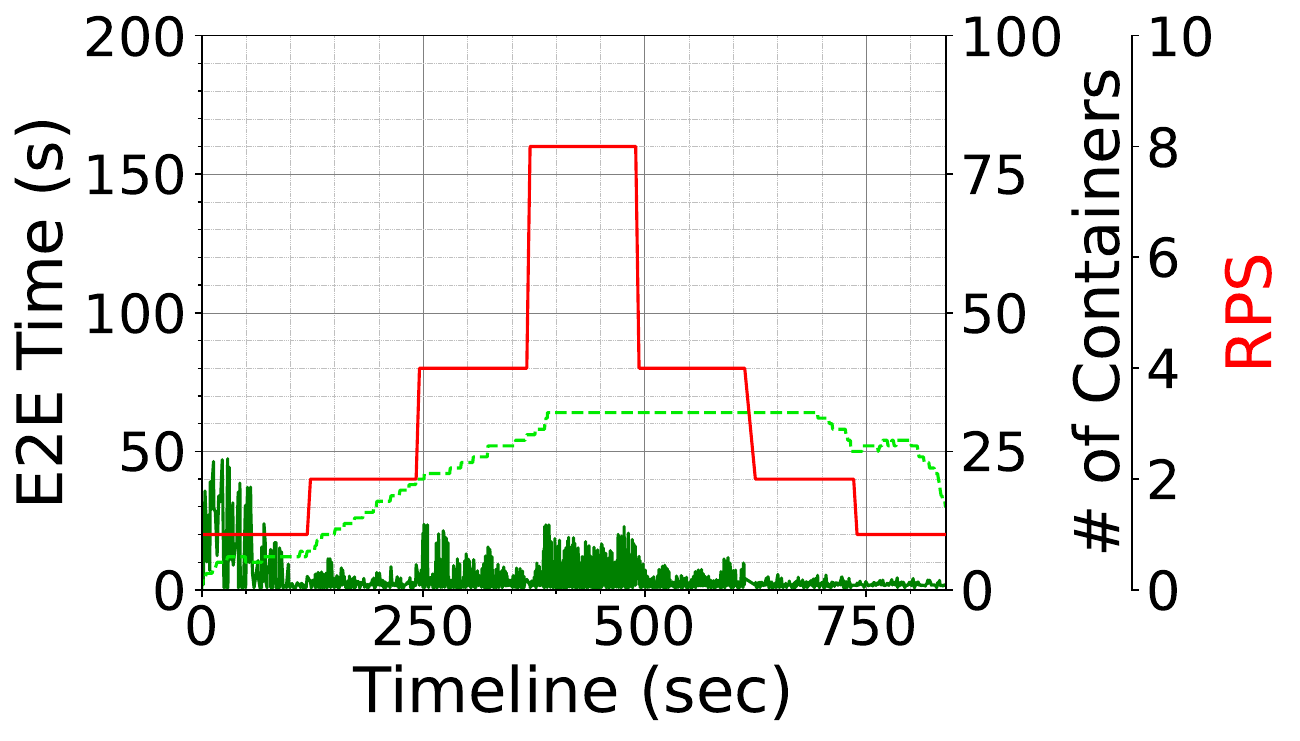}
    \label{fig:graph-az2-step-knobs:mo8-ma1-pc32}%
  }
\caption{AzN for \textit{Graph} Workflow with \textit{Medium} payload size and \textit{Step} Workload, using two max. partition counts (PC) configurations.}
\label{az2-step-knobs}
\end{figure}

\begin{figure}[t!]
\centering%
    \includegraphics[width=0.7\columnwidth]{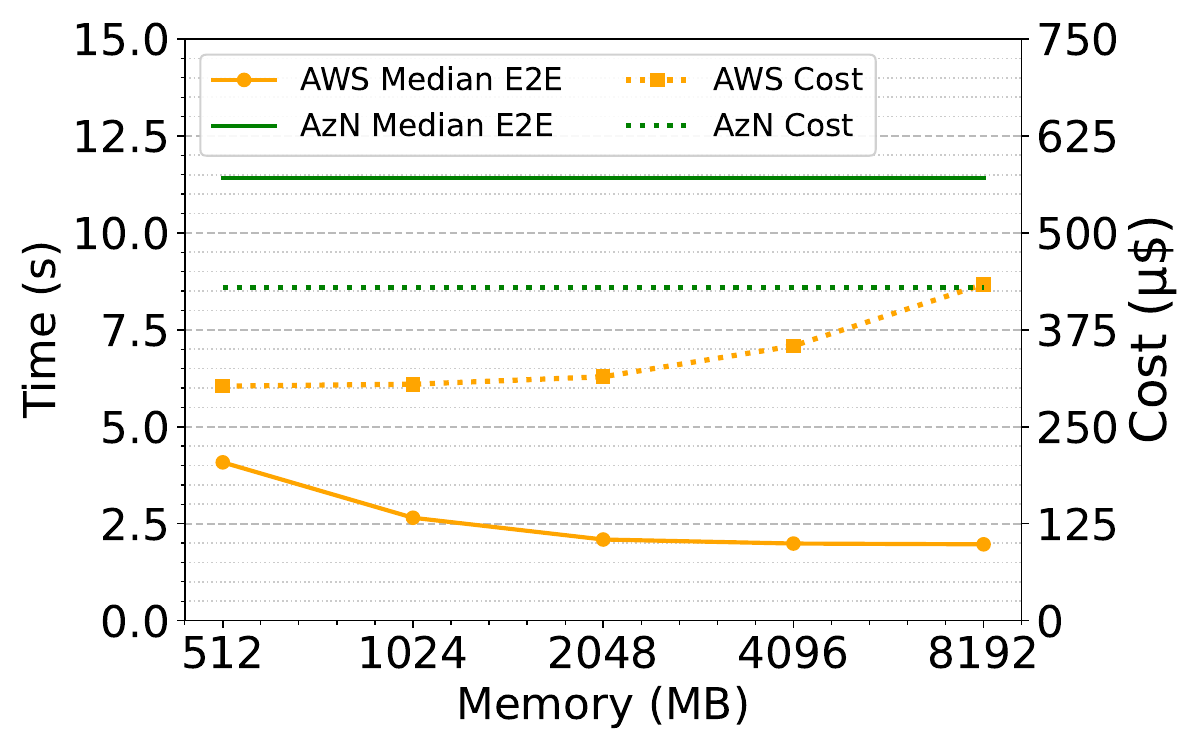}
\caption{\addc{E2E time (left Y axis) and per-invocation cost (right Y axis) as AWS Lambda memory increases, for Image WF, Medium Payload, 1 RPS.}}

\label{fig:summary-aws-knobs}

\end{figure}

\begin{figure*}[t!]
\centering%
  \subfloat[Static 1 RPS]{%
    \includegraphics[width=0.3\textwidth]{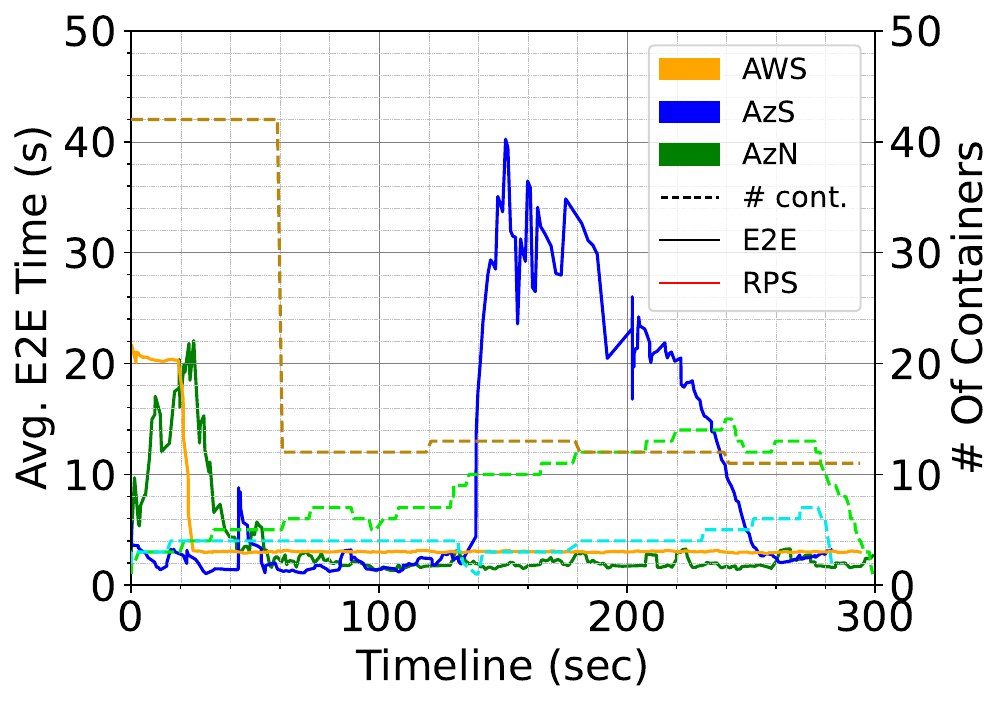}
    \label{fig:graph-1-scaling}%
  }\hfill
  \subfloat[Static 4 RPS]{%
    \includegraphics[width=0.32\textwidth]{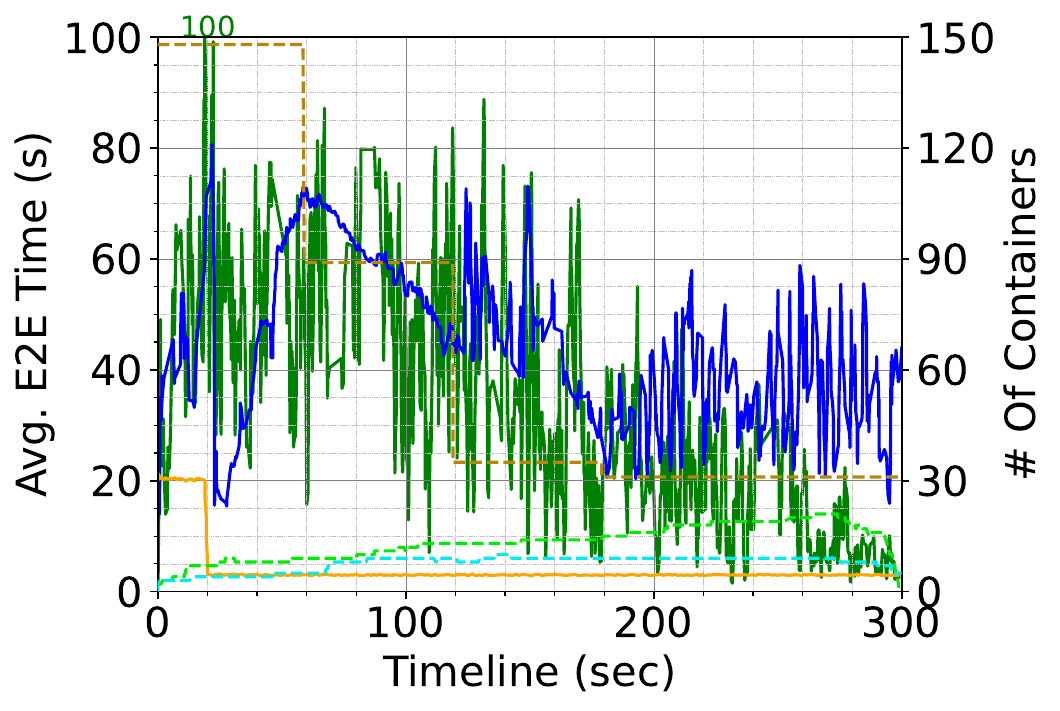}
    \label{fig:graph-4-scaling}%
  }\hfill
  \subfloat[Static 8 RPS]{%
    \includegraphics[width=0.32\textwidth]{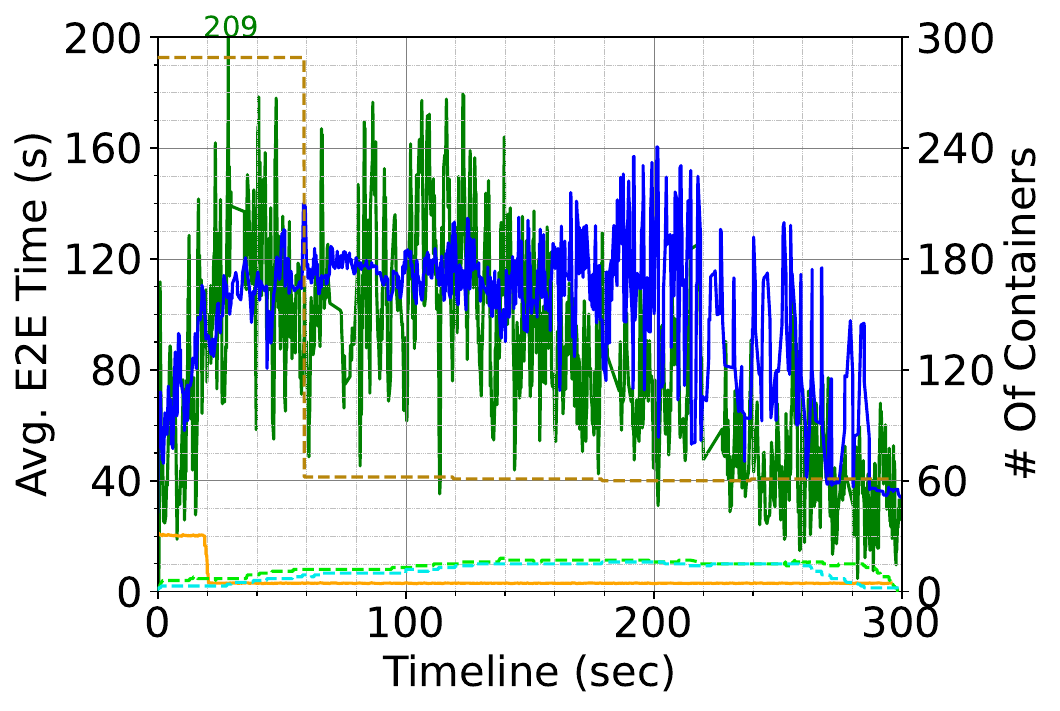}
    \label{fig:graph-8-scaling}
  }\\
  \subfloat[Step]{%
    \includegraphics[width=0.32\textwidth]{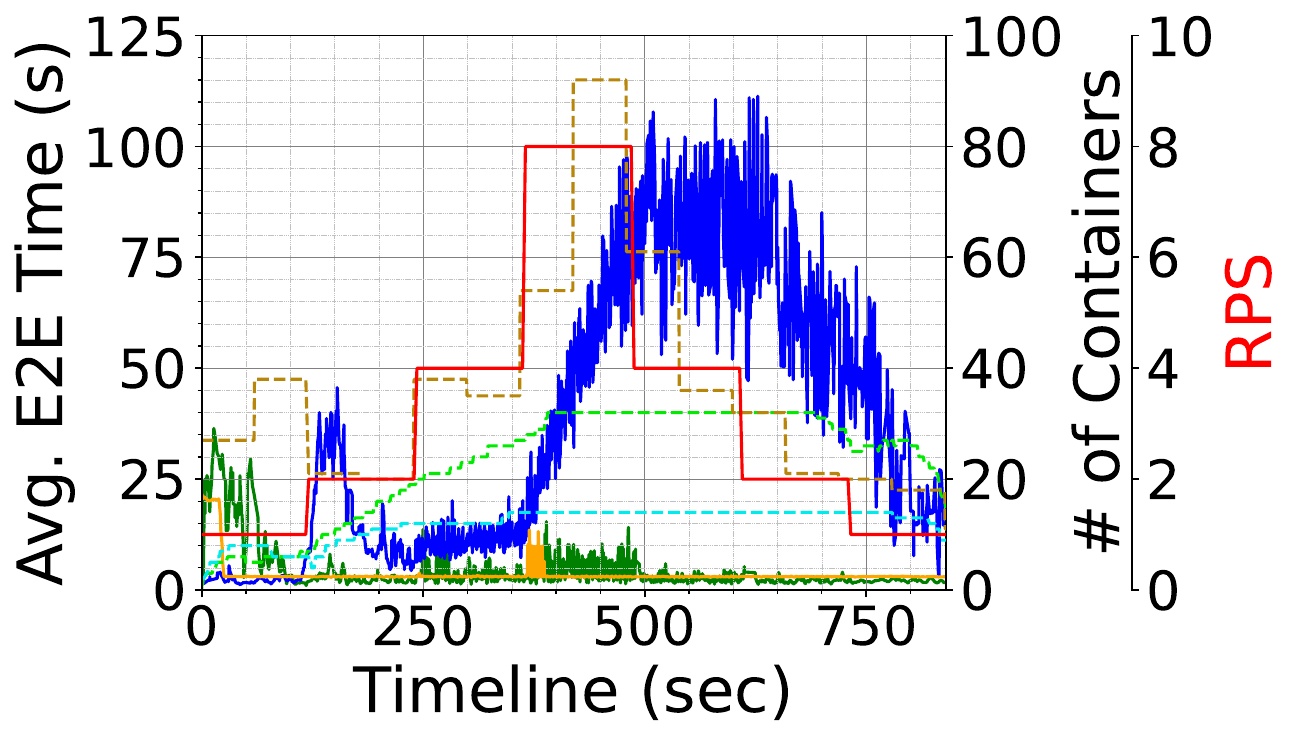}%
    \label{fig:graph-step-scaling}
  }\hfill
  \subfloat[Sawtooth]{%
    \includegraphics[width=0.32\textwidth]{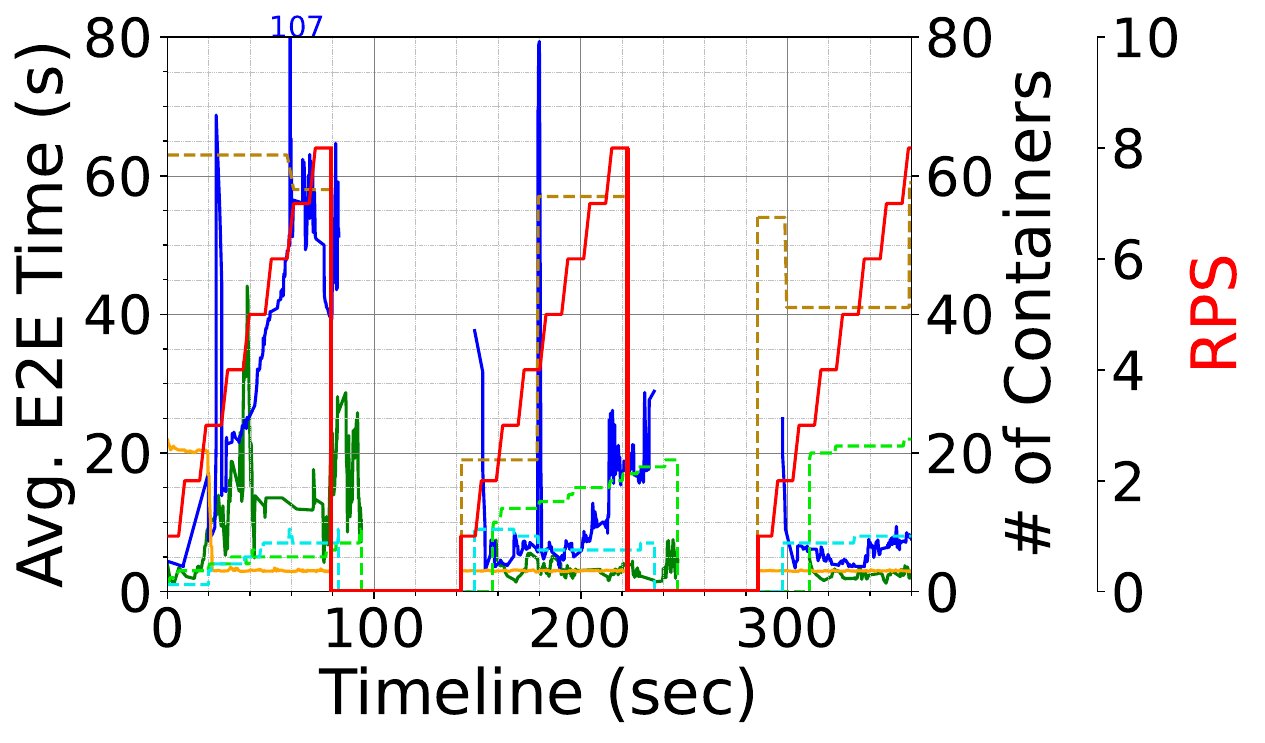}%
    \label{fig:graph-sawtooth-scaling}
  }\hfill
  \subfloat[Alibaba]{%
    \includegraphics[width=0.32\textwidth]{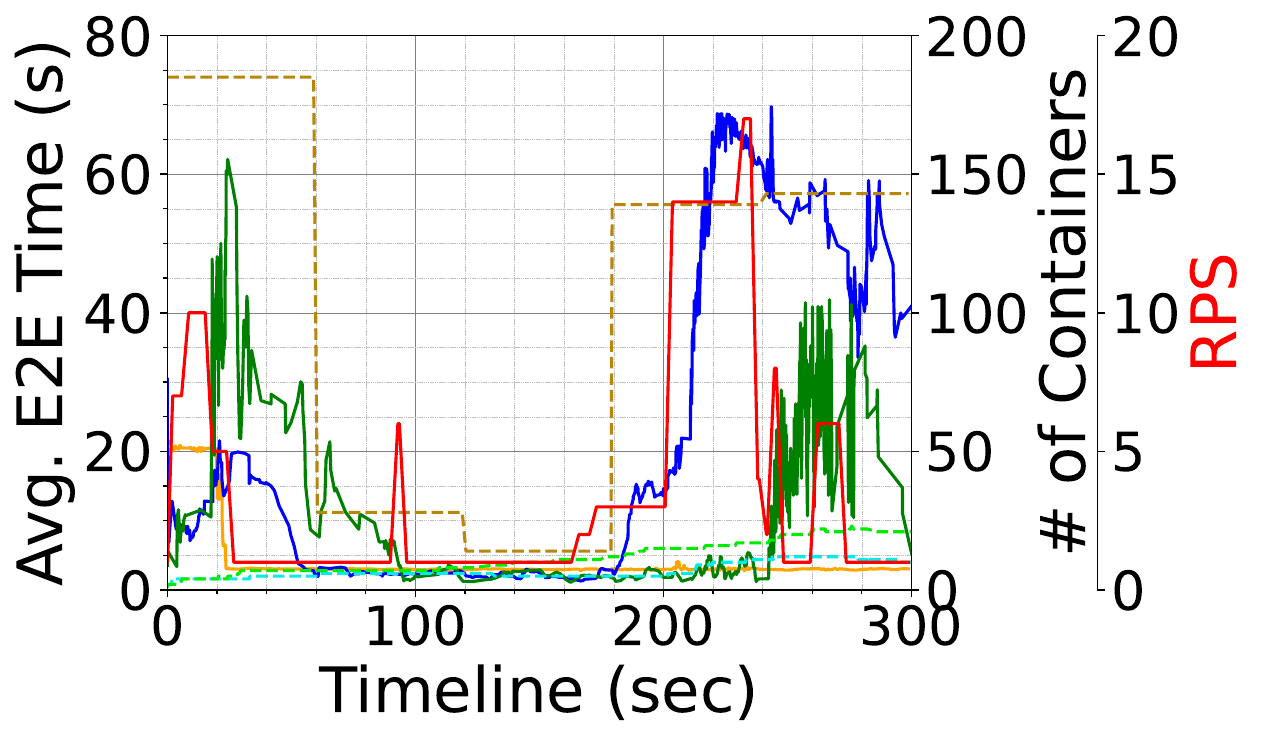}%
    \label{fig:graph-alibab-scaling}
  }%
\caption{Scaling performance of \textit{Graph} workflow with medium payload and different input rates for AWS, AzS and AzN.}%
\label{fig:graph-timeline}

\end{figure*}

All executions for an AzS or AzN workflow share the same set of containers (VMs). 
The default settings to control container scaling use \texttt{maxConcurrent\-Orchestrator\-Functions} ($MO=5$) and \texttt{maxConcurrent\-Activity\-Functions} ($MA=10$), the orchestrator and function logic executions per container.\footnote{\href{https://learn.microsoft.com/en-us/azure/azure-functions/durable/durable-functions-perf-and-scale}{Performance and scale in Durable Functions}, Microsoft, 2024}
However, this causes the Graph workflow to be \textit{unstable} for a Step workload on AzS (Fig.~\ref{fig:graph-az1-step-knobs:mo5-ma10}), 
i.e., the \textit{E2E execution time} (dark blue line, left Y axis) spikes from $40s$ to $200+s$ as we reach \textit{8 RPS} (red line, right outer Y axis) leading to timeouts beyond the $490^{th}s$ (X axis).
At the peak, $10$ containers (light blue dashed, right inner Y axis) are spun up by AzS to handle this load with $MA=10$ \textit{concurrent executions} each. But each container is unable to handle 10 concurrent executions, causing functions to  timeout.

We \textit{customize} these to $MO=8$ and $MA=1$ to have one function per container but $8$ lighter-weight orchestrator executions. 
The scaling for AzS improves and complete the workload with $14$ containers (Fig.~\ref{fig:graph-az1-step-knobs:mo8-ma1}).
The peak latency stays within $110s$, though higher than the $<10s$ seen at a lower rates. 
While AzS can scale to $100$ containers,\footnote{\href{https://learn.microsoft.com/en-us/azure/azure-functions/functions-scale}{Azure Functions hosting options}, Microsoft, 2024} its cold-start behavior limits this growth (\S~\ref{sec:results:cold-start}).
The impact of these knobs is less acute for AzN. In the rest of the article, we default to $MO=8$ and $MA=1$ for AzS and AzN.

The \texttt{maximum partition count} ($PC$) for AzN decides the peak container count, as each container holds one or more partitions.
The default $PC=12$ (Fig.~\ref{az2-step-knobs}, green dashed line, right inner Y axis)~\footnote{\href{https://microsoft.github.io/durabletask-netherite/}{Netherite Configuration, Partition Count considerations}, Microsoft, 2024}
causes high variability in workflow execution times (green solid line, left Y axis), reaching $160s$ at $8$ RPS (red solid line, right outer Y axis). As PC is increased (Fig.~\ref{apx:azN-pc-knobs} in Appendix), this gradually drops for $PC=32$ to a peak latency of $25s$, other than the initial cold-start.
We use $PC=32$ for AzN in the rest of this article.

Overall, the control knobs in Azure affect the concurrency and scaling, which significantly impact the E2E latency. This forces developers to tune the FaaS platform for their workload. In contrast, AWS scales well without such knobs, other than setting the function memory, which also affects its vCPUs~\cite{peak-behind}.

\subsubsection{AWS Memory Tuning Knobs}

\takeaway{AWS provides a controllable trade-off between cost and performance -- higher memory allocation to a function consistently reduces makespan and enables predictable scaling at the expense of higher cost, though with diminishing benefits beyond $\approx2$~GB.\label{ta:aws-knobs}}

\addc{
We perform a sensitivity study on the \textit{Lambda memory allocation} for AWS, which proportionally scales its vCPU, network bandwidth and cost with memory. 
Figure~\ref{fig:summary-aws-knobs} shows the median E2E latency for the \textit{Image WF}, with medium payload at 1~RPS, and its per-invocation cost for five memory configurations (512MB -- 8192MB), with AzN shown as a reference.}

\addc{Increasing a Lambda's memory substantially reduces workflow makespan -- from 4.08s at 512MB to 2.09s at 2GB -- for Image WF, saturating beyond 2GB (1.97s at 8GB). 
This comes at a modest cost increase from $\mud300$ to $\mud430$ per invocation from 0.5GB to 8GB -- costs are reported in micro-US\$ ($\times10^{-6}$).
In contrast, the AzN baseline remains stable at 11.42s and $\mud430$, 
illustrating that AWS offers configurable performance gains even with similar cost budgets.}

\takeaway{AWS Step Functions scale up containers more rapidly and retain more active containers than Azure to quickly stabilize the E2E time.\label{ta:aws-scale}}
\takeaway{AzN is more responsive at scaling up than AzS but is limited by the peak number of containers.\label{ta:azs-scale}}
\takeaway{For stable workloads, AzN achieves comparable performance as AWS, but with fewer containers.%
\label{ta:scaling:up}}

In Fig.~\ref{fig:graph-timeline}, for the Graph workflow using a medium payload running on AWS and Azure, we report the E2E execution time (solid lines, left Y axis), the number of containers
(dashed lines, right inner Y axis) and the RPS if dynamic (red line, right outer Y axis) over the invocation timeline (X axis). The E2E time is a sliding window average over $5s$ to smooth out spikes. AWS reports the number of containers only each $1min$.

\subsubsection{Static Workflow Input Rates}\label{sec:scaling:static}
AWS creates containers faster in response to concurrent invocations than Azure. E.g., comparing static RPS of $1$ and $8$ in Figs.~\ref{fig:graph-1-scaling}--\ref{fig:graph-8-scaling}, AWS spawns $42$ containers at the start (yellow dashed line) as it responds to the initial cold start delay, as seen from the higher execution times (yellow solid line) till the $20^{th}$ second. Once these calls catch up, the number of containers drops steadily but the execution time stays constant. This is seen for static $4$ and $8$~RPS too, where AWS initially spawns $148$ and $286$ containers before stabilizing to fewer container within $60s$.
The E2E execution time for AWS is also much smaller than Azure at $4$ and $8$~RPS.

Azure starts with fewer containers and only gradually spins up more.
For AzS at 1~RPS (Fig.~\ref{fig:graph-1-scaling}), it initially has $3$ containers (light blue dashed line) and this rises to $4$ after $32s$ and stays there, with a partially stable E2E time (solid blue line). However, a scale down at the $125^{th}$ second to $1$ container causes the E2E latency to spike, and it is unable recover till $250s$ by scaling up to $7$ containers. This is even slower for $4$ and $8$~RPS where the peak of $9$ and $16$ containers is reached only at $100s$, with E2E latencies growing till then.
AzN has a similar trend, except that it scales more proactive and stabilizes the E2E time more quickly. E.g., for $4$~RPS (Fig.~\ref{fig:graph-4-scaling}) the number of containers (light green dashed line) grow to $12$ by $100^{th}$sec and rise to $21$ containers by $270^{th}$~sec, even as the latency steadily drops (solid green line).

\subsubsection{Dynamic Workflow Input Rates}
Figs.~\ref{fig:graph-step-scaling}--\ref{fig:graph-alibab-scaling} report the performance of the Graph workflow under dynamic input rates.
AWS closely tracks changes in request rate, with the container count following the workload and stabilizing within the $1$min reporting window. \modc{This is consistent for both Step and Sawtooth, and results in low and stable E2E latency of $\approx3$s after an initial $20$s cold-start period.}

Azure responds more slowly to RPS changes. AzN scales containers faster than AzS and regains a low and stable E2E latency after load increases. E.g., for Step, AzN ramps up to about $32$ containers within few minutes, while AzS reaches only about $14$ containers in that period. Despite using fewer containers than AWS (which peaks near $95$ containers), AzN achieves comparable end-to-end latency ($2.9$--$3.1$s), as Azure scales at the workflow level and shares VMs across functions, unlike AWS which scales individual functions.

For Sawtooth, AzN under-provisions containers during the first load increase, leading to a transient latency spike of up to $44$s. In later waves, AzN adapts and maintains stable latency of $2.9$--$3.1$s with $\approx19$ containers, approaching AWS's performance with $54$--$63$ containers. AzS remains slow to scale and stabilizes at a higher latency of $7.1$s with fewer containers.

For the Alibaba trace with higher variability (peaks of $17$~RPS), Azure struggles to maintain low latency. AzN is limited by its max $32$ containers and reaches an E2E latency of about $14.1$s, while AzS is worse with latencies up to $45$s. AWS scales beyond $150$ containers and maintains latency close to $3$s.
\addc{Scaling results for other workflows is in Appendix~\ref{sec:app-other-scaling}.}

\begin{figure}[t]
\centering%
  
\includegraphics[width=0.9\columnwidth]{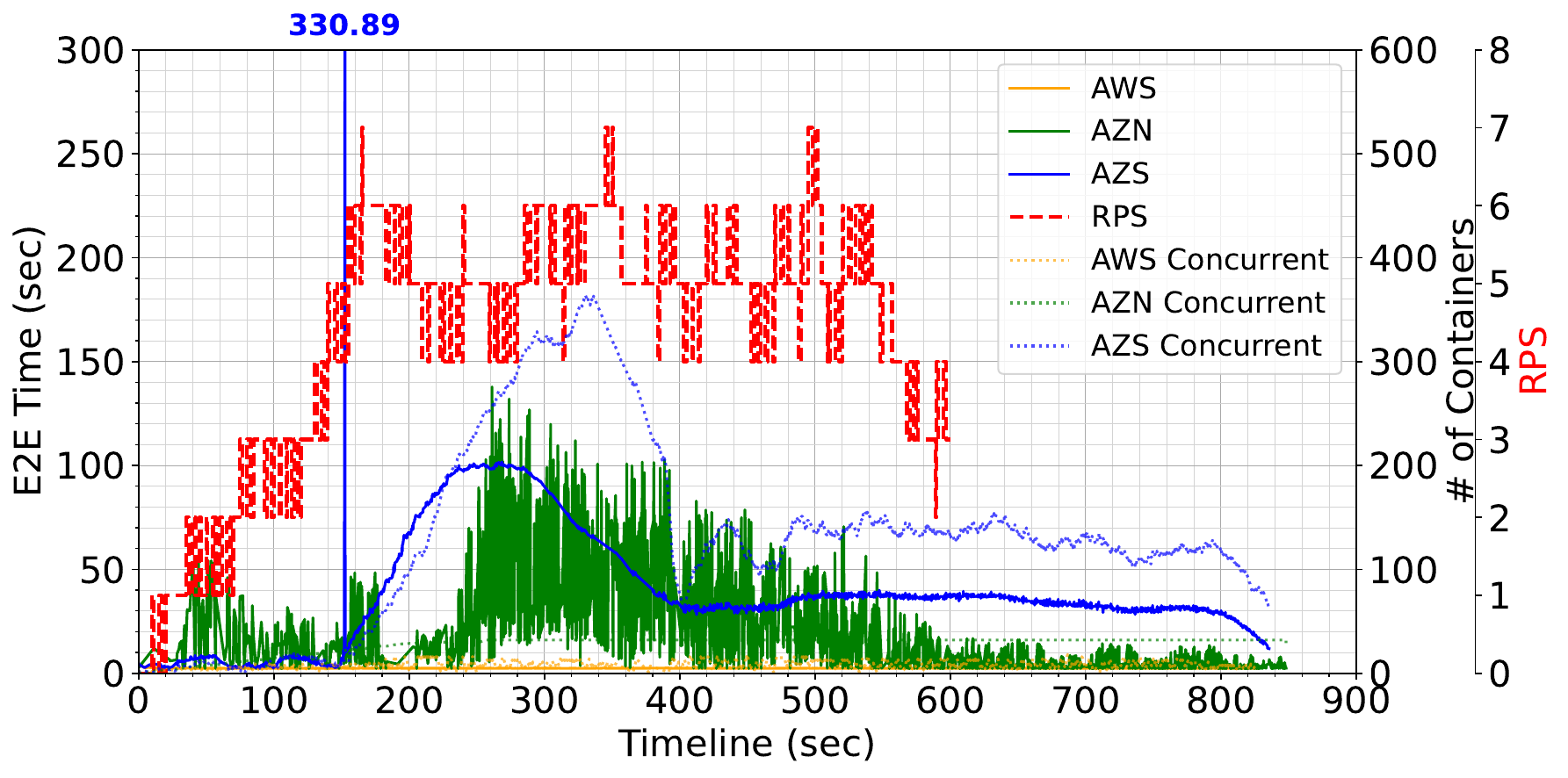}%
\caption{Timeline Plot For UTM Flight Monitoring Workflow.}
\label{fig:utm-timeline}

\end{figure}

\subsubsection{Scaling in UTM Workflow}

\addc{
For the UTM workflow with its custom workload (Fig.~\ref{fig:utm-timeline}) AWS maintains a low and stable E2E latency, with a median of $2.4\,s$ and container concurrency $<20$. In contrast, AzS exhibits higher and more variable latency, with a median of $35.4\,s$ and concurrency $> 150$ VMs during peaks. AzN fall in-between, with a median latency of $13.6\,s$ and faster recovery as load decreases. Its concurrency is limited to 32. 
These reinforce the differing scalability behavior observed earlier.}

\subsubsection{Discussion}
\textit{AWS} scales rapidly within seconds to RPS changes and offers a stable E2E time. 
It also uses more containers, hitting its peak of $1000$, due to per-function scaling rather than per workflow.
\textit{AzN} performs well if RPS growth is slower, over minutes, and its E2E time approaches AWS -- provided $32$ containers can handle the load. This limit prevents it from processing a 
demanding workloads. 
Scaling per workflow 
allows resource pooling across functions and using fewer containers than AWS. 
\textit{AzS} performs worse than AzN and AWS, with slower scaling.
It is unable to use its 100-container limit efficiently and is unstable for several workloads.

\subsection{Cold Start Overheads }
\label{sec:results:cold-start}

One of the novel aspects of our study is to understand the cumulative effect of cold-starts for a workflow rather than just individual function cold-starts overheads studied earlier.

\takeaway{\label{ta:cold-start:vm-start}
Azure Durable Functions (AzS and AzN) exhibit tangible cold-start overheads while it is negligible for AWS Step Functions
}

\subsubsection{AWS Step Functions}
The results from Fig.~\ref{fig:graph-timeline} \modc{(and Fig.~\ref{fig:all-wf-timeline} in Appendix)} show an absence of spikes in the E2E workflow time for AWS, other than in the initial warmup period of $<20s$ as containers are rapidly spun-up to handle the load. As containers are retained for $5~mins$, changes in container counts up to this initial peak and within $5~mins$ do not cause a cold-start penalty. 

E.g., for Sawtooth on the Graph workflow (Fig.~\ref{fig:graph-sawtooth-scaling}), 
the initial wave starts $54$ containers causing an increase in E2E latency due to cold-starts; there is another mild latency spike when this grows to $62$. But when containers drop to zero after wave one, and again grow to $52$ and $62$ for waves two and three, the warm containers from earlier are reused as it is $<5~mins$~\cite{xfaas}.
These low overheads are due to the fast startup of Firecracker microVMs, and optimizations such as sparse loading and deduplication 
of base images~\cite{aws-containers-on-demand}, with $<200ms$ cold-start overheads reported~\cite{bfaas}. 

\begin{figure}[t!]
\centering
\subfloat[AzS]{
    \includegraphics[width=0.43\columnwidth]{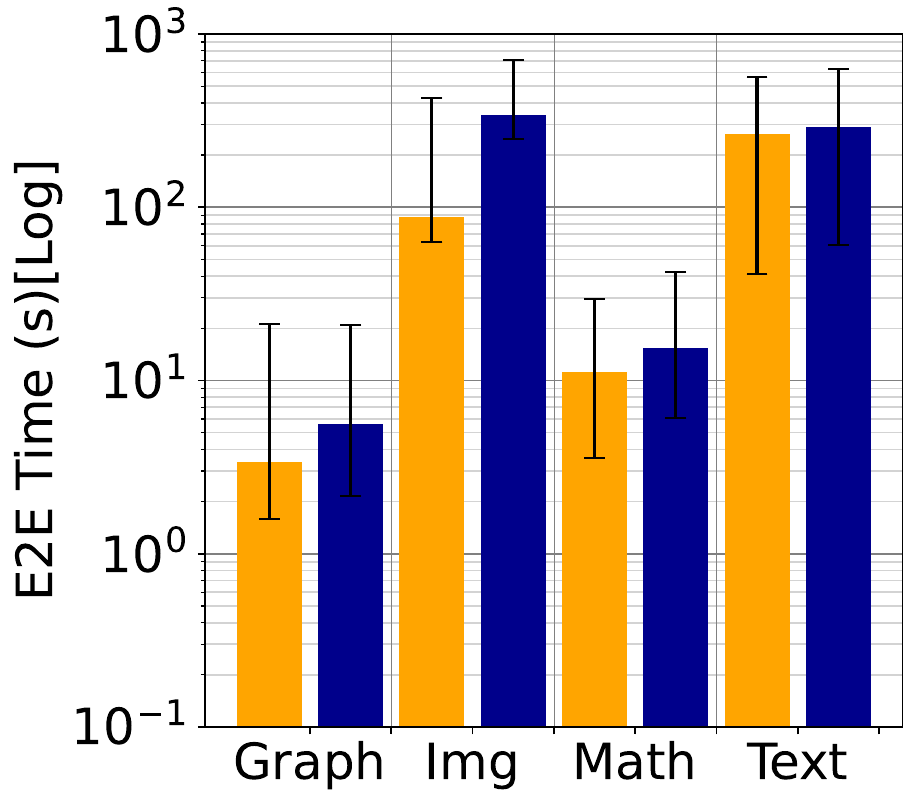}
    \label{fig:wfs-medium-e2e-v1}
  }\hfill
    \subfloat[AzN]{
    \includegraphics[width=0.43\columnwidth]{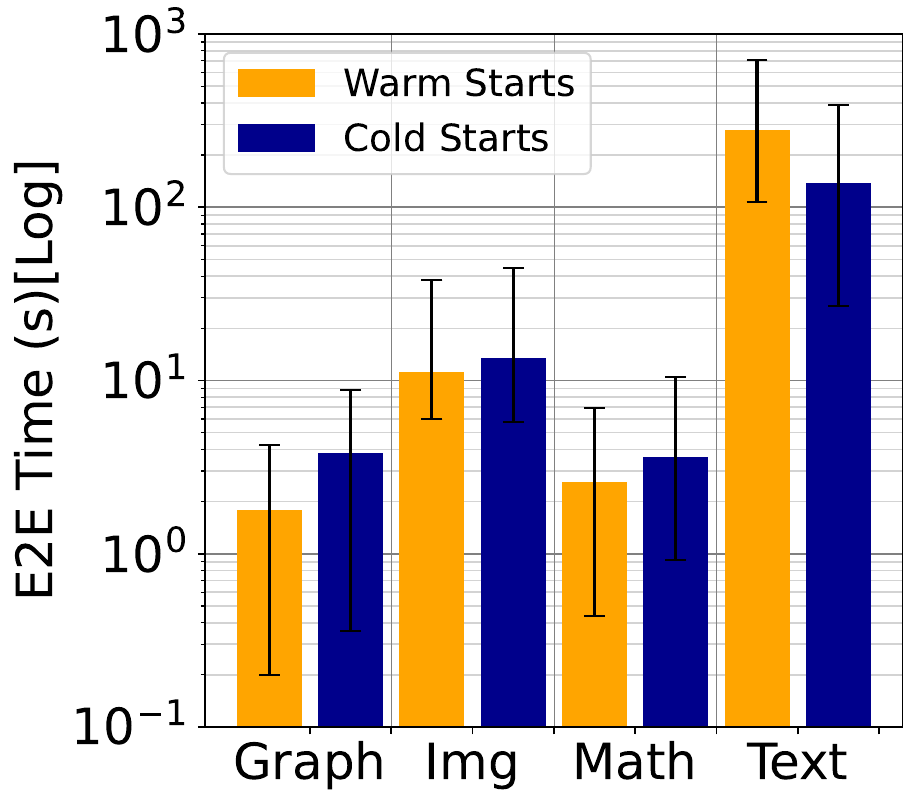}
    \label{fig:wfs-medium-e2e-v2}
    }
\caption{E2E latency \textit{with and without container cold-starts}, for all workflows with medium payload and 1~RPS. Whiskers show Q1--Q3 latency range.}
\label{fig:graph-workload-az-containers-box}
\end{figure}

\begin{figure}[t!]
\centering
\subfloat[AzS]{%
    \includegraphics[width=0.47\columnwidth]{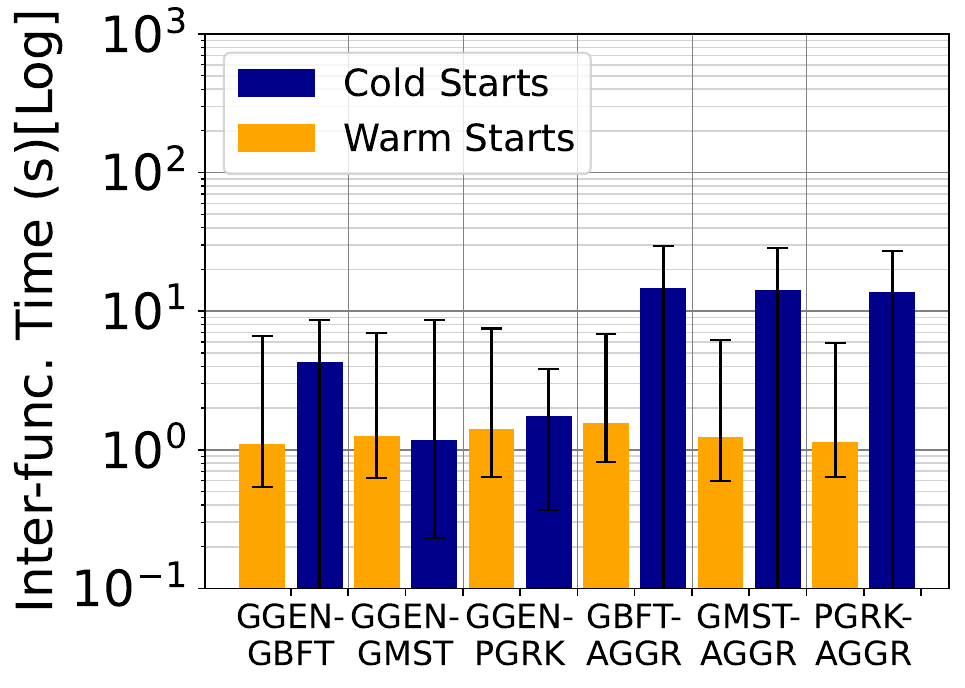}%
    \label{fig:wfs-medium-inter-fn-v1}%
  }%
  \subfloat[AzN]{%
  \includegraphics[width=0.47\columnwidth]{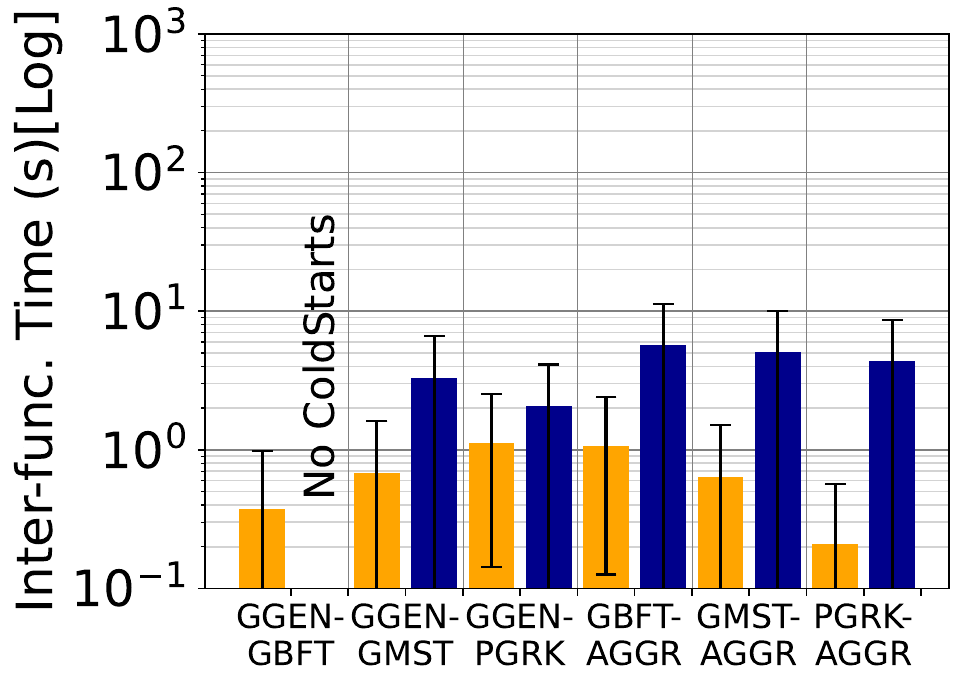}%
    \label{fig:wfs-medium-inter-func-v2}%
  }%

\caption{Inter-function latencies \textit{with and without container cold-starts} for AzS and AzN, for Graph workflow on medium payload and 1~RPS.}

\label{fig:graph-workload-az-containers-box-split}
\end{figure}

\subsubsection{Azure Durable Functions (AzS and AzN)}

Fig.~\ref{fig:graph-workload-az-containers-box} reports the E2E time for those workflow invocations with \textit{no cold-starts} for any of their function invocations (only warm starts, yellow bar) and those that had \textit{one or more cold-starts} (blue bar). This is shown for all four workflows using a medium payload and 1~RPS.
The whiskers indicate Q1--Q3 latency ranges.
Azure's E2E time clearly varies with cold-starts, and is more variable than AWS's as the input rate changes.
Azure workflow's cold-start has three parts: (1) \textit{Container and runtime cold-start}, the time to spawn a new container and load the Python runtime with dependencies from \textit{requirements.txt}; (2) \textit{Function cold-start}, the latency to run a function for the first time on a container, and lazily load the pending dependencies; and (3) \textit{Dataflow cold-start}, the time to initialize data transfer services. These are discussed next.

\takeaway{\label{ta:coldstart_cascading_blob}
Azure has higher cold-start overheads for workflows with larger packages and larger payloads, though the container cold-start overhead is modest.}

In Fig.~\ref{fig:graph-workload-az-containers-box-split}, we zoom into the inter-function latencies between each pair of functions in the graph and image workflows, and split them into warm-starts and cold-starts for the downstream function execution; this is shown only for \textit{function-pairs} having one or more cold-starts. This \textit{inter-function latency} is measured from the time an upstream function completes to the time the downstream function is invoked. This includes multiple components that \textit{we attempt to disaggregate}.

\begin{figure}[t!]
\centering%
  \subfloat[Graph WF on AzN]{%
    \includegraphics[width=0.45\columnwidth]{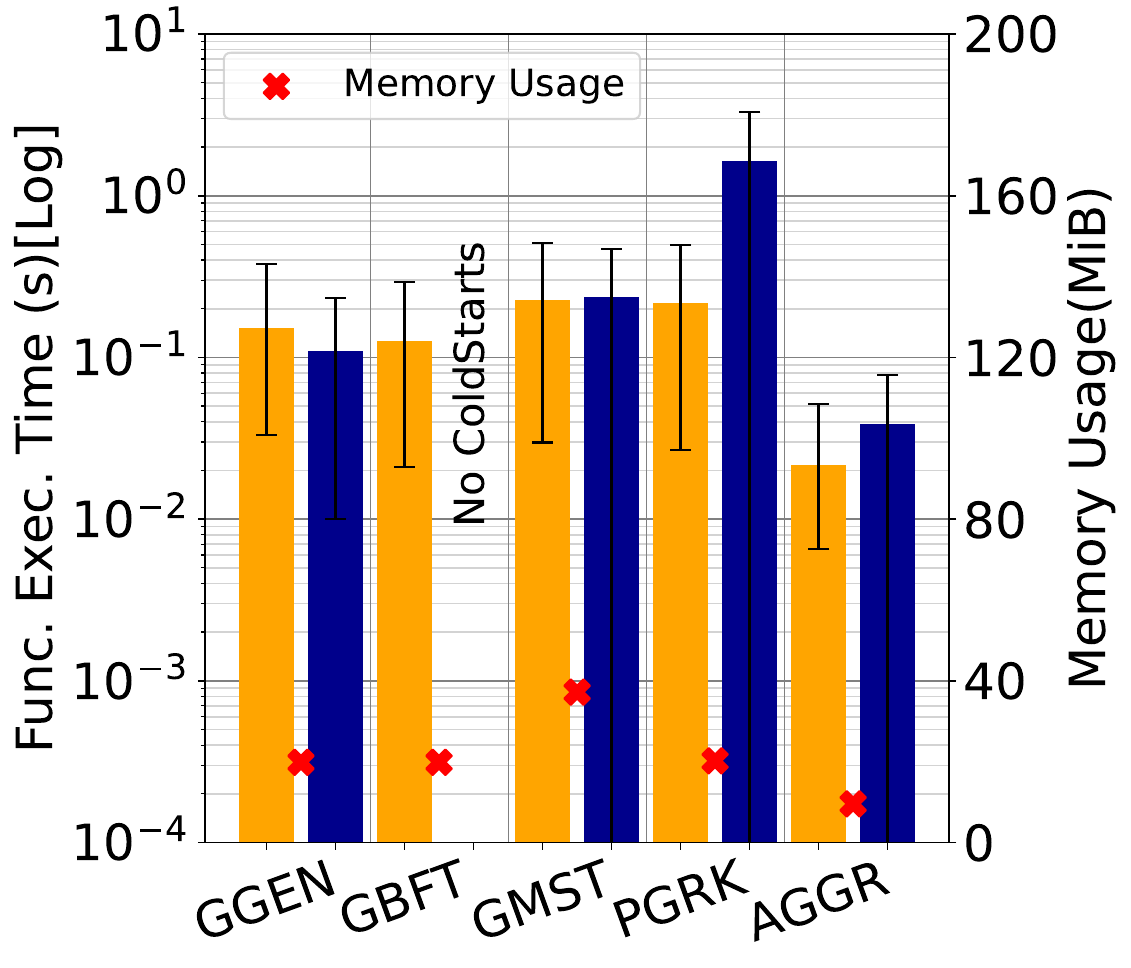}%
    \label{fig:azv2-fn-exec:graph}%
  }%
  \subfloat[Image WF on AzN]{%
    \includegraphics[width=0.55\columnwidth]{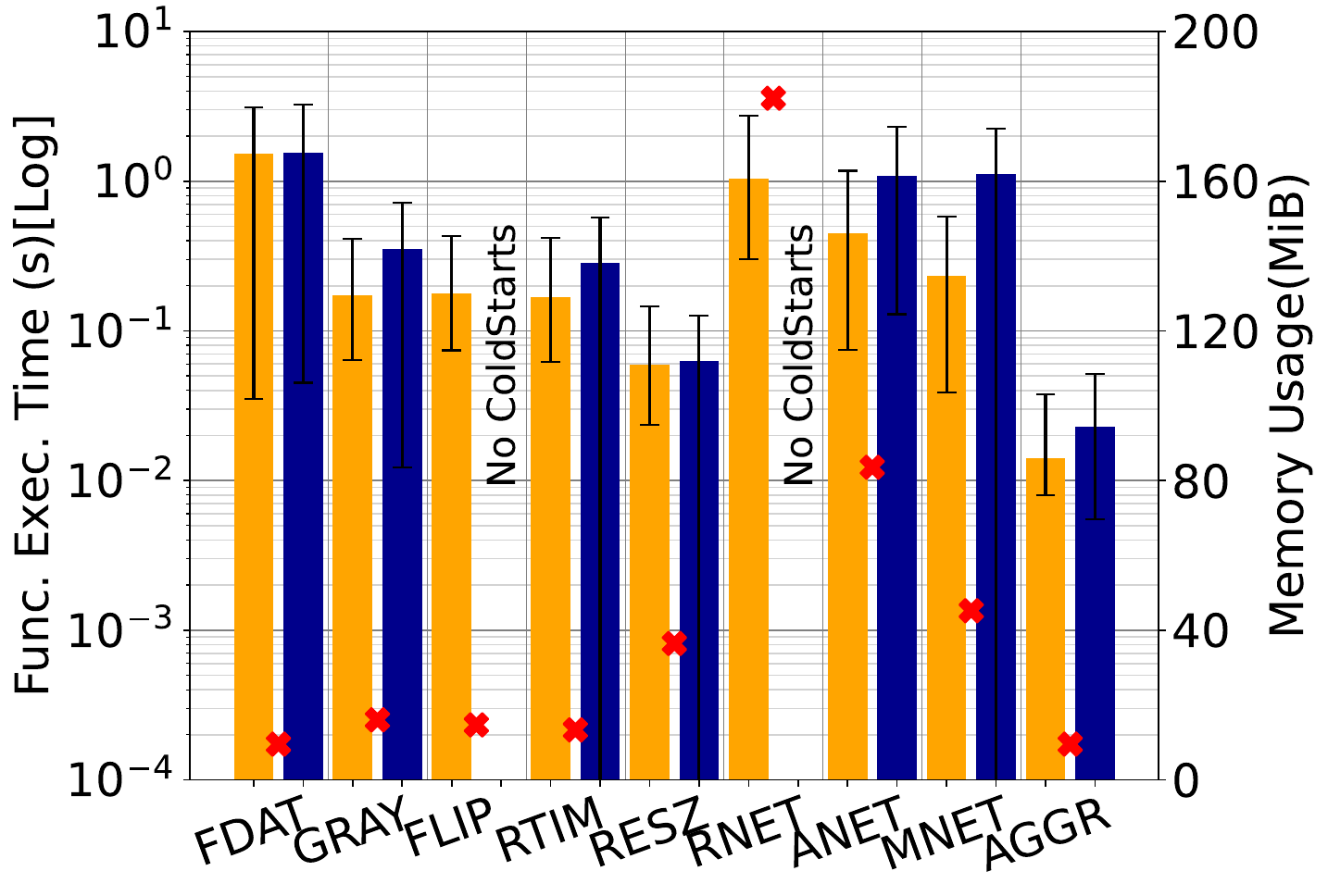}%
    \label{fig:azv2-fn-exec:image}%
  }%
\caption{Function execution times \textit{with and without function cold-starts} with medium payload and static $1$~RPS}
\label{fig:funcion-exec-coldstarts}
\end{figure}

The first time a container is created for the workflow, we incur a \textit{container cold-start} overhead, which is included in this latency. We perform separate micro-benchmarks to measure this and find it to be $\approx 290ms$ for AzS and $\approx 51.5ms$ for AzN \modc{(see Appendix~\ref{app:sec:container-cs-est}).} This overhead is modest.

\begin{figure*}[t!]
\centering%
    \includegraphics[width=1\textwidth]{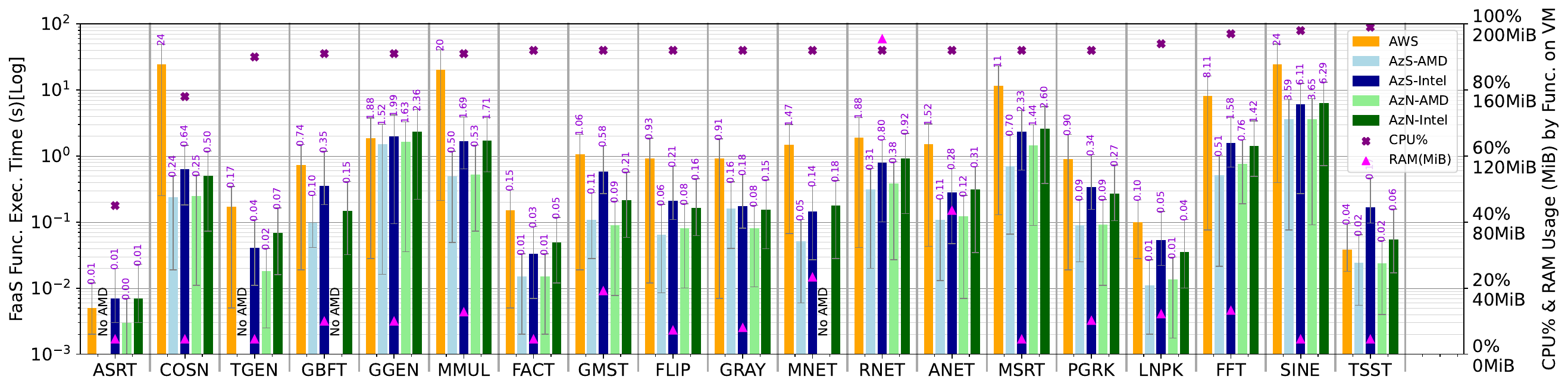}%
  \caption{Function exection time within singleton workflow using medium payload input and static 1~RPS on AWS, AzS and AzN are shown as bars on left axis (log scale). CPU utilization\% and Memory consumption (MiB) on Azure VM are shown as markers ($\times,\blacktriangle$) on right Y axis.}
  \label{fig:singleton-wf-exec-cpu-times:zoom-out}
  
\label{fig:singleton-wf-exec-times}
\end{figure*}

The next component is Python \textit{runtime cold-start} overhead when a function first executes in a new container. This loads \textit{all} the functions of the workflow and their dependent packages into container memory since the container is common to the workflow. It depends on the size of the packages, and should be constant for that workflow.%
So the inter-function overhead (Fig.~\ref{fig:graph-workload-az-containers-box-split}, difference in blue and yellow bars) should be constant for different function-pairs since the container and runtime overheads are constant. However, this is seen to be variable. This is because of the \textit{dataflow coldstart}.

We discern that the first time a message payload passes through the workflow, the system sets up necessary data transfer entities and connections for queues/blobs. For small payloads, this is through queues and for large ones through blobs, with the setup overheads being larger for the blob. E.g., Fig.~\ref{fig:graph-workload-az-containers-box-split} \modc{(and Appendix~\ref{app:cs}, Fig.~\ref{app:fig:funcion-exec-coldstarts})} shows the time for a medium payload of Graph and Image that needs a blob and has higher/variable overheads. The corresponding overheads using queues for a small payload \modc{(Appendix, Fig.~\ref{app:fig:wfs-app-medium-inter-func-new-1:graph} and \ref{app:fig:wfs-app-medium-inter-func-new-2:image})} are smaller and stable at $\approx4.1$s for Graph and $\approx5$s for Image. Based on a microbenchmark (Appendix~\ref{app:cs}), the Graph workflow using queues has $1s$ of dataflow overheads while blobs have $8.4$--$19.6s$. We are the first to report this.

\takeaway{\label{ta:cold-start:func-warmup}
The function cold-start overheads for Azure Durable Functions (AzS and AzN) depend on the size of dynamic packages and models.
}

The \textit{function cold-start} occurs when a function executes for the first time in a container, due to loading dependencies lazily (e.g., Python imports) and models into memory. 
Fig.~\ref{fig:funcion-exec-coldstarts} shows the user-logic execution time with and without cold-starts for the Graph and Image workflows on AzN \modc{(AzS is shown in Appendix~\ref{app:cs}, Fig.~\ref{app:fig:funcion-exec-coldstarts})}. PGRK shows a penalty with cold/warm-start times of $\approx1.1s/0.1s$ as it imports \textit{numpy} and \textit{networkx} packages (Fig.~\ref{fig:azv2-fn-exec:graph}). The other graph functions do not import these.
The ML inferencing functions in the Image workflow load their ONNX models from disk 
and this ONNX session is retained for future warm-starts. 
In Fig.~\ref{fig:azv2-fn-exec:image}, 
RNET, ANET, and MNET have cold-start execution times that are $0.38$--$1.43$~s ($142$--$400\%$) higher than a warm-start.

\takeaway{\label{ta:coldstart_cascading}
Cold starts can impact workflow execution time in sequential workflows with cascading delays, but are less severe for workflows with parallel paths.
}

\textit{Image} has the longest execution path and reports the highest cold start overhead for AzS and AzN due to cascading effects of workflow containers scale-up, with a penalty of $252s/74\%$ for AzS and $2.13s/16\%$ in AzN (Fig.~\ref{fig:graph-workload-az-containers-box}).
In contrast, the cold start overheads are more modest for \textit{Graph}, at $2.2s$ and $2.08s$ for AzS and AzN, 
and $4.01s$ and $1.03s$ for \textit{Math}. 
This suggests that the container may spin-up concurrently during first-time workflow execution for those with parallel paths. The package loading times also affect first-time function execution.
For \textit{Text}, the cold-starts are dwarfed by inter-function overheads due to large payloads, as discussed in~\ref{ta:inter-func:workflow-1}.

\subsubsection{Discussion}
While function cold-starts are well studied~\cite{atlsc,bfaas}, their impact on FaaS workflow are not. Our results indicate the need for more coordinated handling of cold-starts at the workflow level rather than at individual functions. In fact, through the use of shared VMs for functions in a workflow, Azure mitigates some of these overheads, though its absolute time is higher than for AWS. Using function fusion can also potentially mitigate the impact for larger payloads~\cite{xfaas}.

\subsection{Function Execution Times }
\label{sec:results:func-exec}
We benchmark the functions used in the workflows individually within a ``singleton'' workflow that contains only one function.
We also profile the CPU and memory usage of each function on an Azure A1v2 VM with 1 x86 vCPU and 2GB RAM,
as this is not reported from FaaS containers.
Fig.~\ref{fig:singleton-wf-exec-cpu-times:zoom-out} shows the FaaS function execution time (bars, left Y-axis), and their VM CPU\% and memory usage (markers, right Y-axis).

We use 512MB and 2 vCPU cores per AWS container, while AzS and AzN allocate a standard 1 core and 1.5GB of RAM per shared container. 
All three FaaS platforms run on x86 CPUs\footnote{While AWS Lambda offers Graviton2 ARM CPUs, we use the default x86\_64 CPUs.}.
\texttt{cpuinfo} shows AWS Lambda in this region run on older (2014) Intel Xeon E5-2680 v3 (Haswell) CPUs with $12$ cores at $2.5$GHz.
Azure containers run on a mix of CPU models: AMD EPYC 7763 (Milan) with $64$ cores at $2.45$GHz (2021) and Intel Xeon E5-2673 v4 (Broadwell) with $20$ cores at $2.30$GHz (2016).
This has two implications.

\takeaway{\label{ta:func:cpu-mem-intensive}
Azure Durable Functions (AzS and AzN) usually execute functions faster than AWS Step Functions, especially for compute-intensive functions.
}
\takeaway{\label{ta:func:variability-2}
CPU heterogeneity causes variable function performance within the same region for AWS and Azure.
}

Given the newer CPU architecture of Azure containers, functions on Azure perform faster than AWS. This is seen in Fig.~\ref{fig:singleton-wf-exec-cpu-times:zoom-out}, where AWS functions take longer or are comparable to Azure functions. This is more acute for numerical functions such as matrix multiplication (MMUL), Page Rank (PGRK), FFT, and ML inferencing (RNET, MNET, ANET), which benefit from the faster processors, e.g., PGRK on AWS takes $0.90s$, while for AzS and AzN it takes $\approx 0.09s$ on the AMD container and $\approx 0.30s$ on the Intel container.

\begin{figure}[t!]
\centering%
\subfloat[Graph WF on AWS]{%
 \includegraphics[width=0.43\columnwidth]{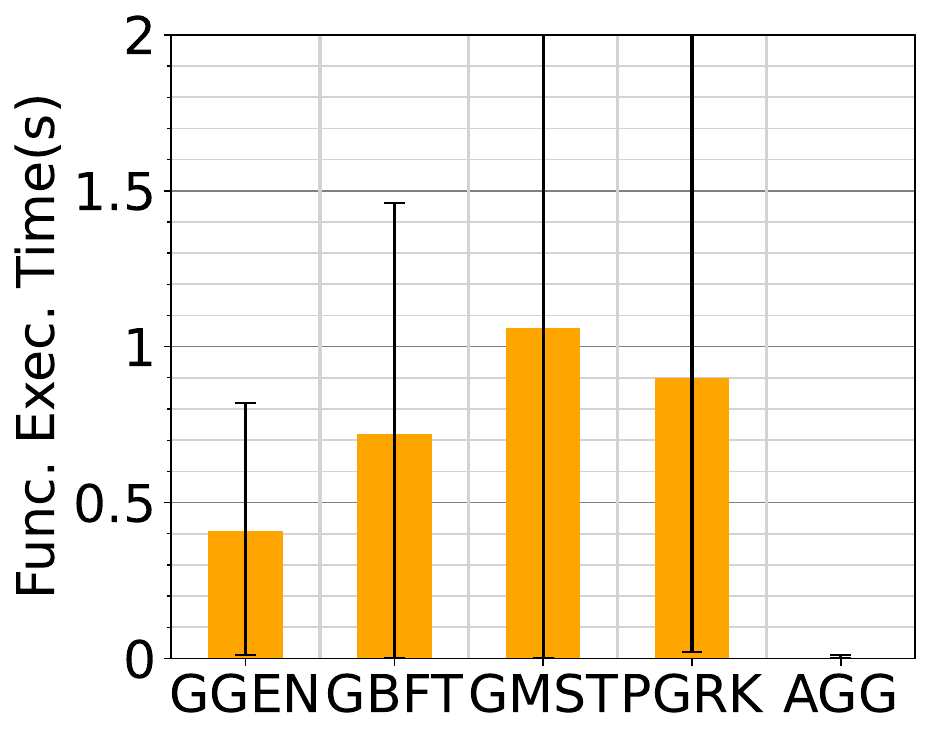}
    \label{fig:aws-intel-amd-arch-graph}
  }\quad
  \subfloat[Image WF on AWS ]{%
 \includegraphics[width=0.46\columnwidth]{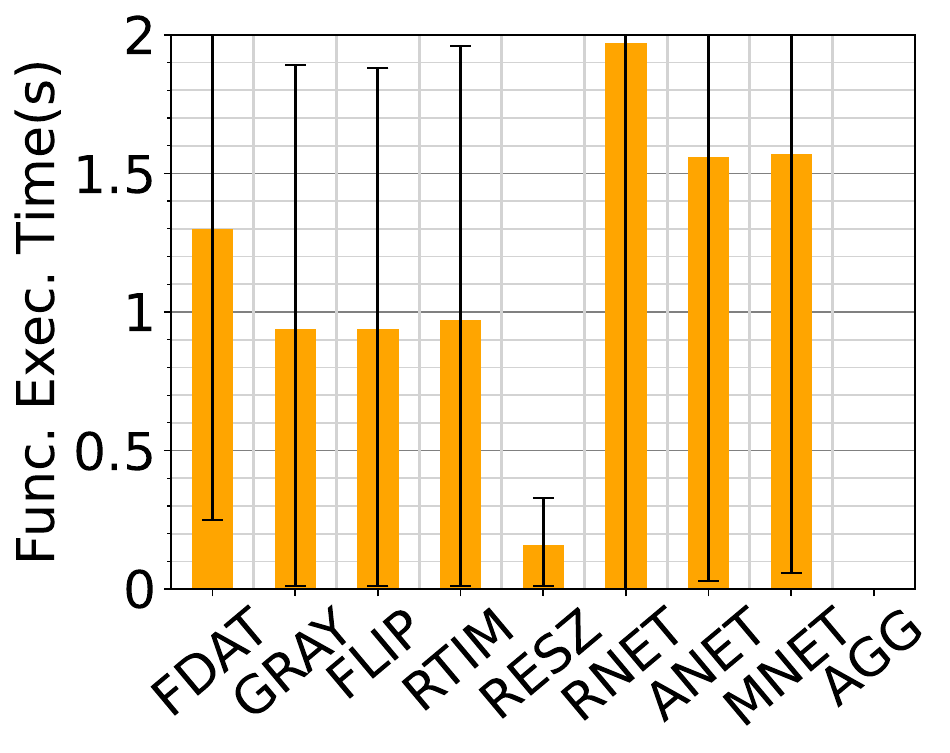}
    \label{fig:aws-intel-amd-arch-img}
  }\\
\subfloat[Graph WF on AzS]{%
 \includegraphics[width=0.49\columnwidth]{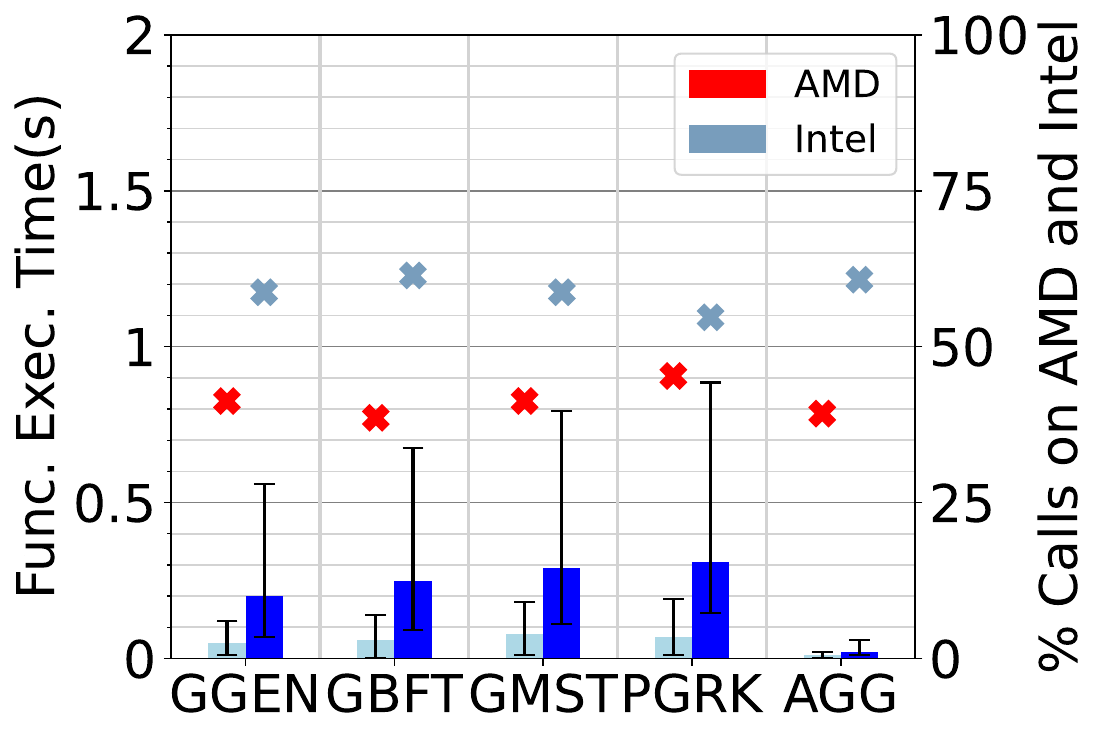}
    \label{fig:azurev1-intel-amd-arch-graph}
  }%
    \subfloat[Image WF on AzS]{%
 \includegraphics[width=0.49\columnwidth]{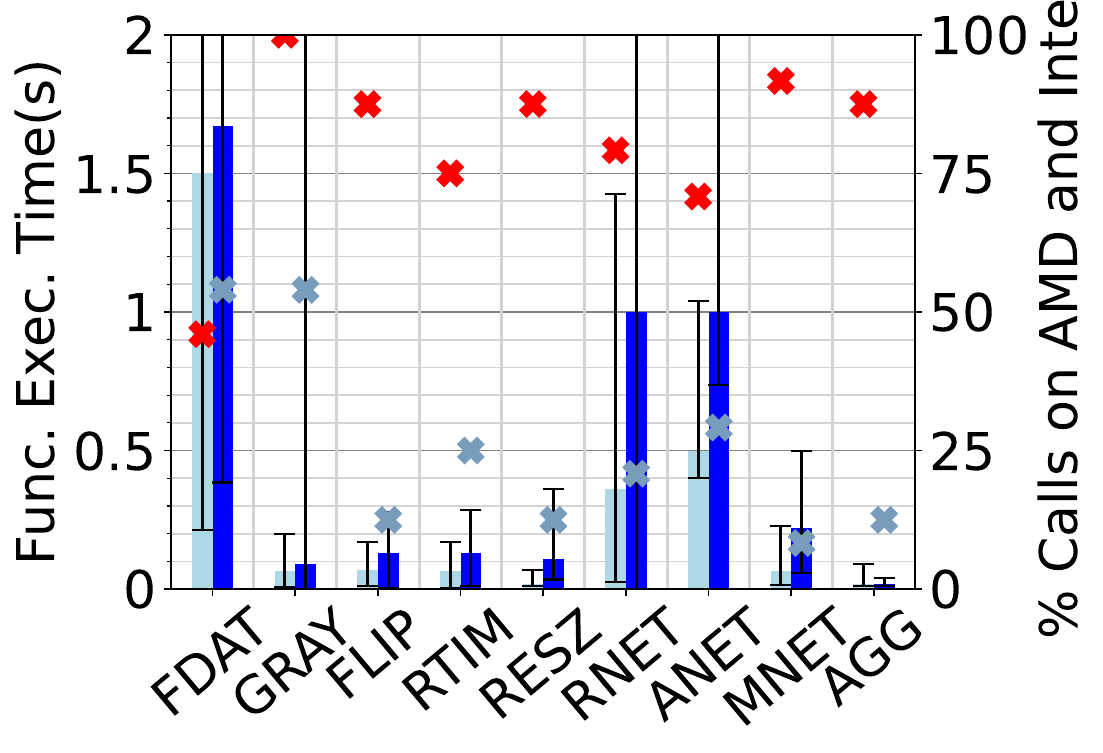}
    \label{fig:azurev1-intel-amd-arch-img}
  }\\
\subfloat[Graph WF on AzN]{%
 \includegraphics[width=0.49\columnwidth]{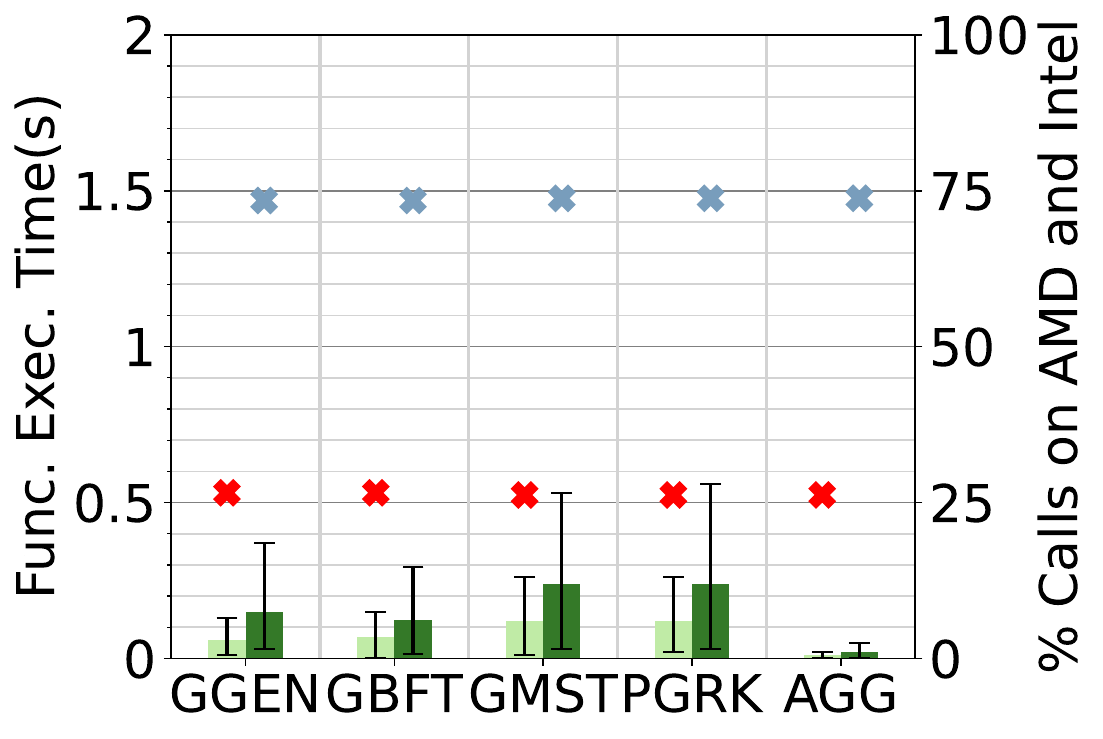}%
    \label{fig:azurev2-intel-amd-arch-graph}
  }%
\subfloat[Image WF on AzN]{%
 \includegraphics[width=0.49\columnwidth]{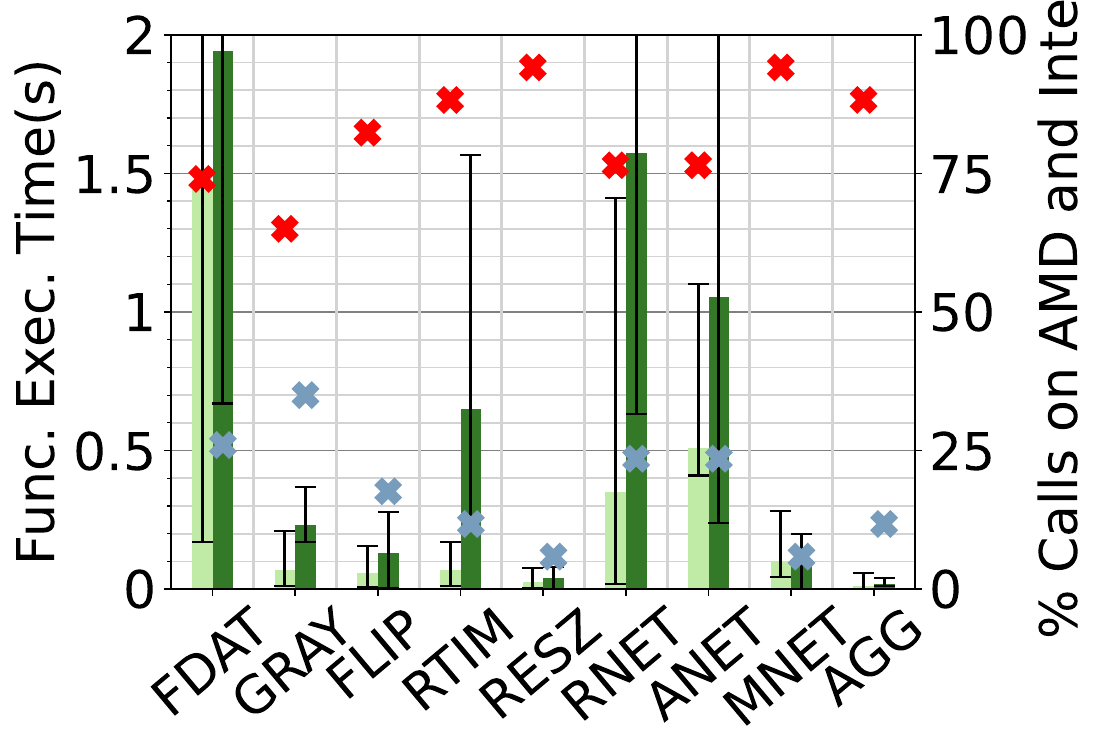}
    \label{fig:azurev2-intel-amd-arch-img}
  }
  \caption{Function Execution Times for Graph (left) and Image (right) workflows on AWS, AzS and AzN, using medium payload and static 1~RPS.
  }
\label{fig:graph-img-wf-exec-times}

\end{figure}

\begin{figure*}[t!]
\centering%
\begin{minipage}{0.67\textwidth}
\centering
      \subfloat[All payload sizes \textit{(Log-Log plot)}]{%
    \includegraphics[width=0.34\textwidth]{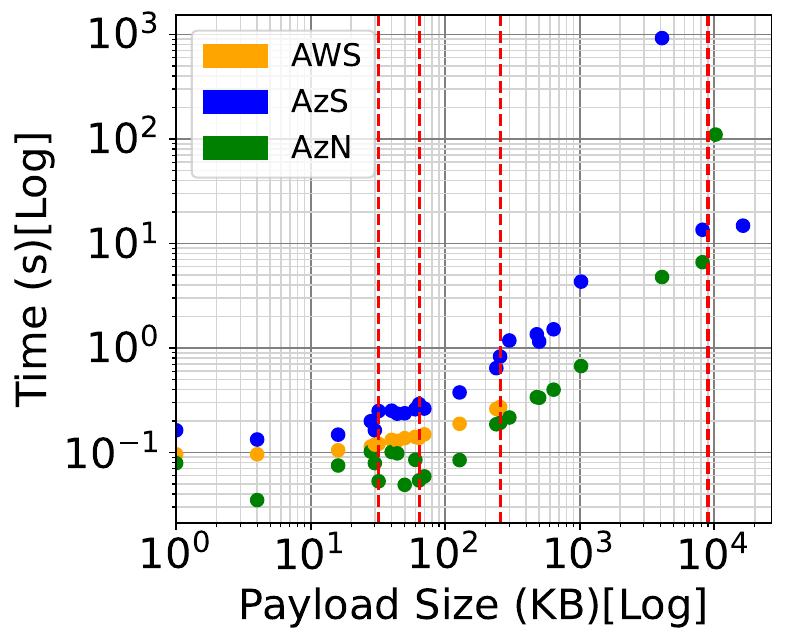}%
    \label{fig:inter-fn-data-transfer:log}%
  }%
  \subfloat[Smaller payload sizes]{%
    \includegraphics[width=0.33\textwidth]{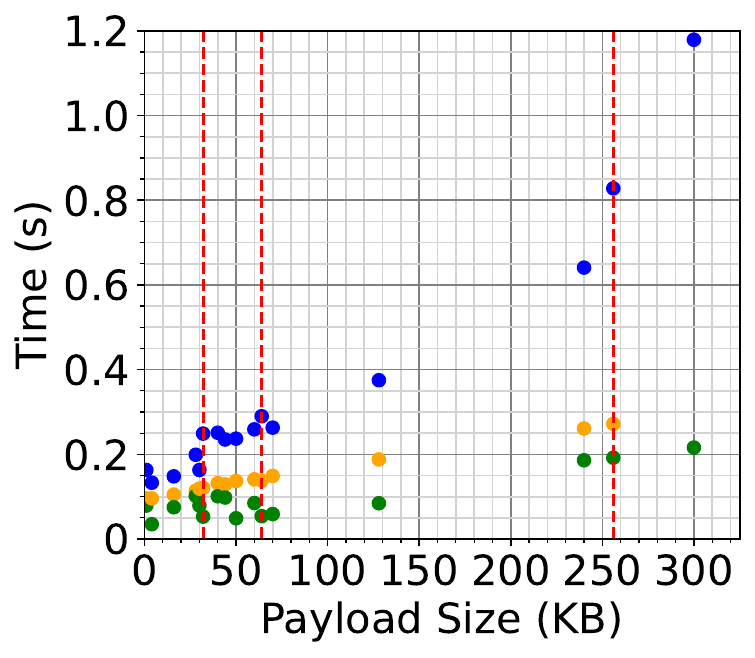}%
    \label{fig:inter-fn-data-transfer:small}%
  }%
  \subfloat[Variability with payload size]{%
    \includegraphics[width=0.32\textwidth]{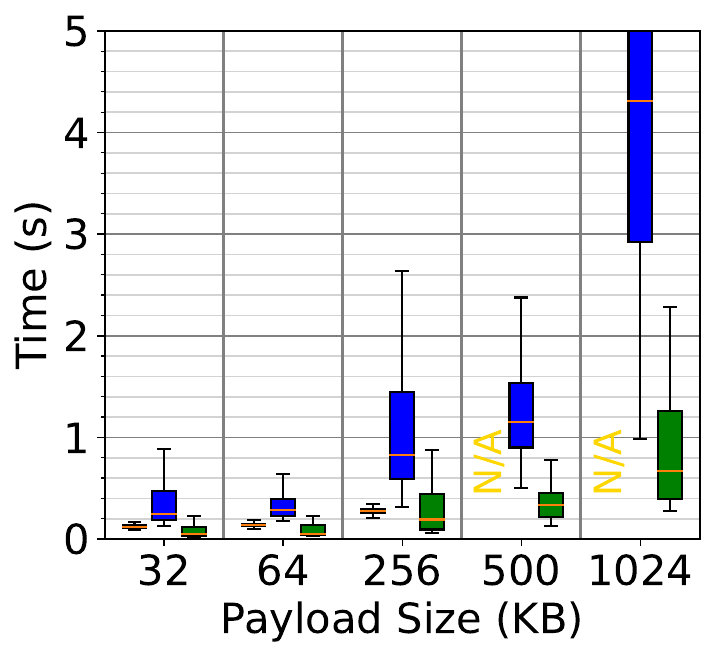}%
    \label{fig:inter-fn-data-transfer:box}%
  }%
  \caption{Inter-function latencies for different payload sizes for micro-benchmark workflow.}
\label{fig:inter-fn-data-transfer}
\end{minipage}~~
\begin{minipage}{0.3\textwidth}
\centering
\includegraphics[width=0.82\columnwidth]{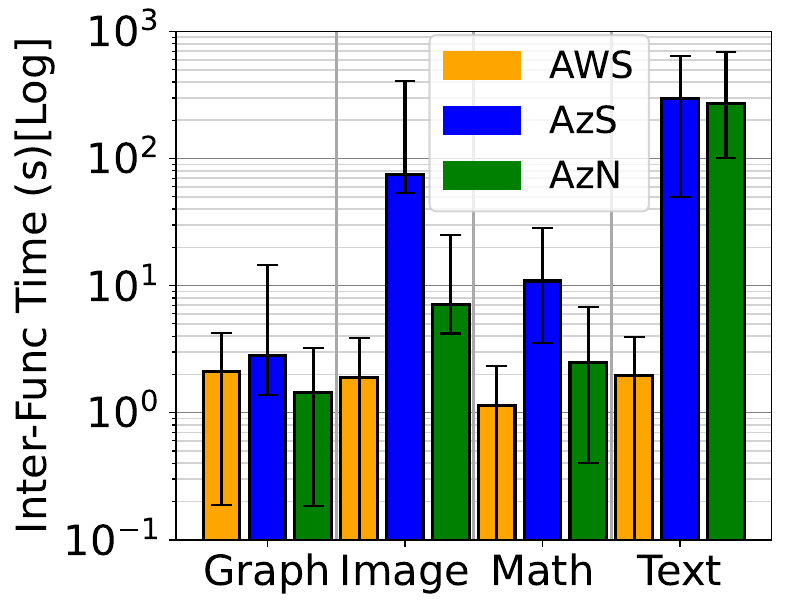}%
    \label{fig:inter-fn-data-transfer-workloads:graph}

  \caption{Inter-function latencies at 1~RPS and medium payload (cold starts omitted)}
\label{fig:inter-fn-data-transfer-wf-all}
\end{minipage}
\end{figure*}

This \textit{hardware heterogeneity} is also present in other global regions, with 3--4 architectures in US East and Central EU for both CSP, which impacts the performance of functions \modc{(Fig.~\ref{fig:appendix-plot-1} in Appendix~\ref{sec:appendix-func-variability-plots})}.
Such function performance directly affects the E2E latencies of workflows. In Fig.~\ref{fig:graph-img-wf-exec-times}, we show \textit{function execution times} for all functions of the Graph and Image workflows. These reaffirm that Azure functions perform faster than AWS functions. 
E.g., RNET in the Image workflow takes $1.99s$ on AWS while AzS and AzN take $\approx0.4s$ on AMD and $1.2s$ on Intel.
These cumulative function execution times also affect the E2E workflow latency, based on functions that lie on its critical path. This critical path functions time  (excluding inter-function time) for Graph and Image workflows is slower on AWS at $2.19s$ and $5.78s$, and faster on Azure at $0.61s$ and $2.42s$ on AMD and $0.95s$ and $3.15s$ on Intel.

CPU diversity also causes variability in the execution times of workflow functions. Further, containers for the same workflow can be from a mix of CPU types.
The markers on the right Y axis of Fig.~\ref{fig:graph-img-wf-exec-times} show the mix of newer AMD and older Intel CPUs. This ranges from $10$:$90$ (Image on AzN) to $60$:$40$ (Image on AzS) for a given workflow run. The execution time difference on these CPUs is as high as $3\times$, consequently affecting the  
workflow E2E times for different runs of the same workflow. 
In this region, AWS happens to use the same CPU model for containers, with stable performance.

\subsection{Inter-function Communication Times}
\label{sec:results:interfunc}
\takeaway{\label{ta:inter-func:small-payload}
The inter-function latency is small for payloads $\leq 100KiB$ between functions on all platforms, with AzN performing the best and AzS the worst.}
\takeaway{\label{ta:inter-func:azure-variability}AzS and AzN exhibit greater variability in transfer latency compared to AWS.
}

We perform micro-benchmarks with a simple workflow having a function that just transfers a fixed-size payload to the next function. 
Fig.~\ref{fig:inter-fn-data-transfer} reports the \textit{inter-function time}, from the end of the first function to the start of the next function's logic. A box plot and median values over $120$ invocations are shown. 
\addc{Data-transfers between functions are handled internally by the FaaS platform and not done explicitly by the application, to ensure platform-native behavior. 
The default mechanism 
for AWS Step Functions is based on Amazon SQS queues, AzS uses Azure Control Queues and Blobs, and AzN uses Event Hubs and Page/Block Blobs.}
We omit cold-start outliers.
The time taken for smaller payloads $\leq 100KiB$ is small (Fig.~\ref{fig:inter-fn-data-transfer:small}): $\leq 100ms$ for AzN, $\leq 150ms$ for AWS and $\leq 300ms$ for AzS. 

For $\leq 256KiB$, AWS Step Functions are expected to use a queue-based mechanism,\footnote{\href{https://docs.aws.amazon.com/step-functions/latest/dg/limits-overview.html}{Step Functions service quotas}, AWS, 2024}
and the transfer times are dominated by the network latency than bandwidth limits.
For larger messages, developers need to use custom means, e.g., by storing and loading from AWS S3 files, which limits the intuitiveness of Step Functions. That said, the inter-function latency for AWS is \emph{linear} for the smaller payload sizes.

AzS 
uses Azure Queues 
for payloads $\leq 45KiB$,\footnote{\href{https://learn.microsoft.com/en-us/azure/azure-resource-manager/management/azure-subscription-service-limits}{Azure subscription and service limits}, Microsoft, 2024} 
which when base64 encoded grows to the $64KiB$ message size limit\footnote{\href{https://github.com/Azure/durabletask/blob/main/src/DurableTask.AzureStorage/MessageManager.cs}{Azure Durable Task, MessageManager.cs}, Microsoft, 2024}.
For larger payloads, it automatically switches to Azure Blobs by compressing and asynchronously uploading the payload and internally passing a reference to the downstream function using Queues, where they are restored.\footnote{\href{https://learn.microsoft.com/en-us/azure/azure-functions/durable/durable-functions-azure-storage-provider}{Azure Storage provider (Azure Functions)}, Microsoft, 2024}
This transition to Blobs affects the performance, and we see the slope change past $45KiB$ (Fig.~\ref{fig:inter-fn-data-transfer:small}); this further degrades at $256KiB$ (Fig.~\ref{fig:inter-fn-data-transfer:log}) due to the way Blobs are handled.

AzN uses Azure Event Hubs, 
with better performance than Queues, for smaller payloads up to $9000KiB$ and Blobs for larger ones.
Interestingly, they split smaller messages into $30KiB$ events and larger payloads into $500KiB$ files, and use parallel and asynchronous transfers.
So
AzN is uniformly faster than AWS and AzS (Fig.~\ref{fig:inter-fn-data-transfer:small}) until $9000KiB$ (Fig.~\ref{fig:inter-fn-data-transfer:log}), when it switches to Blobs and the latency rises.

Also, the variability in inter-function latencies is tight for AWS across its supported payload sizes (Fig.~\ref{fig:inter-fn-data-transfer:box}). In contrast, AzS and AzN exhibit high variability, which deteriorates for larger payload sizes. 
For larger payloads, AzS uses Blob files and AzN splits/rejoins messages, causes variability. 
AzS and AzN exhibit failures for a larger payload transfers of $\geq 4MiB$; 
at $16MiB$, failures during transfers are routine.
\addc{We show effects of payload sizes on workflows in Appendix~\ref{sec:app-int}.}

\takeaway{\label{ta:inter-func:workflow-1}As the workflow length increases, the inter-function latency significantly increases in AzS and AzN; this is less pronounced in AWS.}

Fig.~\ref{fig:inter-fn-data-transfer-wf-all} shows the median inter-function latencies (log scale) on the workflow's critical path, i.e., directly contributes to the E2E time. The error bars indicate Q1--Q3 values over 300 invocations ($300s$).
The length of the workflow significantly affects the inter-function latencies. 
E.g, both Graph and Image have similar fan-out degrees and payload sizes, but Image is longer than Graph (Fig.~\ref{fig:workflows:dag}). \modc{For Graph, AzS has the highest median latency ($2.9s$) and AzN the least ($1.2s$), while AWS is in-between ($2s$). In contrast, Azure latencies for Image are much higher ($73s$ for AzS, $6s$ for AzN),
while comparable for AWS ($2s$).}
In Azure, before every function execution, the entire workflow history is retrieved, and this increases with longer paths. This is observed for Image and Text, where the last two edges on average take $3.2s$, $52.4\%$ more than the first two edges.
AzS uses Azure Storage Tables that are slower than AzN 
which uses the in-memory table for orchestrators.\footnote{\href{https://microsoft.github.io/durabletask-netherite/)}{Durable TaskNetherite Github}, Microsoft 2024}
Math and Text have larger fan-out degrees of 6 and 10. Here, AzS and AzN both incur larger inter-function overheads compared to AWS at $1s$, $10s$, and $2.1s$ for Math on AWS, AzS, and AzN, and $1.95s$, $300s$, and $297s$ for Text.

AWS consistently outperforms both AzS and AzN for all workflows, except Graph, where AzN is slightly faster. For Graph, AWS is 40\% faster than AzS ($2s$ vs. $2.9s$), but AzN is $40\%$ faster than AWS ($1.2s$ vs. $2s$). For Image, AWS is $97\%$ faster than AzS ($2s$ vs. $73s$) and $67\%$ faster than AzN ($2s$ vs. $6s$). This suggests that AWS has more efficient inter-function data transfer, maintaining stable latency even as workflows increase in node count/path length.  
AzS performs consistently worse than AzN, with $2.4\times$ higher latency in Graph, $12\times$ in Image, $\approx 5\times$ in Math, and more so for complex workflows.

\takeaway{\label{ta:inter-func:workflow-3}
Inter-function latencies for AWS is not sensitive to the request rate, while AzS and AzN show an increase with the rate, limiting their scalability.
}

\begin{figure}[t!]

\centering
 \includegraphics[width=0.48\columnwidth]{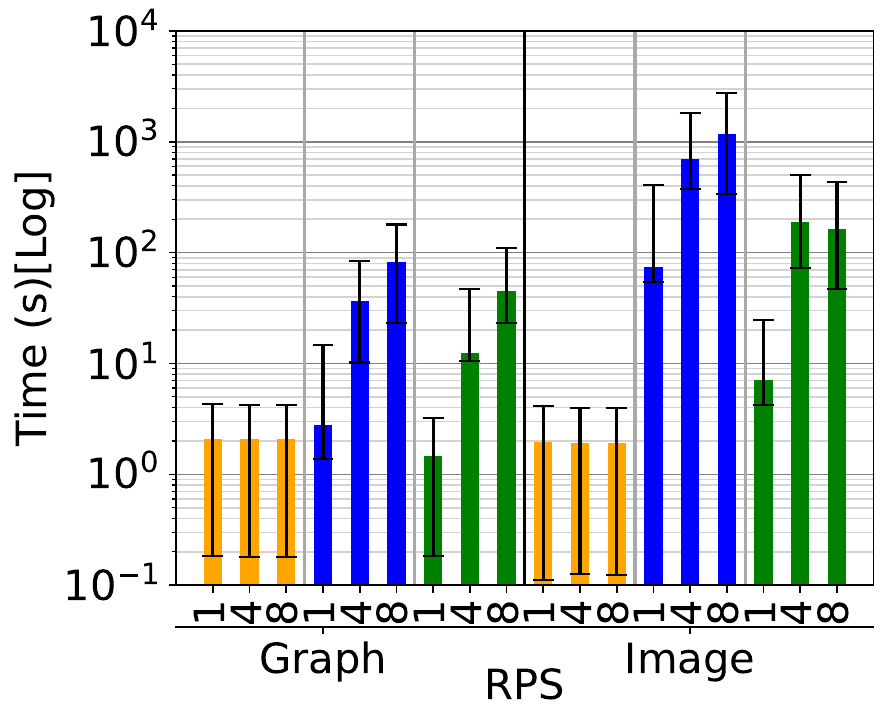}%
 \caption{Inter-function latencies for Graph and Image workflows while varying RPS (omitting cold-starts)}
\label{fig:inter-fn-graph-rps}    

\end{figure}

As the RPS increases $1$ to $8$, there is no increase in the inter-function latencies on AWS from for both Graph and Image (Fig.~\ref{fig:inter-fn-graph-rps}), remaining at $2s$ for Graph and $1.9s$ for Image. Hence, irrespective of the workflow structure, the inter-function performance of AWS scales well with increasing rate. This, along with the favorable cold-start scaling (\ref{ta:cold-start:vm-start}) and sandboxed function performance (\ref{ta:func:variability-2}), leads to overall positive scaling performance of AWS FaaS workflows with input rate (\ref{ta:scaling:up}).

However, for AzS and AzN, a rise in RPS translates to more frequent payload transfers and, importantly, a greater number of orchestrator operations within the same time period. The AzS and AzN orchestrators perform several lookups and updates on the workflow history between two function calls, stressing the storage accounts.
E.g., for Graph, when we go from 1 RPS to 4 RPS, AzS goes from $2.3s$ to $36s$, and AzN from $1.45s$ to $12s$, which are more than $4\times$. This is similar for Image, with AzS increasing by $87.5\times$ and AzN by $28.5\times$.

Overall, the inter-function latencies are lesser for AzN than AzS for both Graph and Image workflows, as it benefits from the partitioning~($PC=32$) of messages to separate queues, one per worker. In contrast, AzS uses Storage Queues to transfer messages and Storage Tables for the workflow state updates, both of which perform and scale poorly.

\begin{figure*}[t!]
\centering%
\subfloat[Trip Booking 1RPS]{
    \includegraphics[width=0.5\columnwidth]{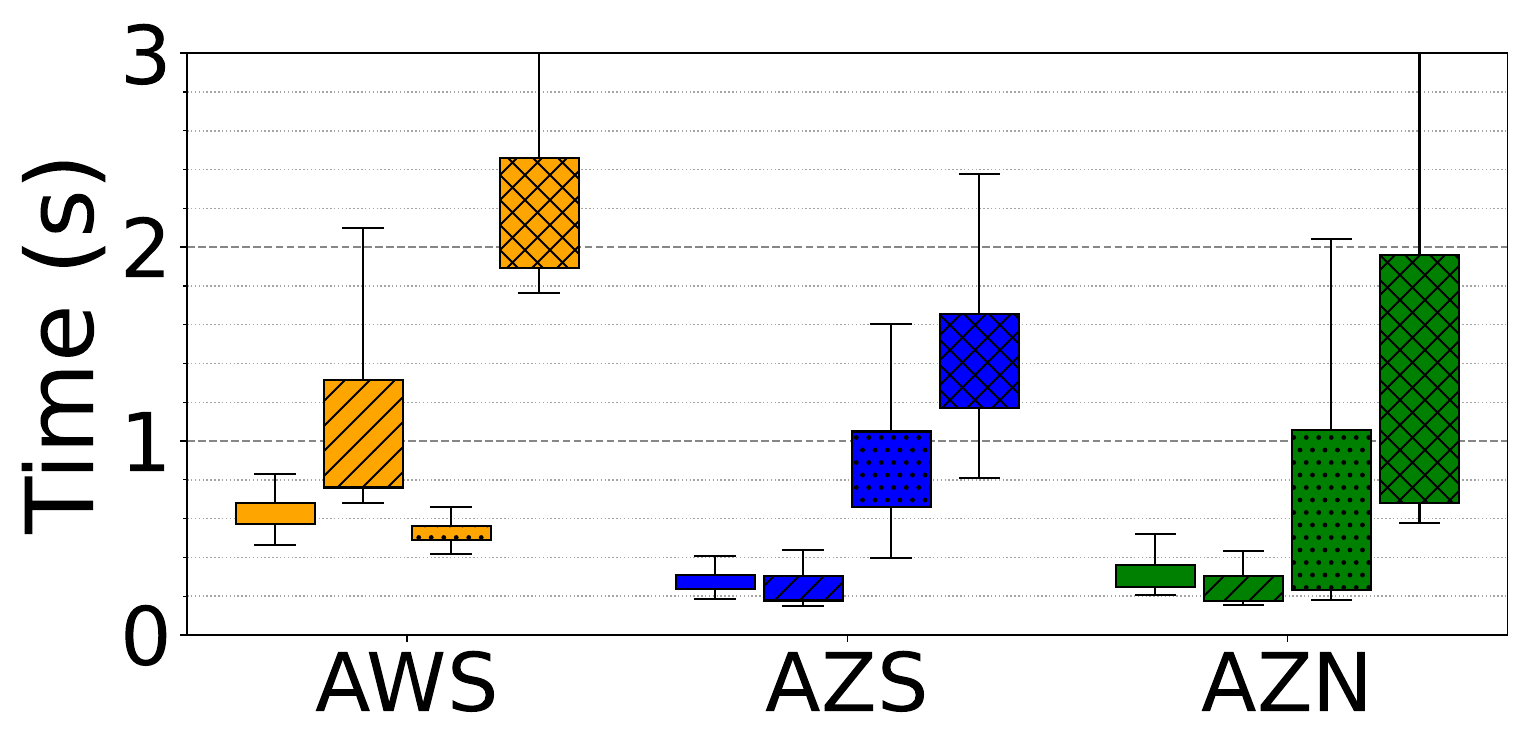}%
    \label{fig:box-saga-1rps}%
  }%
  \subfloat[Trip Booking 4RPS]{%
    \includegraphics[width=0.5\columnwidth]{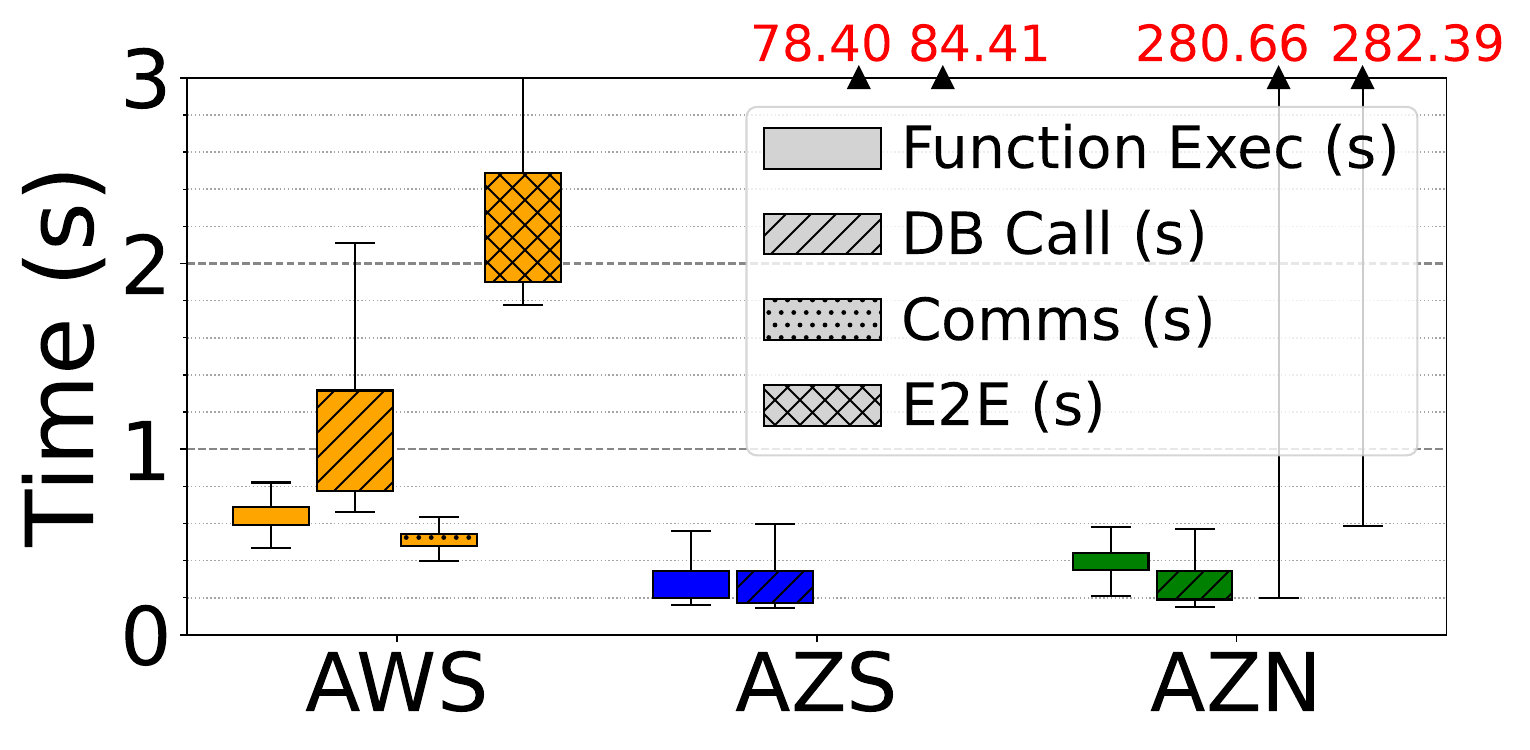}%
    \label{fig:box-saga-4rps}%
  }%

  \subfloat[Trip Booking 8RPS]{%
    \includegraphics[width=0.5\columnwidth]{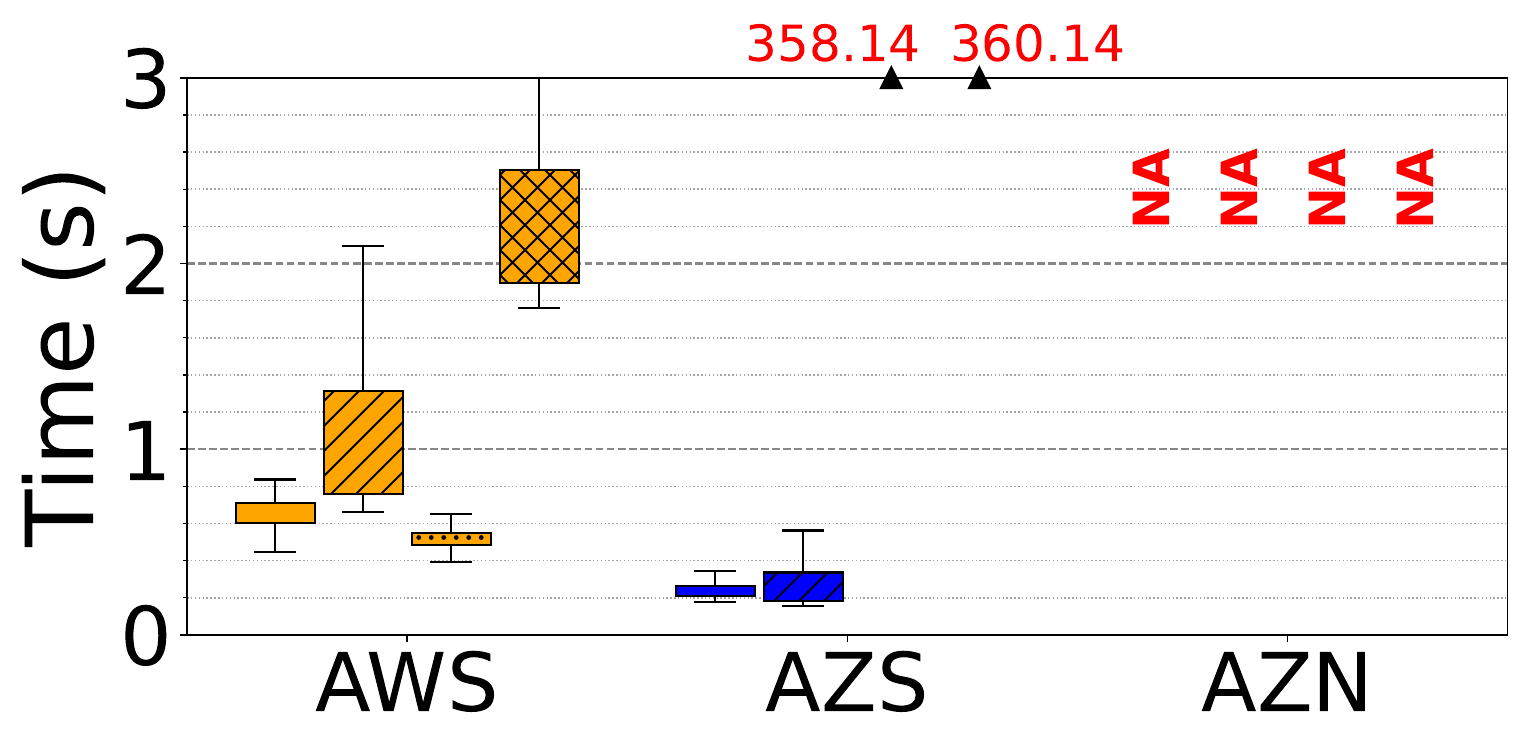}%
    \label{fig:box-saga-8rps}%
  }%
  \subfloat[Trip Booking Alibaba]{%
    \includegraphics[width=0.5\columnwidth]{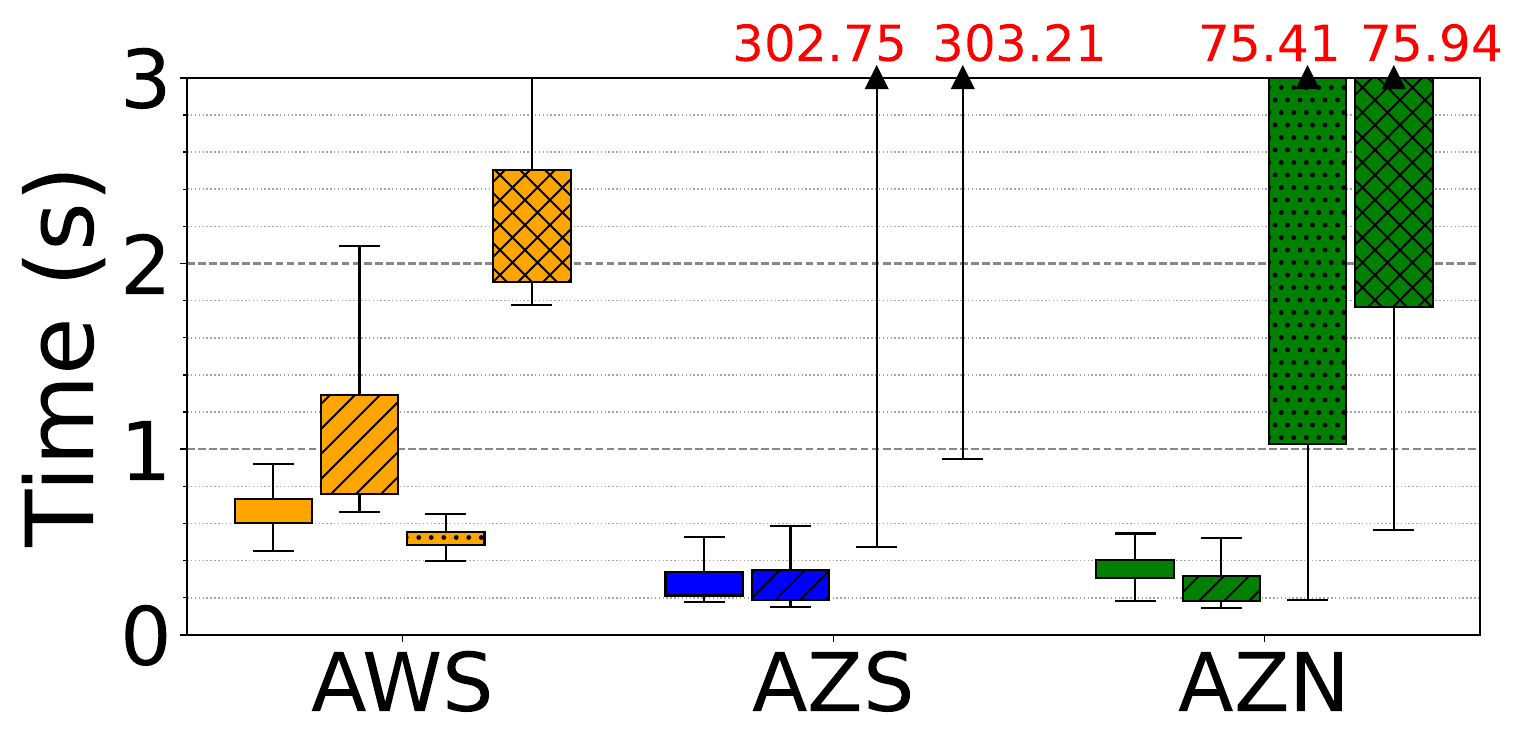}%
    \label{fig:box-saga-alibaba}%
  }%
\caption{Performance of Trip Booking Workflow}
\label{fig:saga-wfs}
\end{figure*}

\takeaway{\addcminor{Azure Durable Functions show lower median per-invocation database call latency than AWS across the evaluated request rates.\label{ta-saga}}}

\modcminor{
We evaluate the \textit{Trip Booking workflow} under four workloads (Fig.~\ref{fig:saga-wfs}).
At low static load (1~RPS), Azure exhibits lower median DB-call latencies than AWS, while inter-function communication remains below $1s$ for all platforms, resulting in comparable E2E latencies. As RPS increases (4~RPS and 8~RPS), Azure continues to show lower per-function and DB-call times, but its inter-function delays grow and dominate the E2E latency. AWS maintains near-constant inter-function times, with stable E2E performance even at higher loads. This trend even sustains for the bursty Alibaba workload.
}

\modcminor{
A plausible explanation is due to differences in execution models. Azure Durable Functions execute on longer-lived VM-based workers, which may reuse database client sessions/connections, contributing to lower observed DB-call medians. AWS Lambda functions execute in more ephemeral containers, where database connections do not persist, and initialization/connection setup overheads are more frequent. While this interpretation aligns with the measured latency differences, the orchestration and inter-function communications ultimately dominate the overall workflow latency at higher loads.
}

\subsection{Workflow Execution Times }\label{sec:results:workflow-exec}
\takeaway{With an increase in the workflow length, the end-to-end latency significantly increases in AzS and AzN both, while the effect of length is less pronounced in AWS, mostly due to the increased inter-function overheads.\label{ta:wf-len}}
\takeaway{An increase in the degree of parallelism has little impact on the end-to-end latency for AWS, while the e2e latency increases in AzS and AzN, both.\label{ta:wf:par}}

\begin{figure}[t!]
\centering%
    \includegraphics[width=0.7\columnwidth]{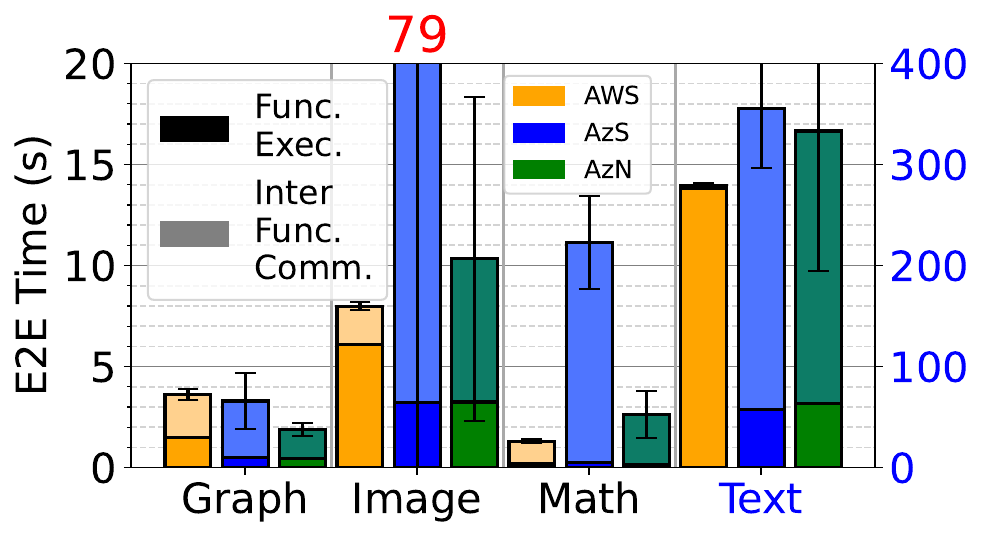}
\caption{E2E WF execution time on Medium Payload, 1 RPS.}
\label{fig:summary}
\end{figure}

Here we examine the time spent along the \textit{critical path} of the workflow, i.e., the path in the workflow from the source function to the sink function that takes the maximum execution time. Fig.~\ref{fig:summary} reports the E2E time of the four workflows at 1 RPS for medium payload. This time is split into the function execution latencies (bottom stack) and inter-function latencies (top) along the critical path. We run these for $5mins$ at 1~RPS for medium payload and omit the coldstart overheads.

For 3 out of 4 workflows -- Image, Math, and Text -- AWS has the least workflow E2E times at
$8.0s$, $1.3s$ and $278s$, and is $22.8\%$, $50.1\%$ and $16.4\%$ faster than AzN, the next best. For all four workflows, AWS has lower inter-function latency, as we saw earlier for medium payload and 1 RPS (\S~\ref{sec:results:interfunc}). 
However, its function execution times are slower than AzN and AzS for Image and Text, and comparable to AzN on Math. The net effect results in these E2E times. AWS is slower than AzS and AzN for Graph, taking $3.6s$ or $91.6\%$ more than AzN, due to much slower function execution time. In general, function execution times tend to dominate for AWS, taking $52$--$96\%$ of the total E2E times for Graph, Image, and Text workflows, with ENCR in Text taking $\approx 280s$ or $99.3\%$ of total time. These are consistent with our results in \S~\ref{sec:results:func-exec}.

On the other hand, Azure runtimes are dominated by inter-function latencies that account for $83$--$96\%$ of E2E times for all workflows.
AzN is faster than AzS for all workflows, by $42.9\%$, $86.7\%$, $50.1\%$, and $6.3\%$ for Graph, Image, Math, and Text. AzN usually has lower inter-function latencies due to better storage performance, and its function execution times are comparable or sightly better than AzS (Fig.~\ref{fig:singleton-wf-exec-cpu-times:zoom-out}).

In summary, our earlier piecewise observations on function execution and inter-function latencies cumulatively affect the E2E workflow latency, and the net result is consistent with earlier expectations. This confirms that our analysis can extend to other workflows and workloads.

\subsection{Cost Analysis}
\label{sec:results:costs}

Cost is a key factor in using cloud services. While AWS Step Functions have a simple and transparent costing, Azure's costs are complex to compute and have not been investigated before. We develop a cost model to help arrive at an accurate costing and analyze these across our workflows.

The cost for a workflow invocation has three parts: \textit{function execution costs} ($\alpha$), \textit{inter-function transfer costs} ($\beta$), and \textit{workflow orchestration costs} ($\gamma$), with the \textit{total cost} $\kappa = \alpha + \beta + \gamma$. Let a given workflow have $n$ functions and $e$ dataflow edges between them. Let the function execution times (seconds) be $t_1,t_2 \cdots t_n$, and their configured memory (GB) be $m_1,m_2...m_n$.
Say the data transfer sizes (GB) on each edge are $D = \{ d_1,d_2...d_e \}$, with the input to the workflow as $d_0$.

\subsubsection{AWS Cost Model}
For AWS, each function execution is billed based on the memory assigned to the container and function execution time. 
The total function execution cost is $\alpha_{aws}=k^{func}_{aws} \times \sum_{1\leq i \leq n}(t_i^{aws} \times m_{i})$. Here, $k^{func}_{aws}$ is the cost per GB-sec of the function execution, and depends on the region and container type (e.g., x86, ARM, etc.).\footnote{\href{https://aws.amazon.com/step-functions/pricing/}{AWS Step Functions Pricing}, AWS 2024} 
AWS does not separately bill for inter-function payload transfers since message sizes have a small upper-bound and only bills for orchestration costs of $k^{xfr}_{aws}$ for every $1000$ state transitions in the Step Function.
Hence, we set $\beta_{aws}=0$ and compute the orchestration costs from the number of workflow state transitions, $\gamma_{aws}=\frac{k^{xfr}_{aws} \times e}{1000}$.
So, the total AWS cost is: 
\[ \kappa_{aws} = \big( k^{func}_{aws} \times \sum_{1\leq i \leq n}(t_i^{aws}\times m_{i}) \big) + \Big( \frac{k^{xfr}_{aws} \times e}{1000} \Big)\]
where $k^{func}_{aws}=\$0.00001667$ per GB-sec and $k^{xfr}_{aws}=\$0.0285$ per state transition for our experiments.

\subsubsection{AzS Cost Model}
The function execution cost for AzS is similar to AWS, $\alpha_{azs}=k^{func}_{azs} \times \sum_{1\leq i \leq n}(t_i^{azs} \times m_{i})$, with a cost coefficient of $k^{func}_{azs}=\$0.000016$ per GB-sec for our region.\footnote{\href{https://azure.microsoft.com/en-in/pricing/details/functions}{Azure functions Pricing}, Microsoft 2024
} \modc{Azure does not use containers per function but instead spins up VMs that may run multiple functions for the same workflow.} Hence, the memory used by functions is calculated from the XFaaS provenance logs.

The inter-function costs for AzS depend on the data transfer size between the two functions, and differ based on whether Queues are used ($<45KB$) or Blobs ($\geq45KB$).
Let the total number of edges in the DAG with less than $<45KB$ of data transfer sizes be $e_q = |d_i|~\forall d_i <45KB \in D$. We are billed for each put and get operations of these messages on the queue at $k_{azs}^{qtxn}$ for every $10k$ operations.\footnote{\href{https://azure.microsoft.com/en-in/pricing/details/storage/queues/}{Azure Storage Queues Pricing}, Microsoft 2024}
Let edge with sizes $\geq45KB$ be $e_b = e - e_q$.
We are billed for each Blob put and get operations  at $k_{azs}^{bput}$ and $k_{azs}^{bget}$ for every $10k$ operations.
So, the inter-function transfer costs is:
\[\scalebox{0.9}{$\beta_{azs} = \Big(k_{azs}^{qtxn}\times \frac{2 \times e_q}{10000} \Big) + 
\Big( (k_{azs}^{bput}+k_{azs}^{bget}) \times \frac{e_b}{10000} \Big) 
$}\]
where
$k_{azs}^{qtxn}=\$0.0004$, %
$k_{azs}^{bput}=\$0.065$ and $k_{azs}^{bget}=\$0.005$.

Lastly, the orchestrator uses Azure Storage Tables to write the input and output of every function. The total data (GB) written and read is:
$d^{orch}_{tbl}=d_0 + \big( 2 \times \sum_{1 \leq i < n}d_i\big) + d_n$, with the initial ($d_0$)/final ($d_n$) payloads only being written/read, with a cost of $k_{azs}^{trw}$ paid for every GB.\footnote{\href{https://azure.microsoft.com/en-us/pricing/details/storage/tables/}{Azure Storage Table Pricing}, Microsoft 2024}
There are also read and write costs for the tables at each of the $n$ DAG vertices, billed for every $10k$ transactions.
The orchestration cost is:
\[ \gamma_{azs} = k_{azs}^{trw} \times d^{orch}_{tbl} + k_{azs}^{ttxn} \times \frac{2 \times n}{10000}\] 
For our setup, we have $k_{azs}^{trw}=\$0.045$ and $k_{azs}^{ttxn}=\$0.0004$.

\subsubsection{AzN Cost Model}
The function execution cost for AzN is identical to AzS, $\alpha_{azn} = \alpha_{azs}$.
Inter-function data transfers in AzN occur through Event Hubs, with a maximum size of $9000KiB$. Its pricing depends on the send and receive bandwidth usage, billed at $k^{ehbw}_{azs}$ per throughput unit (TU) used.\footnote{\href{https://azure.microsoft.com/en-in/pricing/details/event-hubs/}{Azure Event Hubs Pricing}, Microsoft 2024} Each TU offers an average of $1 MB/s$ of receive and $2 MB/s$ of send bandwidth in an hour.
The total payload sent and received (in GB) across all edges of the DAG is $d^{xfr}_{ehub} = \sum_{0 \leq i \leq n} d_i$.
If $t_{e2e}$ is the E2E workflow execution time for the invocation (in seconds), the average receive TUs used for the workflow is $\frac{d^{xfr}_{ehub} \times 1000}{t_{e2e} / 3600}$, and the send TUs is $\frac{d^{xfr}_{ehub} \times 1000}{2 \cdot t_{e2e} / 3600}$.
So the data transfer cost is the larger of the two, i.e., the former: 
\[ \beta_{azn}=k^{ehbw}_{azn} \times \frac{d^{xfr}_{ehub} \times 1000}{t_{e2e} / 3600} \]
In our experiments, $k^{ehbw}_{azn}=\$0.015$ per TU.

AzN uses Page Blobs to maintain the orchestration state of the workflow execution.\footnote{\href{https://azure.microsoft.com/en-us/pricing/details/storage/page-blobs/}{Azure Page Blobs Pricing}, Microsoft 2024} Their operations are billed at $k^{pbtxn}_{azn}$ per $10k$ reads/writes of $64KB$ blocks with a limit of $1MB$, after which no further charges are incurred. 
Let $\mathcal{D}_{pb}(d_i)=\frac{\min((d_i \times 1024 \times 1024),~~1024)}{64}$ give the number of billed operations for a data item of size $d_i$ GB.
So the total number of operations for the DAG is $d^{txn}_{pb} = \mathcal{D}_{pb}(d_0) + \big( 2 \times \sum_{1\leq i < n} \mathcal{D}_{pb}(d_i) \big) + \mathcal{D}_{pb}(d_n)$.
There is also a storage cost per GB on the Page Blob of $k^{pbrw}_{azn}$, with a total transfer size for the DAG of 
$d^{orch}_{pb}=d_0 + \big( 2 \times \sum_{1 \leq i < n}d_i\big) + d_n$.
So, the orchestration cost is:
\[ \gamma_{azn} = k^{pbtxn}_{azn} \times d^{txn}_{pb} + k^{pbrw}_{azn} \times d^{orch}_{pb} \]
In our setup, $k^{pbtxn}_{azn}=\$0.00036$ and $k^{pbrw}_{azn}=\$0.045$.

\subsubsection{Analysis using Cost Model}

The total cost of a workflow invocation ($\kappa$) has three components: \textit{function execution costs} ($\alpha$), \textit{inter-function transfer costs} ($\beta$), and \textit{workflow orchestration costs} ($\gamma$).  While the costs for short workflow runs (10s of mins) are tiny, they add up over time. So, we report costs in $\mud$ (US$\$\times10^{-6}$).

\addc{We obtain actual costs from cloud provider APIs. 
The function execution costs ($\alpha$) come from AWS CloudWatch metrics, Azure monitoring APIs and our execution logs. 
The orchestration costs ($\gamma$) are sourced from AWS billing and Azure storage account transaction APIs. 
For Azure, the portal provides the combined cost of inter-function transfer ($\beta$) and orchestration ($\gamma$). We split $\beta$ and $\gamma$ using transaction counts from storage account APIs: for AzS, we use the ratio of queue/blob transactions (for $\beta$) to table transactions (for $\gamma$); for AzN, we report $\beta+\gamma$ combined as inter-function costs.}

\begin{figure}[t!]
\centering%
\centering
  \subfloat[All Workflows]{
    \includegraphics[width=0.45\columnwidth]{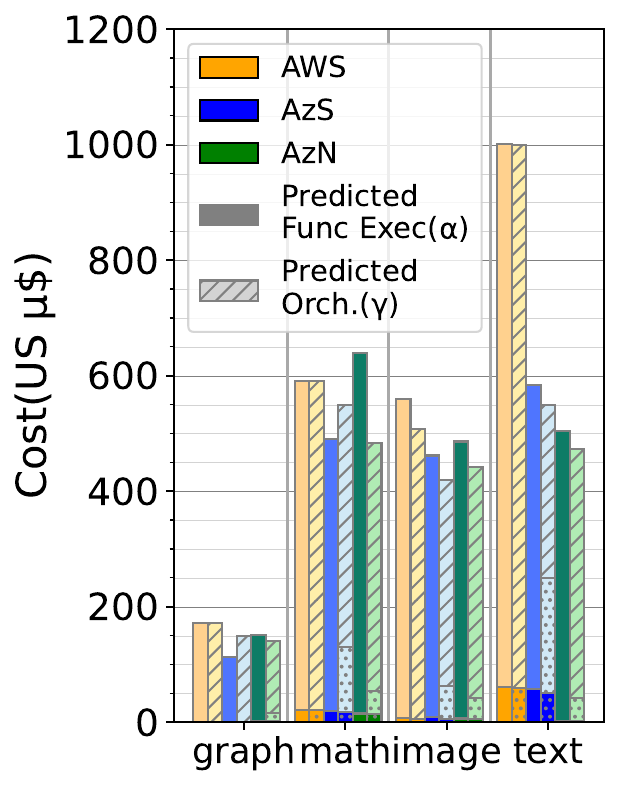}
    \label{fig:all-wf-cost}
  }~
  \subfloat[Graph Workflow]{
  \includegraphics[width=0.45\columnwidth]{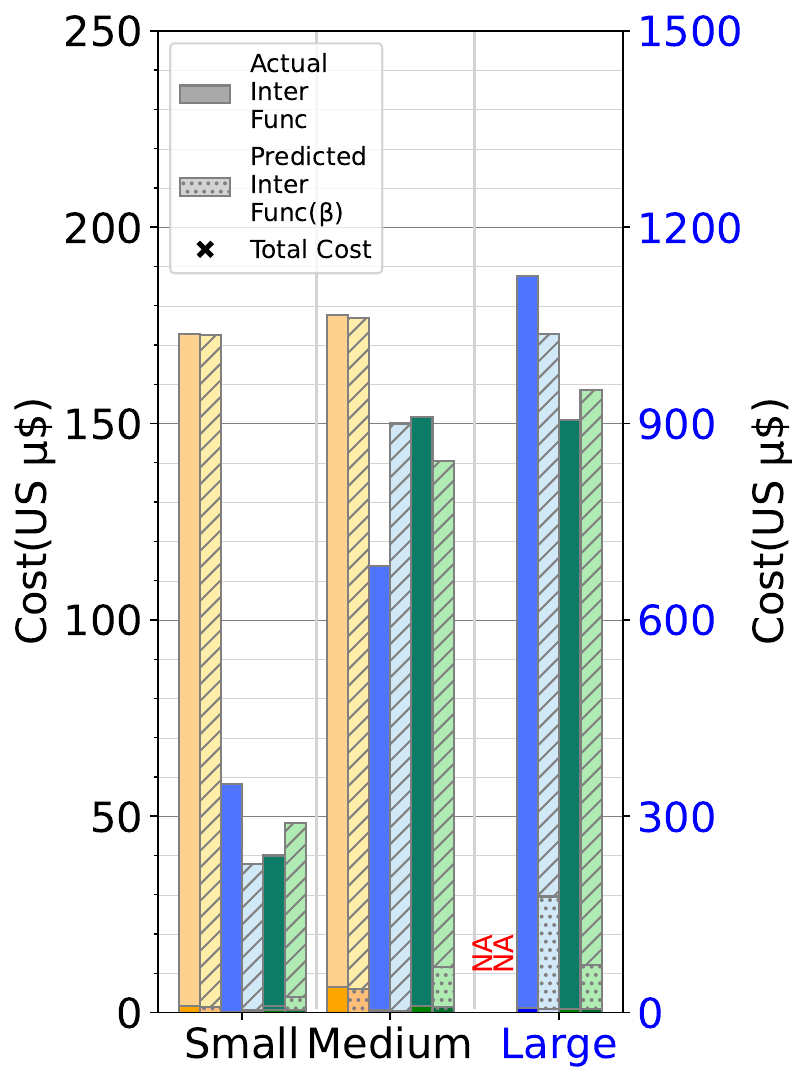}
    \label{fig:graph-cost}
  }
\caption{Cost Analysis for Graph with different payload sizes, and for different workflows on medium payload size with static 1~RPS. \addc{Costs on Y axis are reported in micro-dollars ($10^{-6}$).}}
\label{fig:cost-analysis}
\end{figure}

In Fig.~\ref{fig:all-wf-cost}, we report \addc{actual costs (left bar, hatched) and the predicted costs using our cost model (right bar, dark)} per execution, averaged over $5mins$ for each workflow, with a medium payload at 1~RPS on all three platforms. We also drill down into the Graph Workflow with three payload sizes at 1~RPS (Fig.~\ref{fig:graph-cost}). \addc{Both figures show stacked bars with function execution ($\alpha$) at the bottom, inter-function transfer ($\beta$) in the middle, and orchestration ($\gamma$) at the top.}
\addc{For AzS, we split the actual $\beta$ and $\gamma$ using the same ratio as the predicted costs to ensure comparability. For AzN, we report $\beta+\gamma$ combined as inter-function costs since Event Hubs throughput units are shared between orchestration control messages and data transfer.}
`NA' indicates unsupported payload size (AWS), timeouts (Text and Image on AzS and AzN), or negligible cost that is not reported.

\takeaway{\label{ta:cost-analysis:0}Our cost model provides an accurate estimate of the workflow execution costs for AWS and AzS, is reasonable even for complex Azure billing.}

\modc{The $R^2$ values for our cost models for AWS, AzS and AzN are $0.98$, $0.80$ and $0.86$, when considering all $(\alpha+\beta+\gamma)$.}.

\addc{Most mispredictions in Azure are due to hidden polling costs not seen in logs or easily observable. Durable Functions continuously poll storage queues (in AzS) and Event Hubs (in AzN) to detect new work, incurring read/receive operations that are opaque to the application. Also, the Azure cost APIs takes $\approx24h$ to update and have been unreliable, making ground truth costs partly hidden. This variation is higher in AzS as it uses more queues that are continuously billed.}

\modc{There are no errors in function execution costs since we use the documented pricing models based on memory usage and execution time for AWS, AzS, and AzN.
}

\takeaway{\label{ta:cost-analysis:1}For workflows with a large number of functions and edges, AWS is costly, while for those with larger payloads, AzN and AzS are costly. AzN balances cost and performance.
}

The Graph workflow is the smallest of all, and we report the costs for running it on AWS, AzS, and AzN in Fig.~\ref{fig:graph-cost} for 1~RPS and all 3 payload sizes. 

\modc{For small payloads, AWS, AzS, and AzN respectively cost $\mud173$, $\mud38$, and $\mud40$; 
while for medium payloads they cost $\mud177$, $\mud150$ and $\mud140$.}
AzN is the cheapest due to the
\textit{per hour} throughput unit costing. 
\modc{For the Graph workflow with medium payloads, inter-function transfer costs $(\beta)$ in AzS are $\approx 250\%$ higher than AzN.}
\modc{The total cost for AWS is $\approx 252\%$ higher than AzN for small payloads.}

When we increase the \textit{payload size} from small to medium, \textit{AWS} has a negligible increase in cost, by $0.006\%$. 
The function execution time increase is small, and its impact ($\alpha$) on the total cost is tiny. 
Changing the payload does not impact the number of state transitions in AWS, and $\gamma$ remains the same.

\modc{For \textit{AzS}, as we go from small to medium to large, the costs increase from $\mud38$ to $\mud140$ to $\mud1020$.}
The workflow orchestration costs remain constant
as it depends on the workflow structure and is a large part of the total cost.
\modc{For \textit{AzN}, the costs increase much more, from $\mud40$ to $\mud220$ to $\mud960$ (predicted), mostly due to higher data transfer costs.}

As the workflow becomes \textit{more complex} with more functions and edges, AzN becomes costlier than AWS while AzS is the cheapest (Fig.~\ref{fig:all-wf-cost}). 
For \textit{Math} (15 funcs, 20 edges) and \textit{Image} (9 funcs, 10 edges), 
\modc{AzN is $10.5\%$ and $53.5\%$ (predicted) costlier than AWS due to higher inter-function transfer and orchestration costs from increased accesses to the history table.}
In contrast, 
\modc{AzS is $25.4\%$ and $163\%$ cheaper than AzN for Math and Image workflows.}

AWS is the costliest for \textit{Text} (25 funcs, 33 edges), having the most functions and edges. 
While AzS and AzN costs depend both on data transfers and state transitions across nodes and edges in the workflow, AWS is only affected by the latter. 
So, more complex workflows are costlier for AWS, while those with larger payloads are costlier for Azure.
AzN is cheaper for small workflows and yet offers performance comparable to AWS for non-intense workloads.

\takeaway{\label{ta:cost-analysis:2}The bulk of the workflow execution cost is from the orchestration costs $\gamma$ for all three platforms.}

The dominant cost is the workflow orchestration costs (Fig.~\ref{fig:all-wf-cost}). These amount to $59\%$ to $92\%$ of the total costs for AzS and AzN, and over $99\%$ for AWS.
The next highest for AzS and AzN is the inter-function data transfer costs at $8.5\%$ to $41\%$. Interestingly, \textit{the (predicted) function execution costs contribute the least,} and stay well within the free tier.
This means that the \textit{structure} of the workflow (number of functions and edges) and \textit{payload sizes} are the key cost factor for FaaS workflow, unlike for FaaS functions where the function execution costs dominate.

\section{\addc{Summary of Design Insights}}
\label{sec:insights}

\addc{Our characterization highlights several recurring system-level observations that help understand ``the good'', ``the bad'' and ``the ugly'' regarding performance, stability, and cost differences between AWS and Azure FaaS workflow platforms. Together, these suggest several practical considerations for designing and deploying serverless workflows across clouds.}

\addc{\subsubsection{Elasticity and Stability Trade-offs}}

\addc{AWS Step Functions generally exhibit responsive scaling behavior and maintain stable performance across a range of request rates (\ref{ta:aws-scale}). In contrast, Azure Durable Functions (AzS and AzN) show more variability when the incoming rate increases sharply (\ref{ta:scaling:complex}). This is related to the storage-backed orchestration queues used by Azure's scale controller, which leads to delayed or uneven scaling decisions before the system stabilizes.  
AzN is more responsive at scaling up than AzS, though it remains limited by the peak number of containers it can provision (\ref{ta:azs-scale}). For stable, predictable workloads, AzN can achieve comparable performance to AWS but with fewer active containers (\ref{ta:scaling:up}).}

\addc{\textit{\underline{Implication:}} AWS is better suited for bursty or latency-sensitive workloads, while Azure Netherite can be a reasonable choice for steady workloads with moderate request rates, provided that the configurations are carefully tuned. Developers can also trade-off between cost and performance in AWS by tuning memory allocation, enabling predictable scaling behavior and latency optimization (\ref{ta:aws-knobs}).}

\addc{\subsubsection{Cold-start Cascades and Latency Effects}}

\addc{Azure Durable Functions exhibit tangible cold-start overheads across container, runtime and function initialization phases, while AWS Step Functions show negligible impact (\ref{ta:cold-start:vm-start}). These are particularly pronounced for workflows with larger deployment packages and payloads (\ref{ta:coldstart_cascading_blob}), and the function-level cold-start overhead itself depends on the size of dynamic packages and models (\ref{ta:cold-start:func-warmup}). Cold-start overheads in Azure can propagate through sequential workflow chains, as downstream functions wait for upstream container and runtime initialization (\ref{ta:coldstart_cascading}). Parallel execution paths help reduce this cascading effect, but do not eliminate it entirely.}

\addc{\textit{\underline{Implication:}} When using Azure, developers may benefit from limiting the length of workflows and considering techniques such as fusing functions~\cite{xfaas,wisefusesigm} or pre-warming~\cite{cs1,cs2} for latency-sensitive applications. Workflows with multiple parallel branches can mitigate cold-start penalties, as containers may spin up concurrently.}

\addc{\subsubsection{Function Execution Performance Trade-offs}}

\addc{No single platform consistently outperforms the others on function execution time. AzS and AzN typically execute individual functions faster than AWS Step Functions, especially for compute-intensive workloads  (\ref{ta:func:cpu-mem-intensive}). However, CPU heterogeneity within the same region causes variable function performance on both AWS and Azure (\ref{ta:func:variability-2}), which affects application predictability.}

\addc{\textit{\underline{Implication:}} Workflows with compute-dominant functions can benefit from Azure. Developers should be aware of the architectural diversity of CPUs that can affect deterministic performance for latency-sensitive applications~\cite{mahgoub2022orion}.}

\addc{\subsubsection{Orchestration \& Inter-function Communication Overheads}}

\addc{Inter-function latency remains small (under $100ms$) for payloads $\leq100KiB$ on all platforms, with AzN performing best and AzS worst (\ref{ta:inter-func:small-payload}). However, Azure's queue- and storage-based orchestration introduces higher variability (\ref{ta:inter-func:azure-variability}), whereas AWS's state-machine execution path tends to be more consistent.}

\addc{As the workflow length increases, inter-function latency grows significantly in both AzS and AzN, while remaining stable in AWS  (\ref{ta:inter-func:workflow-1}, \ref{ta:wf-len}). Similarly, increasing the degree of parallelism has little impact on AWS's end-to-end latency, but causes noticeable increases in both Azure variants (\ref{ta:wf:par}). Under higher request rates, AWS's inter-function latency remains stable, while AzS and AzN show proportional increases that limit their scaling (\ref{ta:inter-func:workflow-3}). This is evident even in database-heavy workloads: Azure exhibits good scaling for database calls with stable per-invocation latency, yet inter-function communication latency grows rapidly with load, affecting end-to-end performance (\ref{ta-saga}).}

\addc{\textit{\underline{Implication:}} Workflow designs that reduce the number of coordination points, limit unnecessary function boundaries and keep payload sizes manageable can reduce inter-function communication overhead and improve performance. Minimizing workflow length and avoiding excessive parallelism can prevent inter-function communication from becoming a scalability bottleneck in Azure~\cite{xfaas,wisefusesigm,yu2023following,faasflow2022}.}

\addc{\subsubsection{Cost Model Accuracy and Optimization Strategies}}

\addc{Our cost model provides accurate estimates for AWS and good approximations even for Azure's complex billing structure (\ref{ta:cost-analysis:0}). The bulk of workflow execution cost comes from orchestration  (\ref{ta:cost-analysis:2}). However, the cost sensitivity differs by platform and workload characteristics: AWS is costly for workflows with many functions and edges due to per-state-transition charges, while Azure is costly for workflows with larger payloads due to storage and data transfer overhead (\ref{ta:cost-analysis:1}). Among them, AzN balances cost and performance (\ref{ta:cost-analysis:1}).}

\addc{\textit{\underline{Implication:}} Cost optimization strategies should focus on reducing the number of functions and edges for AWS deployments, while Azure deployments benefit from minimizing payload sizes and data transfer volume. When both cost and performance matter, AzN may offer the best balance.}

\section{Discussions and Conclusions }
\label{sec:conclusion}

\modcminor{In this article, we have performed a comprehensive experimental study of three popular cloud \textit{FaaS workflow platforms} from AWS and Azure. Our suite of realistic workflows with diverse functions, composition patterns and workloads were invoked for \modc{$\approx 139k$ workflow instances} and $\approx 5.8M$ function calls across three regions to reveal performance and scalability characteristics across these platforms. 
Our observations are supported by detailed experiments, micro-benchmarks and platform design analysis. These expose the \textit{good} (e.g., rapid stable scaling on AWS, faster per-function execution on Azure), the \textit{bad} (e.g., message size limits in AWS, execution variability in Azure), and the \textit{ugly} (e.g., hardware heterogeneity across regions, complex pricing structures).}

\modcminor{\modc{Our design insights offer practical value for \textit{developers}}
to reason about platform selection based on workload intensity, workflow structure, payload size and tolerance to latency variability. While AWS provides consistent end-to-end stability, Azure Netherite can benefit compute-bound scenarios with limited orchestration overheads.}
\modcminor{\textit{Researchers} can build upon these findings to design mitigation strategies for the platform gaps, such as automated CSP selection, region-aware deployment, hybrid workflow partitioning~\cite{xfaas}, and intelligent pre-warming to reduce cascading cold-starts. 
\textit{Cloud Service Providers} may also use these insights to address variability, scaling behavior and pricing transparency.}

\addcminor{\textit{Limitations.} This study focuses on AWS Step Functions and Azure Durable Functions (AzS and AzN) and does not include other workflow services such as Google Workflows. The experiments were conducted primarily using Python runtimes at the specific regions during the said measurement periods; results may vary across programming languages, regions or future platform versions as cloud services evolve. While we employ both realistic and synthetic workflows, no benchmark suite can capture all possible production scenarios. These factors should be considered when generalizing the findings.}

\addcminor{As future work, we can extend the evaluation to additional runtimes such as Node.js, Go and Java to study cross-language behavior, examine premium versus basic offerings for deeper cost--performance trade-offs, and incorporate additional workflows to broaden cross-cloud comparison.}

\bibliographystyle{IEEEtran}
\bibliography{arxiv.v2}

\clearpage
\appendix
\subsection{Scaling with Input Rates}
\begin{figure}[t!]
\centering%
  \subfloat[Default, \textit{PC=12}]{%
     \includegraphics[width=0.245\textwidth]{arxiv-v2-figures/takeaway-wise-plots/scaling-and-coldstarts/graph_medium_observation1_step_step_azv2_12.pdf}
    \label{app:fig:graph-az2-step-knobs:mo8-ma1-pc12}%
  }
  \subfloat[\textit{PC=16}]{%
    \includegraphics[width=0.245\textwidth]{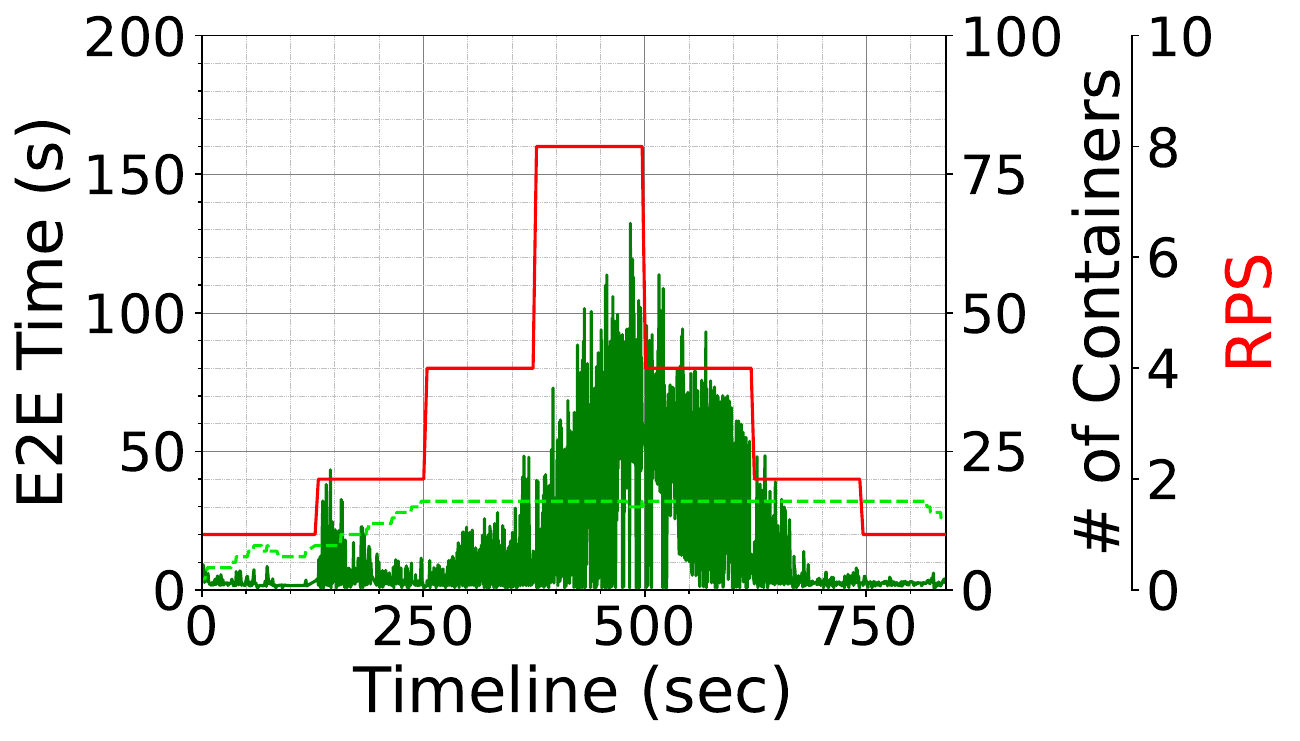}
    \label{app:fig:graph-az2-step-knobs:mo8-ma1-pc16}%
  }\\
   \subfloat[\textit{PC=24}]{%
   \includegraphics[width=0.245\textwidth]{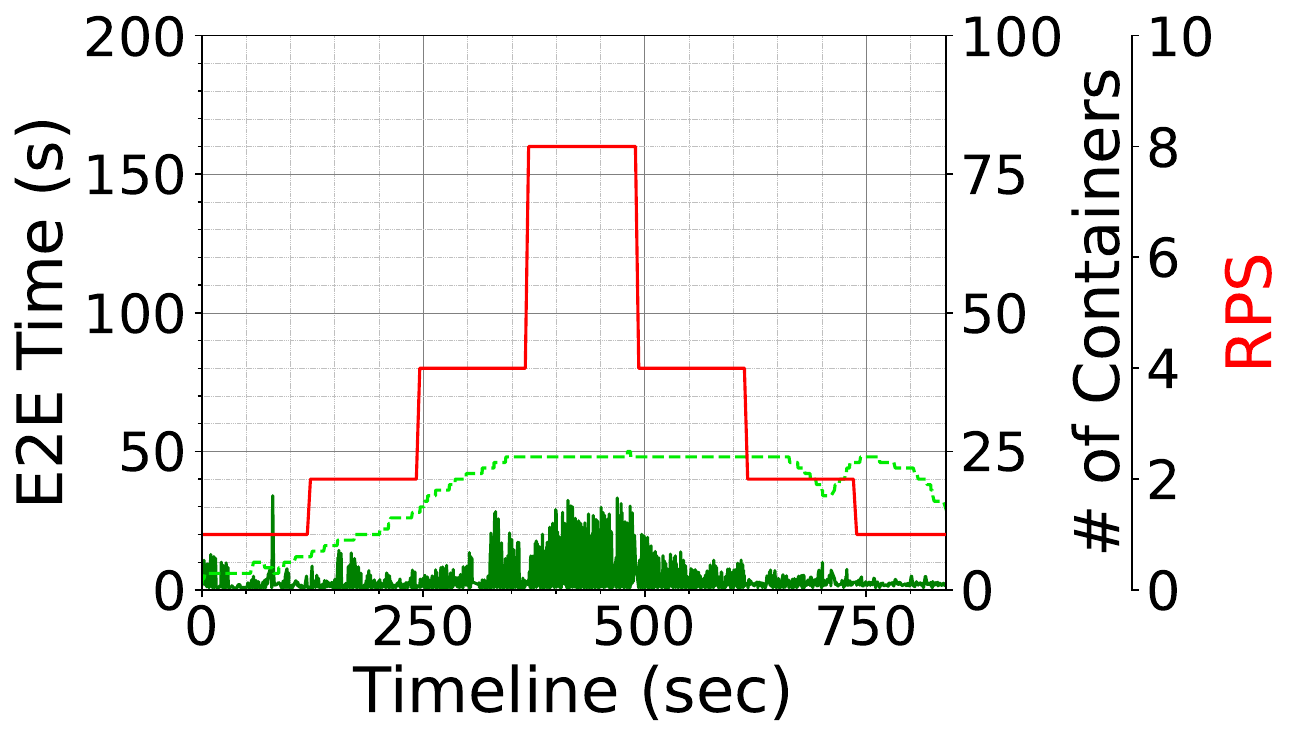}
    \label{app:fig:graph-az2-step-knobs:mo8-ma1-pc24}%
  }
  \subfloat[Article, \textit{PC=32}]{%
    \includegraphics[width=0.245\textwidth]{arxiv-v2-figures/takeaway-wise-plots/scaling-and-coldstarts/graph_medium_observation1_step_step_azv2_32.pdf}
    \label{app:fig:graph-az2-step-knobs:mo8-ma1-pc32}%
  }
\caption{AzN for \textit{Graph} Workflow with \textit{Medium} payload size and \textit{Step} Workload, using different max. partition counts configurations.}
\label{apx:azN-pc-knobs}
\end{figure}

Fig.~\ref{apx:azN-pc-knobs} reports the performance of Graph workflow with an increase in the maximum partition count (PC) on AzN, and complements Fig.~\ref{az2-step-knobs} in the main article. This supports the analysis in \S~\ref{sec:results:scaling}.

Fig.~\ref{app:fig:all-wf-timeline} reports additional plots for our experiments that could not fit within the main article. It shows the performance of the other workflows for 1 RPS and medium payload. This complements Fig.~\ref{fig:graph-timeline} in the main article that is focused on Graph workflow. It supports the analysis in \S~\ref{sec:scaling:static}.

\begin{figure}[h]
\centering%
  \includegraphics[width=0.45\textwidth]{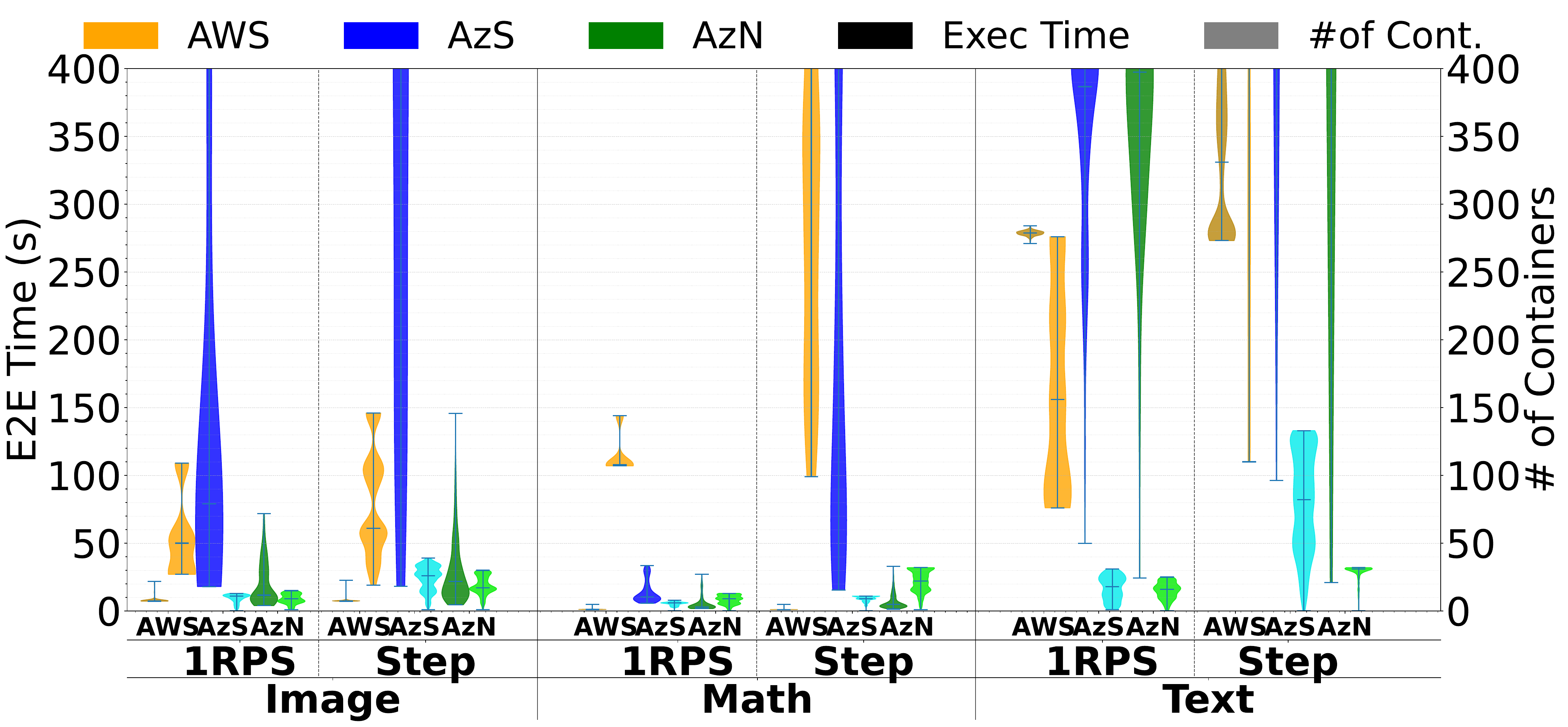}
  
\caption{Scaling of Math, Image and Text workflows with medium payload for static 1~RPS and Step workloads.}

\label{fig:all-wf-timeline}
\end{figure}

\subsection{Other Workflows Scaling}
\label{sec:app-other-scaling}
Lastly, Fig.~\ref{fig:all-wf-timeline} summarizes the scaling of the three other workflows for static 1~RPS and Step dynamic workload (detailed plots in Appendix Fig.~\ref{app:fig:all-wf-timeline}). In the legend, each cloud provider is shown using a distinct color, where the darker shade represents execution time and the lighter shade represents the number of containers (e.g., dark orange/light orange for AWS, dark blue/light blue for AzS, and dark green/light green for AzN).
The Math, Image and Text workflows at \textit{static $1$~RPS} are similar to earlier trends; AWS and AzN perform well, while AzS shows a more variable E2E time in the violin plot, especially for Image.
For \textit{Step}, AzS is unstable for all workflows beyond $2$~RPS and starts to timeout. It starts $11$ containers for Math workflow and uses $40$--$150$ for the more demanding Image and Text. Though AzS has a higher 
container limit than AzN, they are less effective. AWS and AzN handle Math and Image well for Step, though AzN shows a higher variability in time at the peak $8$~RPS even with $32$ containers. AWS spawns $40$--$120$ containers for these.

Text is the most demanding workflow. AzN is unstable beyond $2$~RPS for Step due to container limits. Even AWS shows variability at $8$~RPS as it reaches its limit of $1000$ containers
\footnote{\href{https://docs.aws.amazon.com/lambda/latest/dg/burst-concurrency.html}{Lambda scaling behavior}, AWS, 2024},
and
its E2E latency doubles from $\approx 300s$ to $\approx 600s$ as multiple requests are concurrently sent to the same container, causing queuing.

\subsection{Cold Starts}
\label{app:cs}
\subsubsection{Additional Plots on Coldstarts}
\begin{figure}[t!]
\centering
\subfloat[AzS | Graph]{%
    \includegraphics[width=0.42\columnwidth]{arxiv-v2-figures/takeaway-wise-plots/scaling-and-coldstarts/cold_warm_azure_graph_edges_upd.pdf}
    \label{app:fig:wfs-app-medium-inter-fn-v1:graph}
  }
  \subfloat[AzS | Image]{%
    \includegraphics[width=0.54\columnwidth]{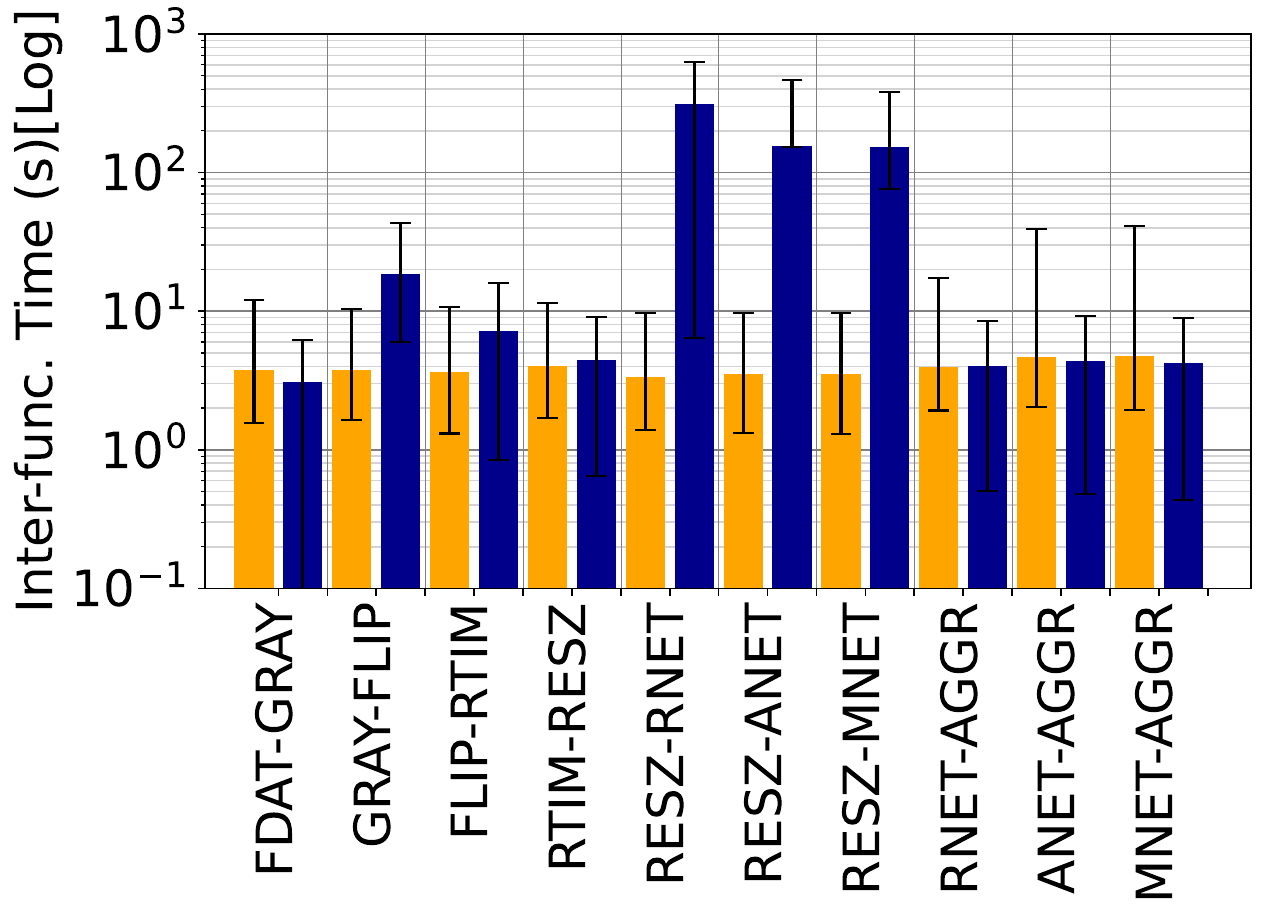}
    \label{app:fig:wfs-app-medium-inter-fn-v1:image}
  }\\
  \subfloat[AzN | Graph]{%
  \includegraphics[width=0.42\columnwidth]{arxiv-v2-figures/takeaway-wise-plots/scaling-and-coldstarts/cold_warm_azure_v2_graph_edges_upd.pdf}
    \label{app:fig:wfs-app-medium-inter-func-v2:graph}
  }
  \subfloat[AzN | Image]{%
  \includegraphics[width=0.54\columnwidth]{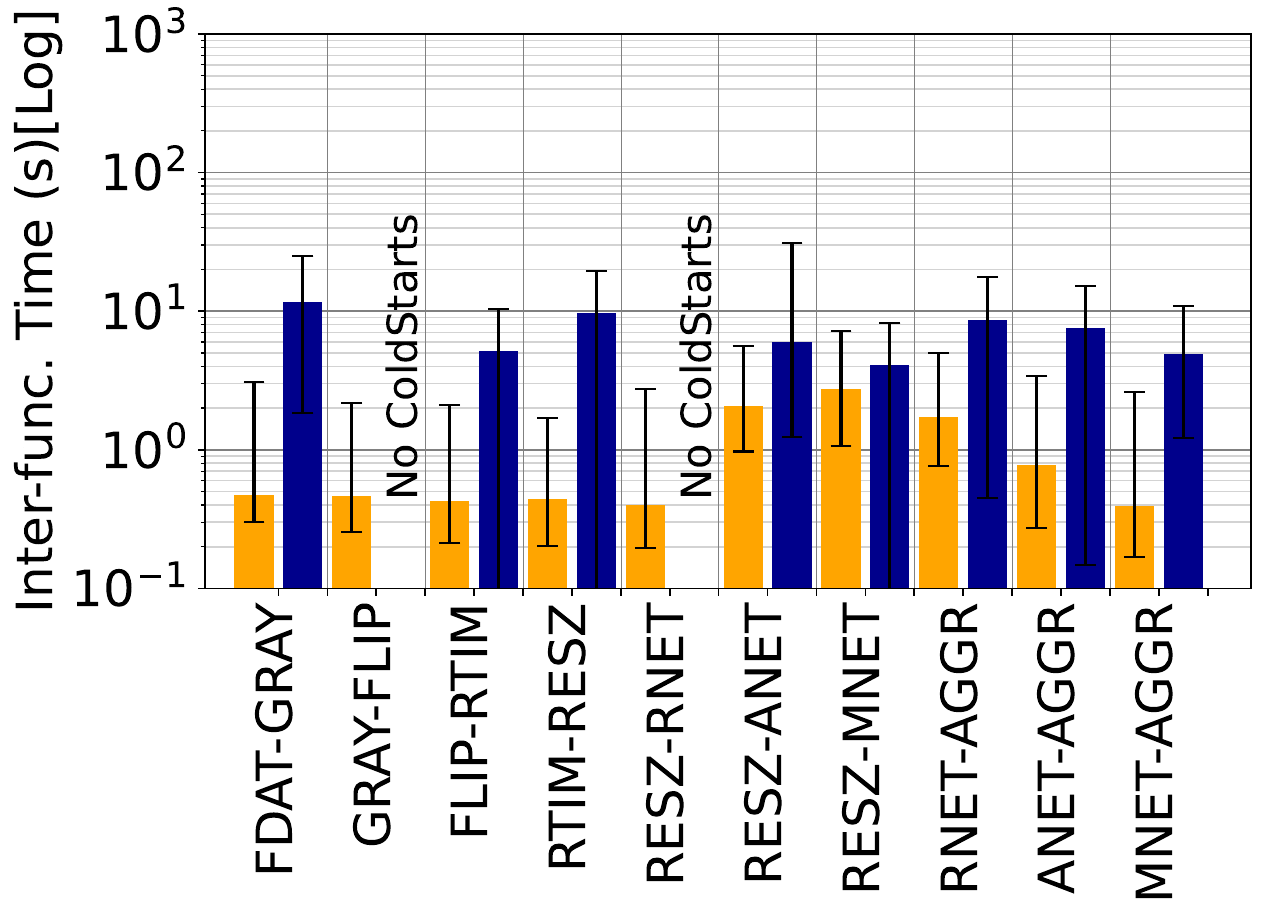}
    \label{app:fig:wfs-app-medium-inter-func-v2:image}
  }\\

  \subfloat[AzS | Small  | Graph]{%
  \includegraphics[width=0.42\columnwidth]{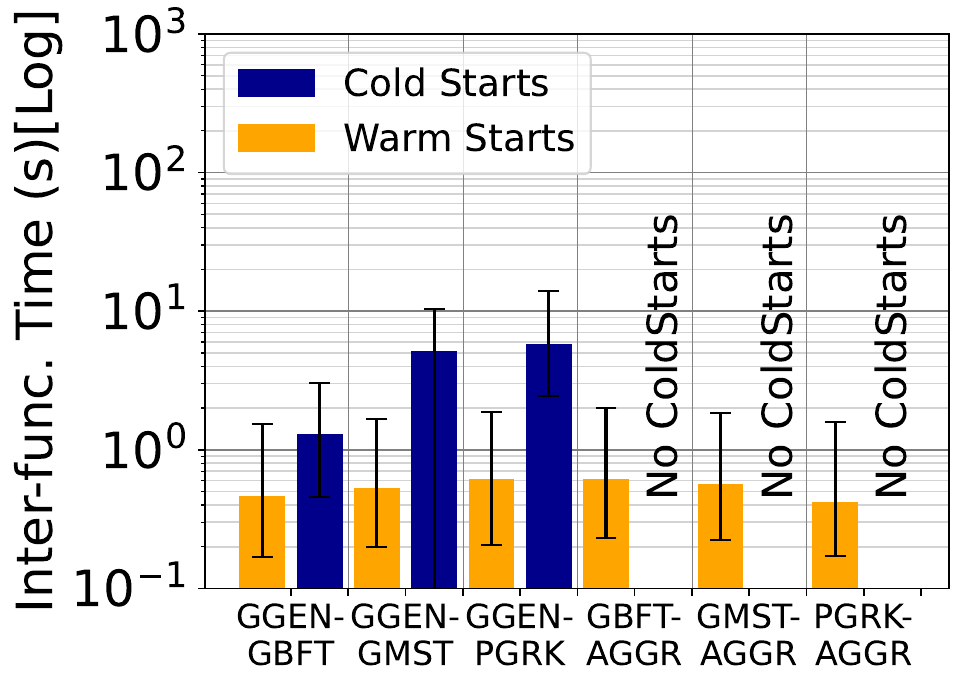}
    \label{app:fig:wfs-app-medium-inter-func-new-1:graph}
  }
  \subfloat[AzS | Small | Image]{%
  \includegraphics[width=0.54\columnwidth]{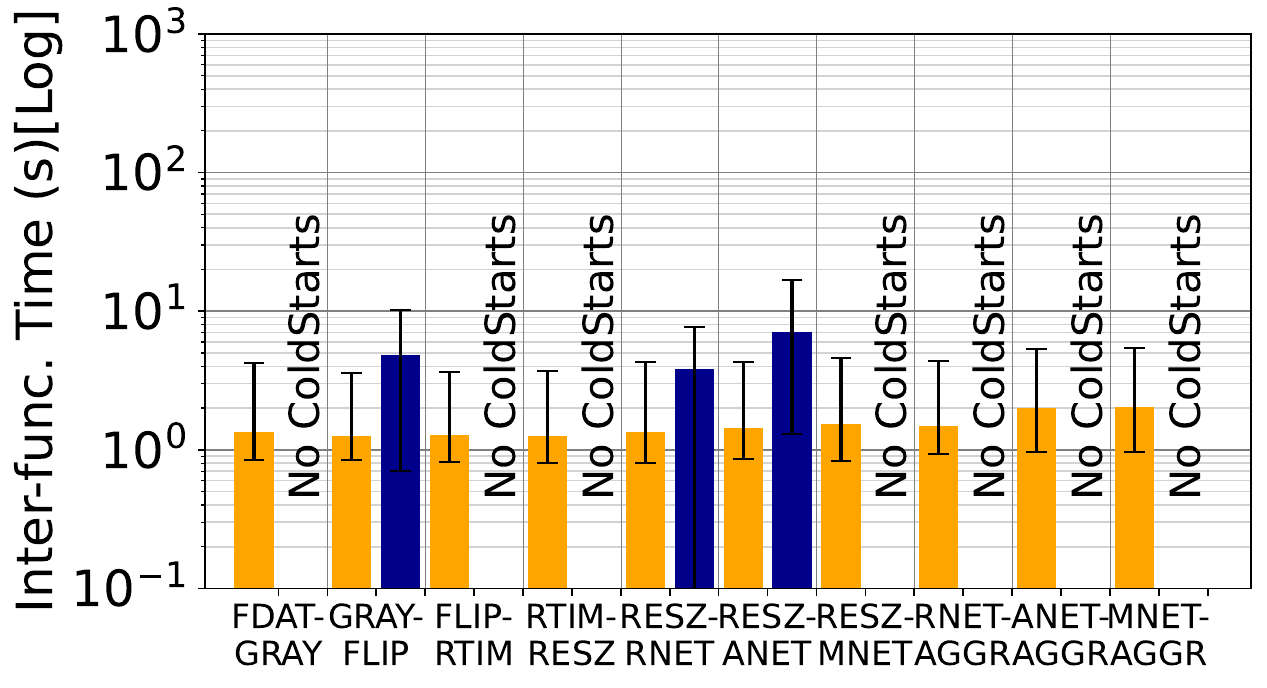}
    \label{app:fig:wfs-app-medium-inter-func-new-2:image}
  }\\

  \subfloat[AzS (Queue plus Blob transfers) | Graph]{%
  \includegraphics[width=0.42\columnwidth]{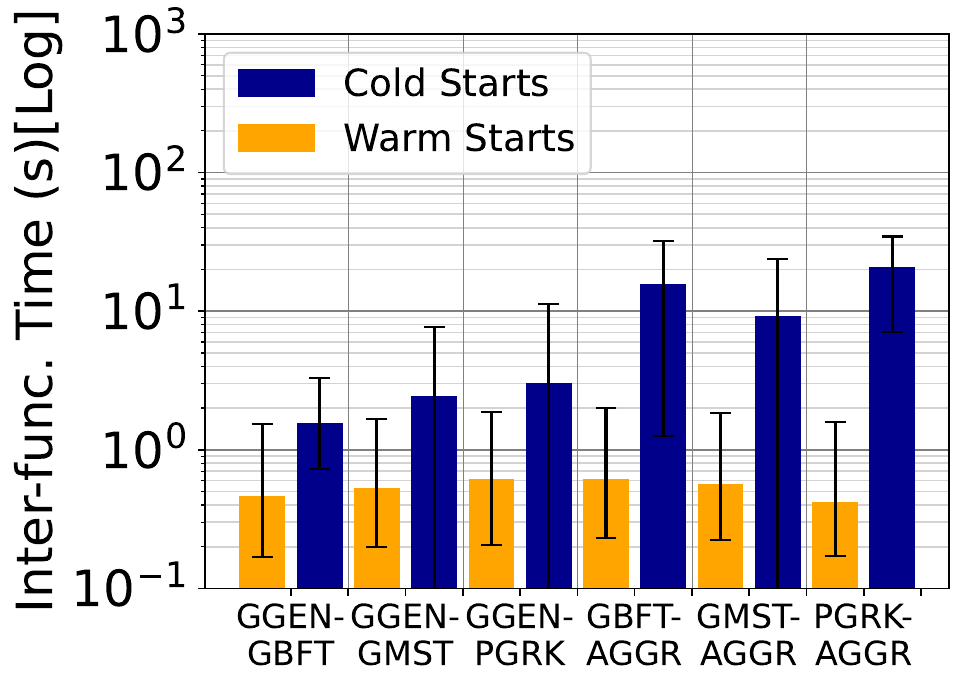}
    \label{app:fig:wfs-app-medium-inter-func-new:graph}
  }

\caption{Inter-function latencies \textit{with and without container cold-starts} for AzS and AzN, for Graph and Image workflows on medium payload and 1~RPS.}
\label{app:fig:graph-workload-az-containers-box-split}
\end{figure}

The container cold-start latency is captured in the inter-function cold-start latency, as shown in Fig.~\ref{app:fig:graph-workload-az-containers-box-split}. This expands upon Fig.~\ref{fig:graph-workload-az-containers-box-split} in the main article and supports \ref{ta:coldstart_cascading_blob}.
AzS and AzN load the entire workflow in memory when a coldstart occurs. So the expected behavior is that the Python runtime overhead should be constant across all the edges, but this is not what we observe. The inter-function latency also includes data transfer setup latency. Now this happens through Azure blobs for the medium size payload shown in the plot. The plot for small payload on Graph and Image (Figs.~\ref{app:fig:wfs-app-medium-inter-func-new-1:graph} and  \ref{app:fig:wfs-app-medium-inter-func-new-2:image}) show that this coldstart overhead time is constant for small payloads ($\approx4.1$s for graph and $\approx5$s for image), i.e., when they are transferred via queues. This suggests that there is also some time that is required for setting up resources for Azure Blobs.

To confirm this we run an alternate experiment for the Graph workflow in which first three edges transfer data using queues (small payload) and last three via blobs (medium). We see in Fig.~\ref{app:fig:wfs-app-medium-inter-func-new:graph} that the cold start overheads (\textit{ColdStartTime} minus \textit{WarmStartTime}) for queue transfer edges are constant $\approx1s$, while they vary for Blob transfer edges: $11.4s$, $8.4s$ and $19.6s$, respectively.

\begin{figure}[t!]
\centering%
  \subfloat[Graph WF on AzS]{
    \includegraphics[width=0.45\columnwidth]{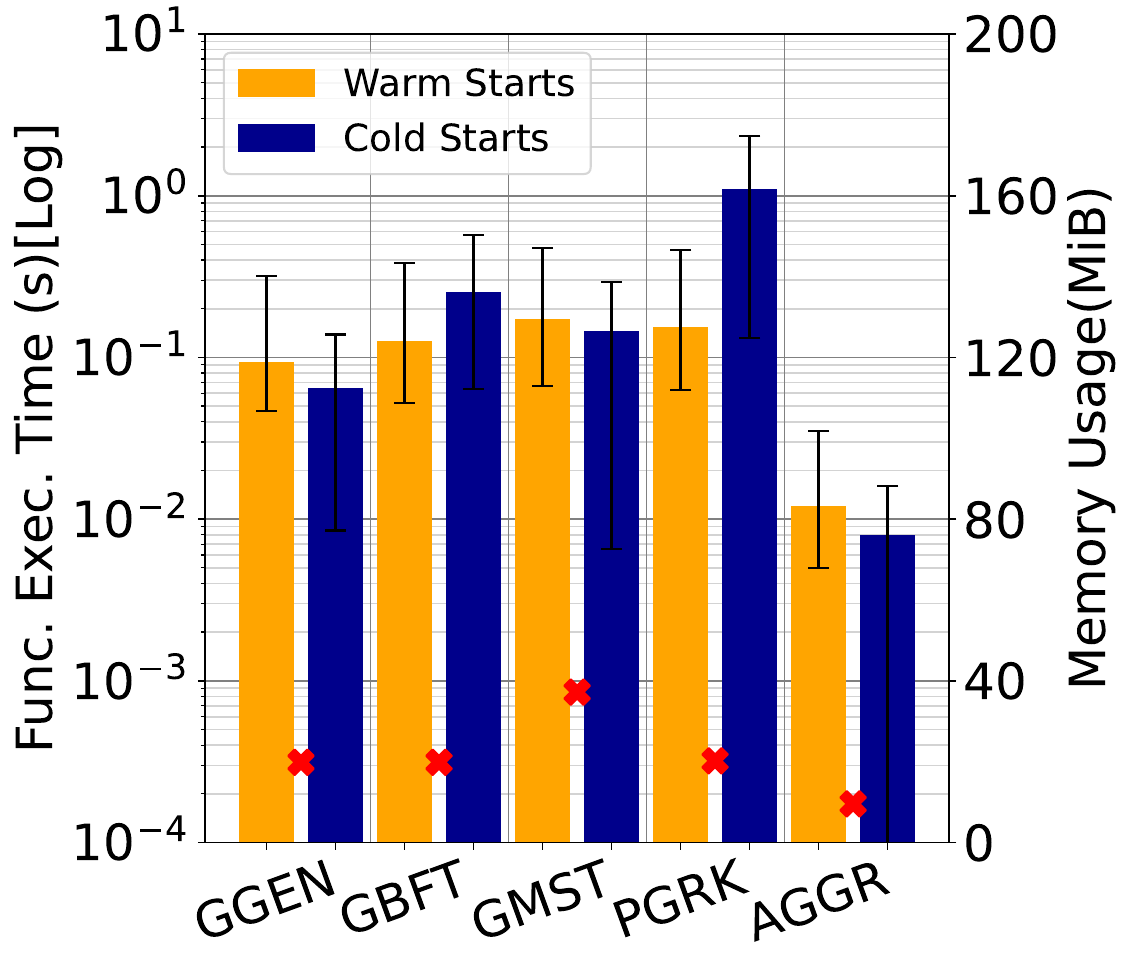}
    \label{app:fig:az-fn-exec:graph}
  }
  \subfloat[Graph WF on AzN]{
    \includegraphics[width=0.45\columnwidth]{arxiv-v2-figures/takeaway-wise-plots/scaling-and-coldstarts/cold_warm_azure_v2_graph_functions_upd.pdf}
    \label{app:fig:azv2-fn-exec:graph}
  }\\
  \subfloat[Image WF on AzS]{
    \includegraphics[width=0.48\columnwidth]{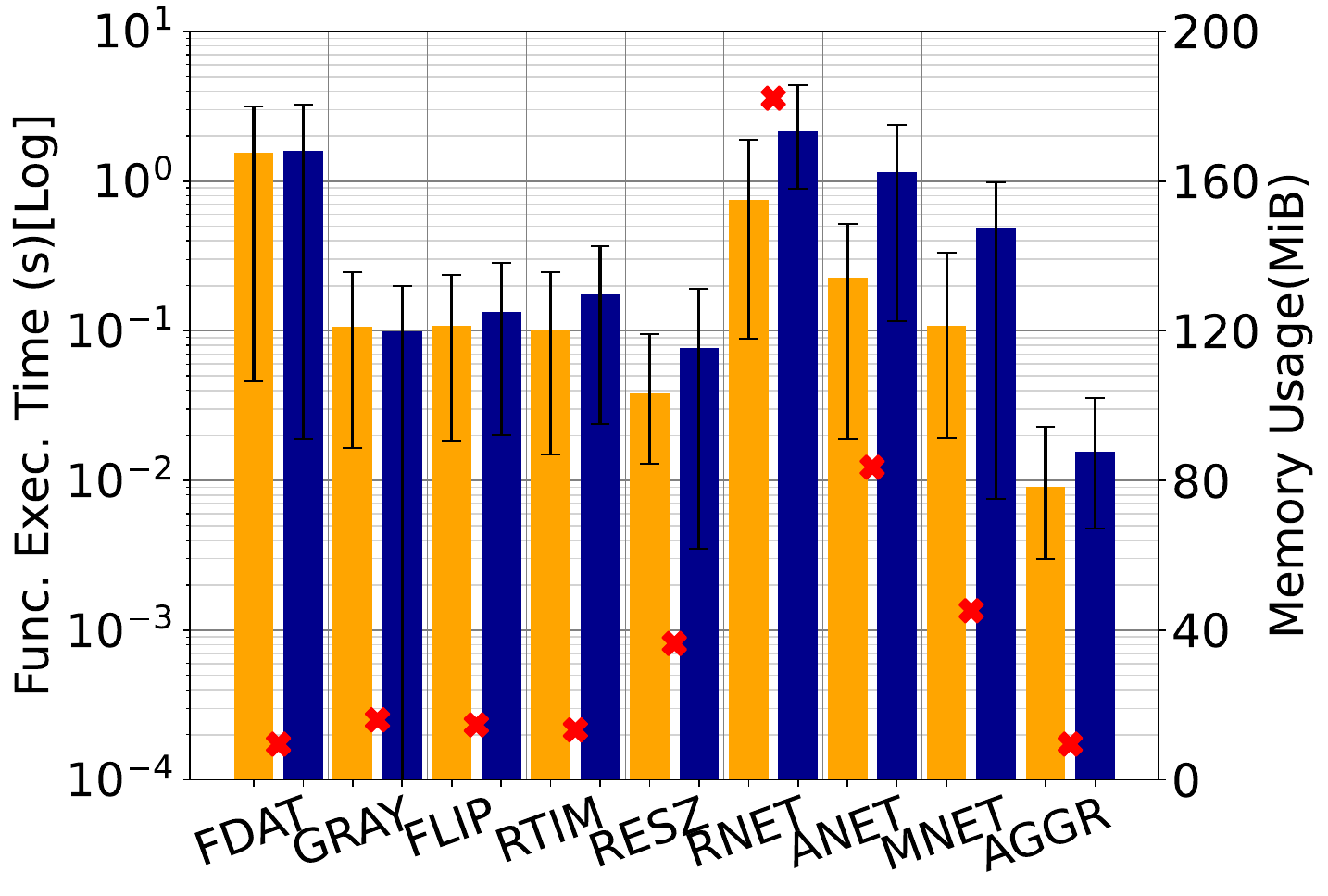}
    \label{app:fig:az-fn-exec:image}
  }%
  \subfloat[Image WF on AzN]{
    \includegraphics[width=0.48\columnwidth]{arxiv-v2-figures/takeaway-wise-plots/scaling-and-coldstarts/cold_warm_azure_v2_image_functions_upd.pdf}
    \label{app:fig:azv2-fn-exec:image}
  }%
\caption{Function execution times \textit{with and without function cold-starts}, for Graph and Image workflows with medium payload and static $1$~RPS}
\label{app:fig:funcion-exec-coldstarts}
\end{figure}

Fig.~\ref{app:fig:funcion-exec-coldstarts} shows the warmstart (yellow) and coldstart(blue) latencies for AzS and AzN on Graph and Image for \textit{medium payload}, as discussed in \ref{ta:cold-start:func-warmup}. It complements Fig.~\ref{fig:funcion-exec-coldstarts} that is limited to AzN.

\begin{figure}[t!]
\centering%

    \includegraphics[width=.65\columnwidth]{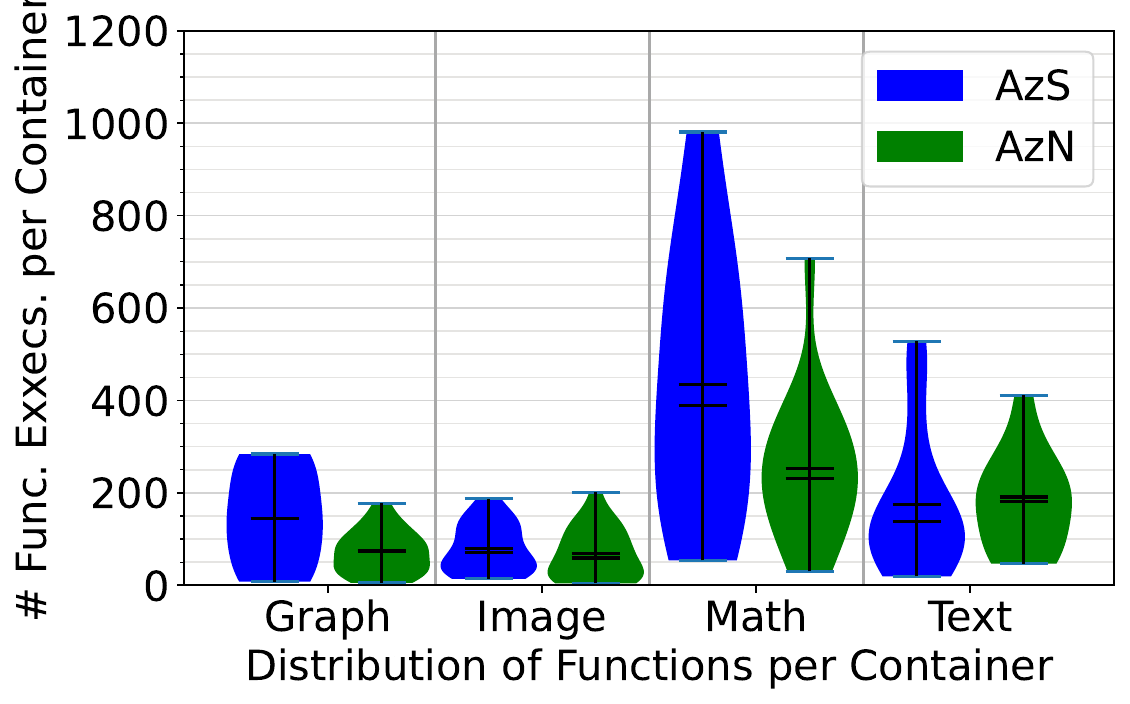}    
\caption{Distribution of the number of functions executing per container in a 300~sec window for AzS and AzN for Medium payload and static $1$~RPS.}\label{fig:az-vs-azv2-variability}

\end{figure}

\subsection{Variability in Function Execuyion Times in AzS and AzN}

Besides CPU heterogeneity, Azure has variability due to the allocation of function executions to each container of a workflow.
AzS uses a common activity queue that containers poll for the next task while AzN partitions the tasks across queues maintained for each partition.
Fig.~\ref{fig:az-vs-azv2-variability} shows a violin distribution of the number of functions executed per container during a $5mins$ window for each workflow. The \textit{tighter} this violin, the more balanced the function executions are across containers, and \textit{lower} the value, the less load each container has. AzS shows a higher variability for Graph than AzN, while for Image, both are comparably lower at a median of $\approx 75$~tasks. This suggests that AzN achieves better load-balancing, reducing the variability of function execution times. The distribution is even more dispersed for complex workflows like Math and Text. 
In contrast, AWS executes only one function per container and avoids such load-balancing skews, albeit using more container resources.

\begin{figure}[t!]
\centering%
  \subfloat[Math Workflow | \textit{1~RPS}]{%
  \includegraphics[width=0.47\columnwidth]{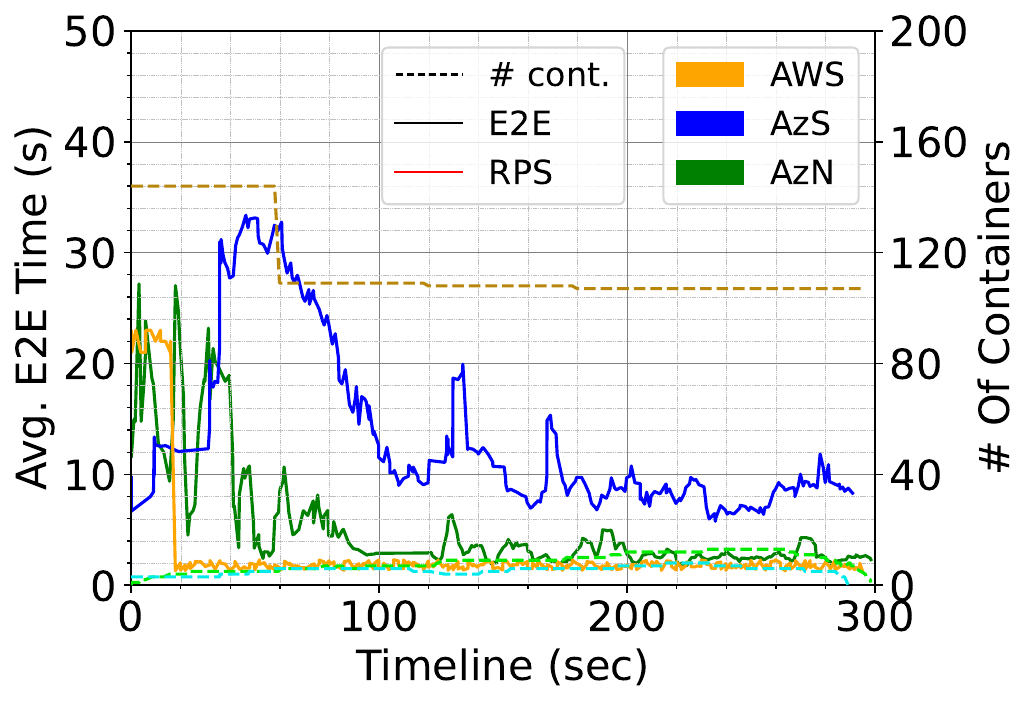}%
    \label{app:fig:math-medium-scaling-1rps}
  }%
  \subfloat[Math Workflow | \textit{Step}]{%
  \includegraphics[width=0.5\columnwidth]{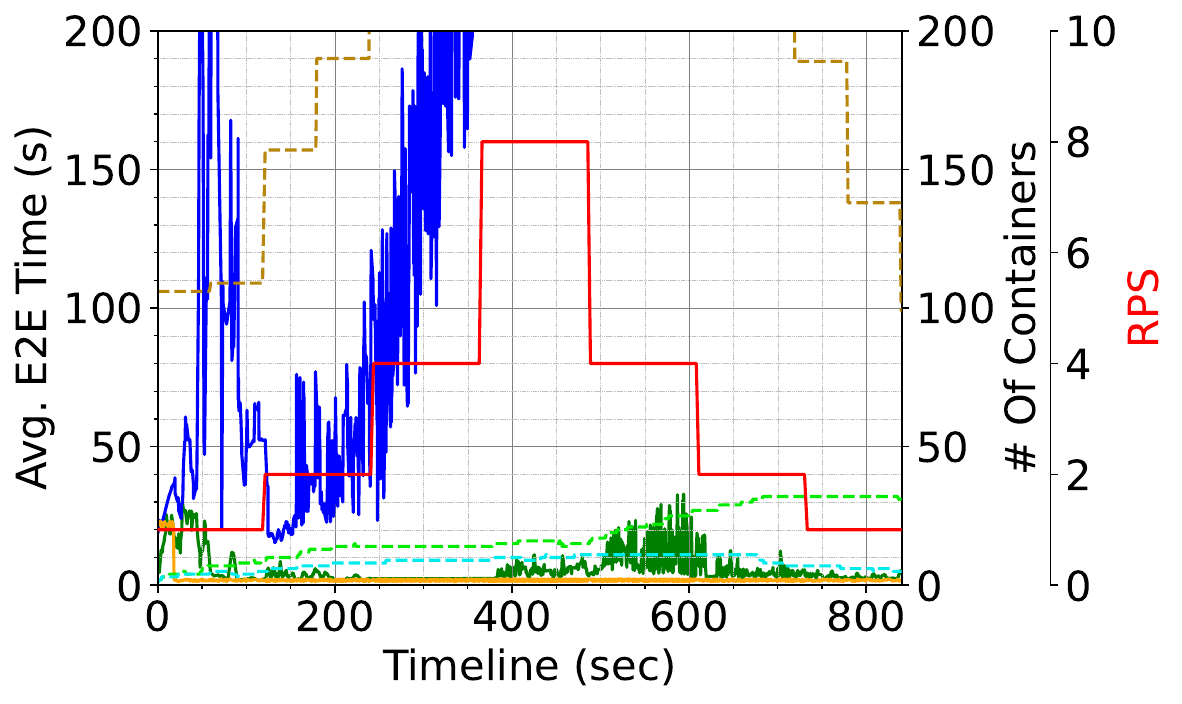}%
    \label{app:fig:math-medium-scaling}%
  }\\
  \subfloat[Image Workflow | \textit{1~RPS}]{%
    \includegraphics[width=0.47\columnwidth]{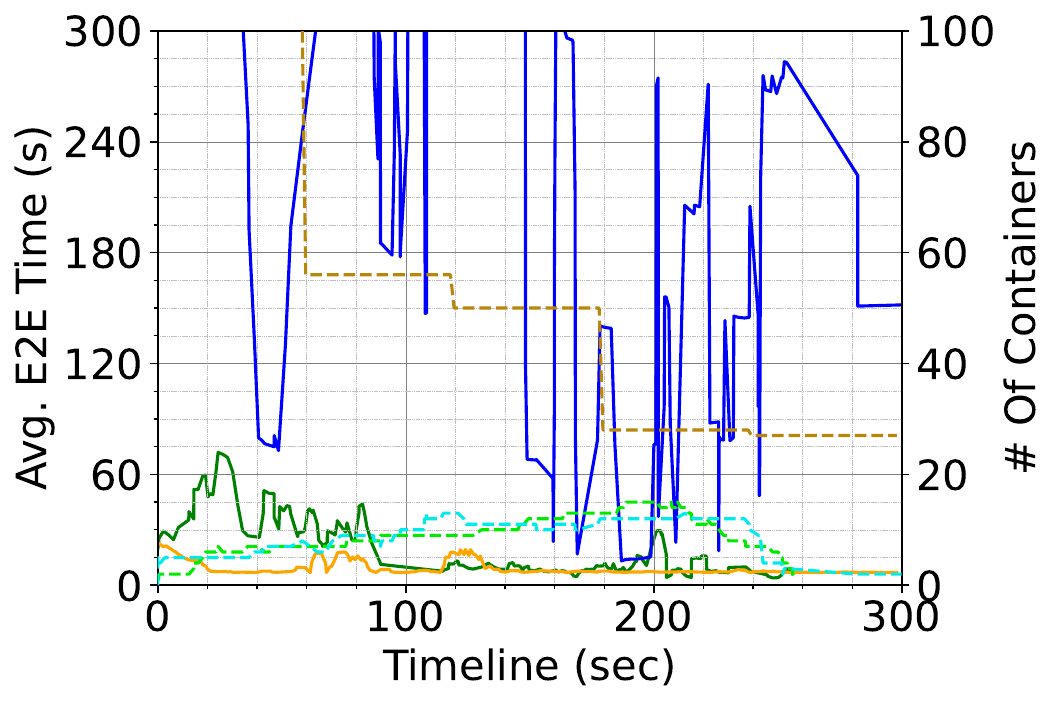}%
    \label{app:fig:image-medium-scaling-1rps}%
  }%
  \subfloat[Image Workflow | \textit{Step}]{%
    \includegraphics[width=0.5\columnwidth]{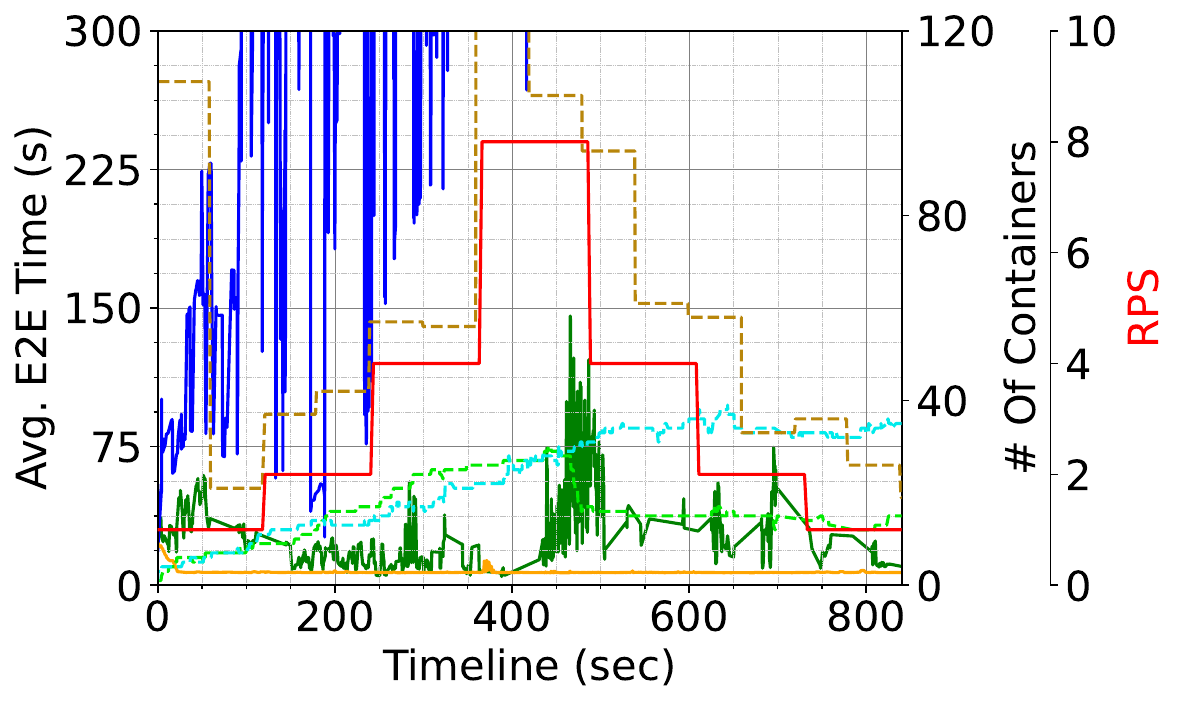}%
    \label{app:fig:image-medium-scaling}%
  }\\
\subfloat[Text Workflow | \textit{1~RPS}]{%
    \includegraphics[width=0.47\columnwidth]{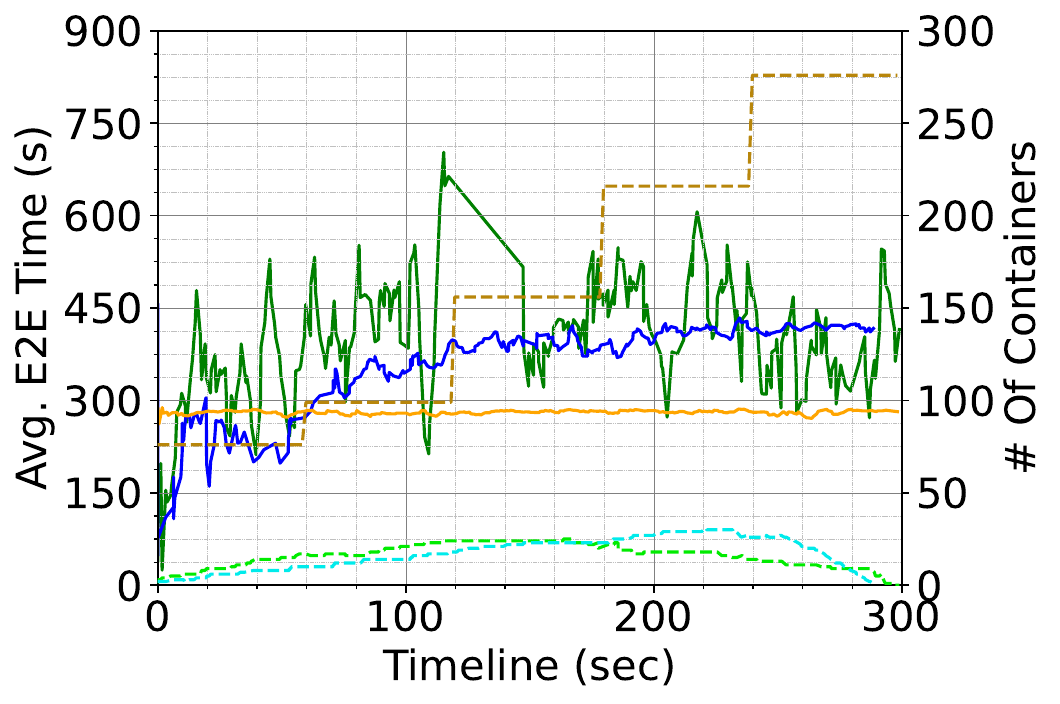}%
    \label{app:fig:text-medium-scaling-1rps}%
  }%
  \subfloat[Text Workflow | \textit{Step}]{%
    \includegraphics[width=0.5\columnwidth]{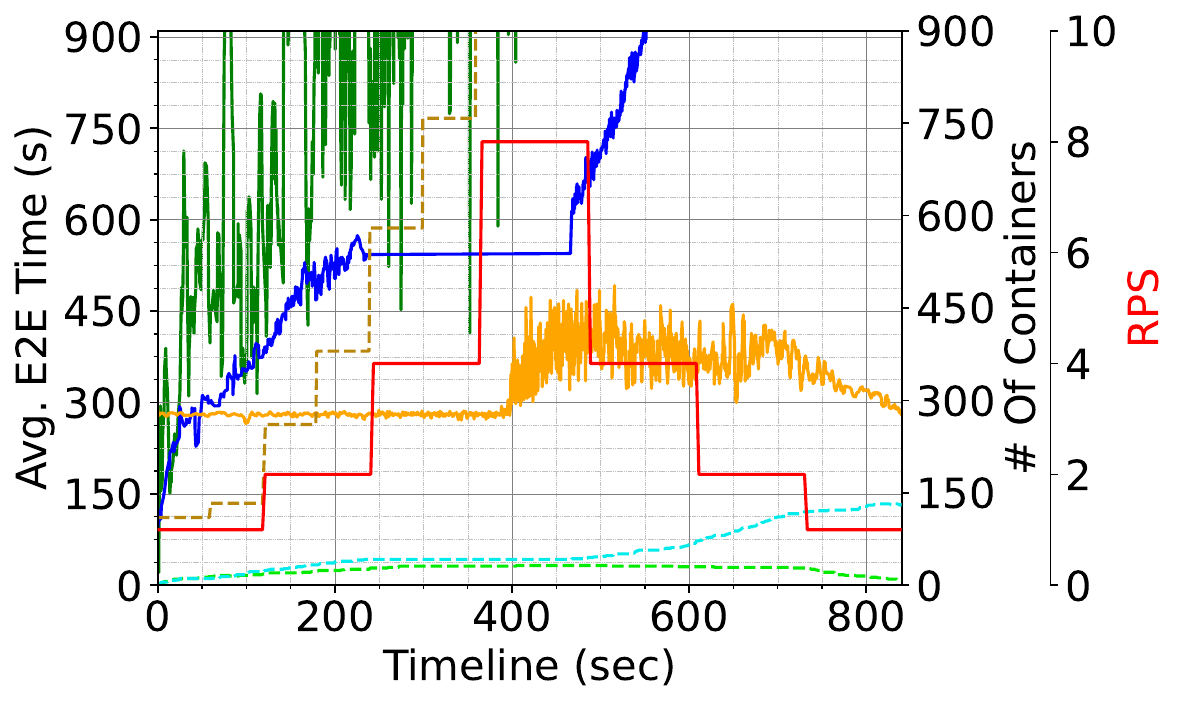}%
    \label{app:fig:text-medium-scaling}%
  }
\caption{Scaling of Math, Image and Text workflows with medium payload. Top row is for static 1~RPS workload while bottom row for Step workload.}
\label{app:fig:all-wf-timeline}
\end{figure}

\begin{figure}[t!]
\centering%
    \includegraphics[width=0.48\columnwidth]{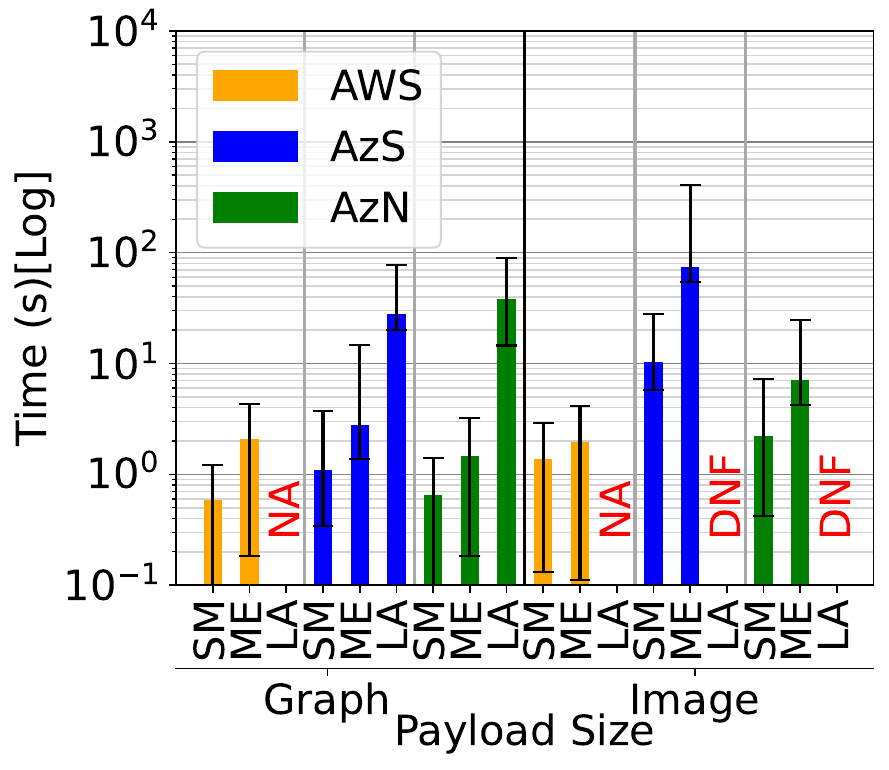}%
    \label{fig:inter-fn-payloads:graph}
  \caption{Inter-function latencies for Graph and Image workflows while varying Payloads(omitting cold-starts) 
  }
\label{fig:inter-fn-data-transfer-workloads}

\end{figure}

\subsection{Effect of Paylod size on inter-fn zlatency on Different workflows}\label{sec:app-int}

We examine the effect of using different payload sizes and RPS on the inter-function latency. Fig.~\ref{fig:inter-fn-data-transfer-workloads} shows the median inter-function latency (with error bars indicating Q1--Q3 values) for the Graph and Image workflows when we vary the payload~(retaining $1$~RPS) or vary the RPS~(retaining a medium payload). 
For the large payload, AWS does not support these sizes while Azure gives timeout errors for Image (DNF).

In the Graph on \textit{AWS}, the inter-function latencies increase from $0.59s$ to $2.09s$ as we switch from \textit{small} ($47KiB$) to \textit{medium} payload ($220KiB$), which is a $3.5\times$ increase, while for Image this increases from $1.37s$ to $1.94s$ for small $39KiB$ to medium $245KiB$: a $1.41\times$ increase. On average the increase is about $\approx 2.4\times$. This is consistent with the microbenchmarks (Fig.~\ref{fig:inter-fn-data-transfer:small}), where we see a $\approx2.07\times$ increase when the size increases from $45KiB$ to $250KiB$.

For Graph/Image on \textit{AzN}, going from \textit{small} ($47KB$/$39KB$) to \textit{medium} ($220KB$/$245KB$) is an $\approx 5\times$ increase in payload and causes a somewhat linear increase in time for AzN ($0.55s\rightarrow1.45s$/$2.23\rightarrow7.09s$) which is about $2.63\times/3.17\times$. This too is consistent with the microbenchmarks (Fig.~\ref{fig:inter-fn-data-transfer:small}) where we see an increase of $\approx2.85\times$ in inter-function time from $45KiB$ to $250KiB$.
For \textit{AzS}, this payload size change causes a switch from Queues to Blobs. Here, the impact 
is worse for Image ($10s$ to $74s$, $7.4\times$)
due to its longer path, causing more requests per second to the common Blob storage.
AzN's use of Event Hubs even for medium sizes mitigates this.

When we move from \textit{medium to large} payload ($1.1MB$, $\approx 5\times$ increase), the increase in time is super-linear and sharper for Graph on both \textit{AzS} ($27s$) and \textit{AzN} ($38s$), and worse than the latency increase seen in the micro-benchmarks for comparable payload sizes.
Larger payload sizes cause greater stress on the storage accounts in writing and reading the inter-function messages. This is exacerbated as the workflow size and latency growth since more operations happen concurrently for multiple executing workflows. 
The large payload for Image workflow also causes frequent failures. 
Hence, for larger and more complex workflows the inter-function latency growth is much worse than seen in the micro-benchmarks for AzS and AzN (Fig.~\ref{fig:inter-fn-data-transfer}), limiting workflow scalability.

\subsubsection{Estimating Container Coldstarts}
\label{app:sec:container-cs-est}

As discussed in \ref{ta:coldstart_cascading_blob}, whenever there is a cold-start in Azure, multiple steps take place.
To accurately decompose the cold-start latency in Azure, we design a controlled micro-benchmark that isolates key overheads in function execution. The cold-start process consists of multiple sequential steps: (1) worker selection, (2) function application file system and settings mount, (3) runtime environment setup (e.g., Python, Node.js), (4) function loading into memory, and (5) function execution. The first three steps contribute to the \textit{container cold-start time} ($t_c$), while the latter two steps together form the \textit{runtime cold-start time} ($t_r$). Further, step (4) specifically corresponds to the \textit{inter-function cold-start overhead} ($t_{i}$), which is influenced by the function package size $p$. 

To quantify these values, we deploye a minimal workflow consisting of two empty functions that solely record execution time and container identifiers. No external imports were included to eliminate extraneous loading overheads ($t_f \approx 0$) and there is no other user logic. We execute this setup for three package sizes (50MB, 100MB, and 200MB) across AzS and AzN, measuring the total cold-start latency $T_{cp}$. Given the equation $t_c + t_i(p) + t_f = T_{cp}$, we estimate $t_c$ by regressing $T_{cp}$ against $p$, assuming a linear dependence of $t_i(p)$ on $p$. The resulting model yielded $t_i(p) \approx \alpha p + \beta$, where $\alpha$ and $\beta$ were empirically derived. 

\begin{figure}[t!]
\centering%
\subfloat[East USA Graph]{
    \includegraphics[width=0.5\columnwidth]{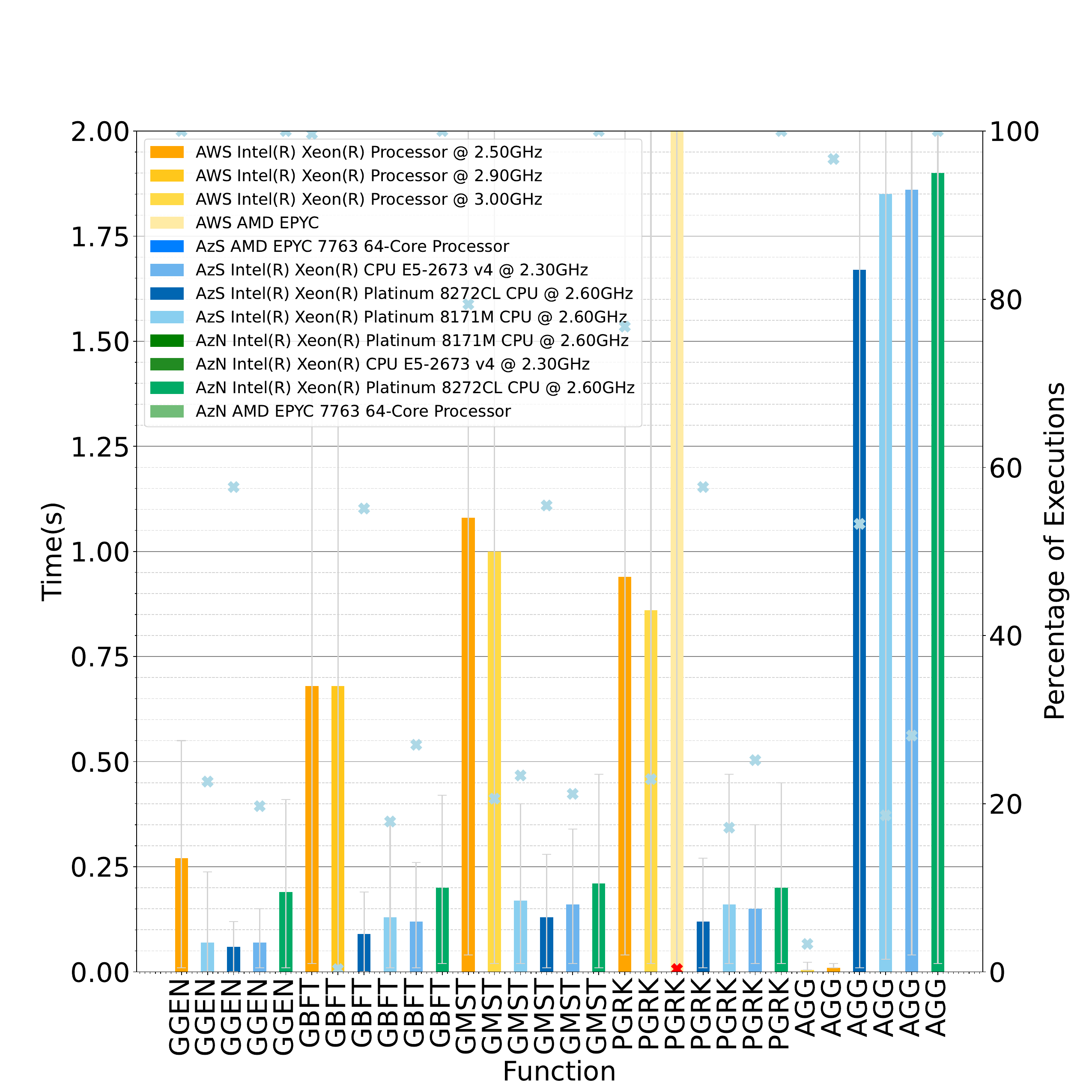}
    \label{fig:appendix-east-usa-graph}
    }
    \subfloat[East USA Image]{
    \includegraphics[width=0.5\columnwidth]{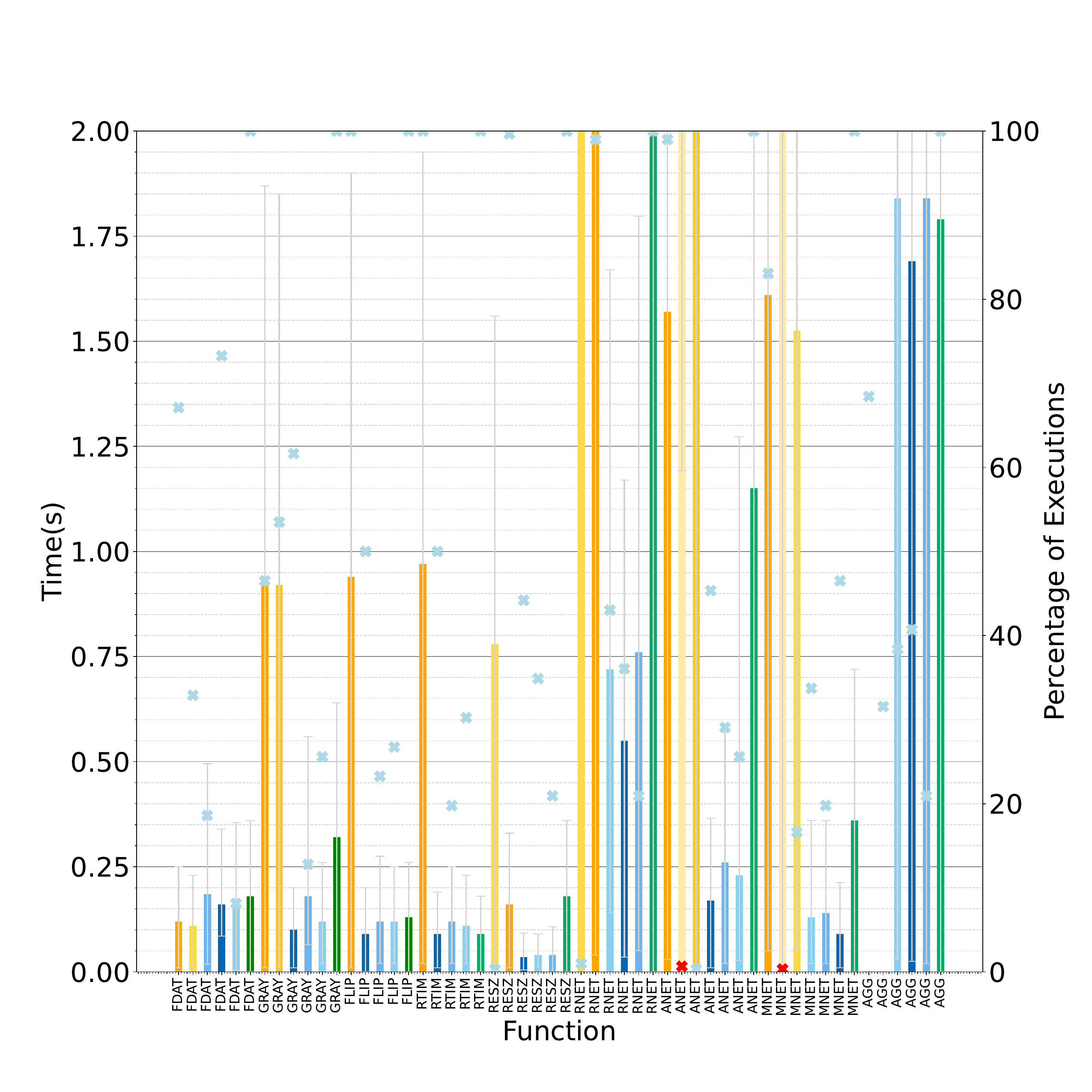}
    \label{fig:appendix-east-usa-image}
    }
    
    \subfloat[Central Europe Graph]{
    \includegraphics[width=0.5\columnwidth]{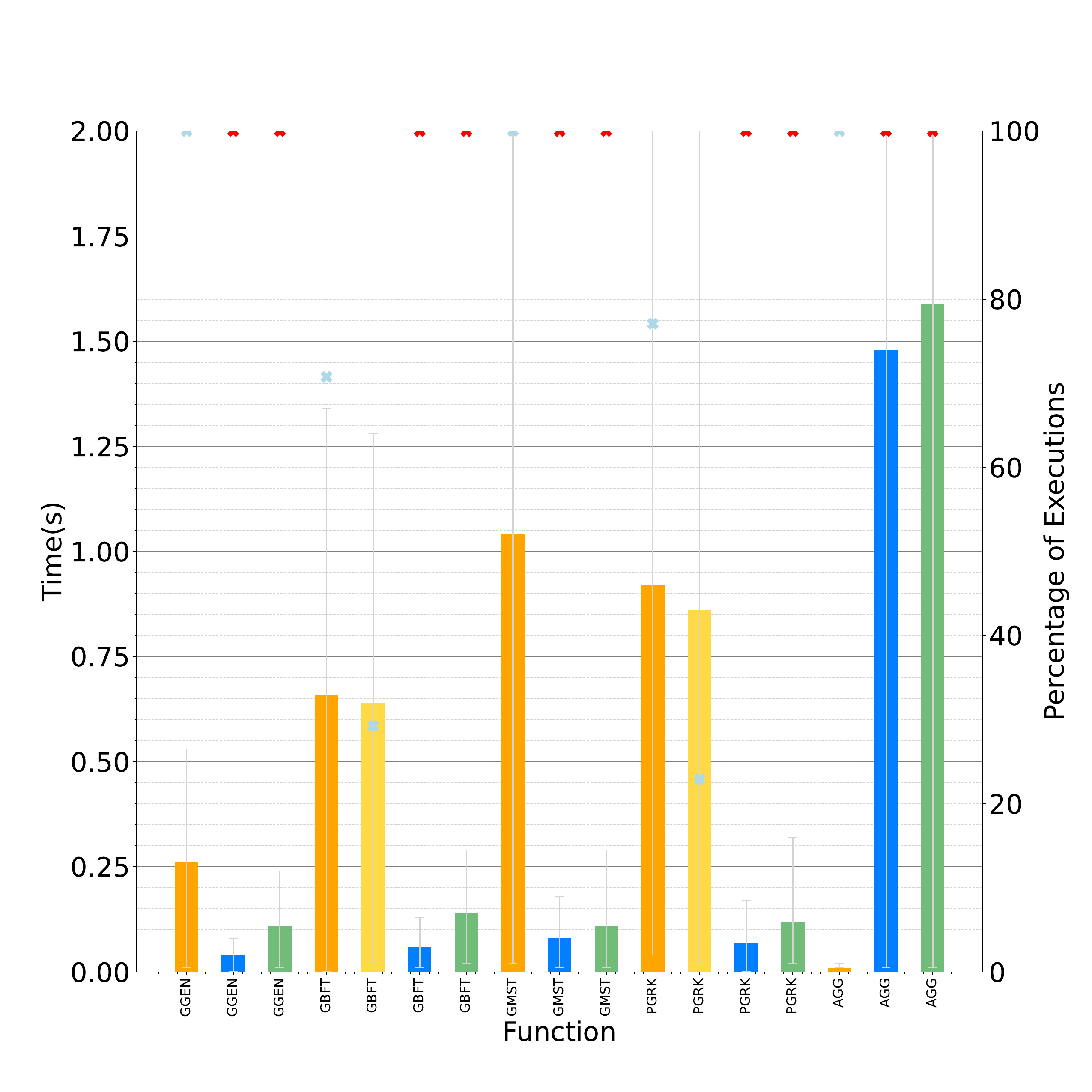}
    \label{fig:appendix-central-europe-graph}
    }
    \subfloat[Central Europe Image]{
    \includegraphics[width=0.5\columnwidth]{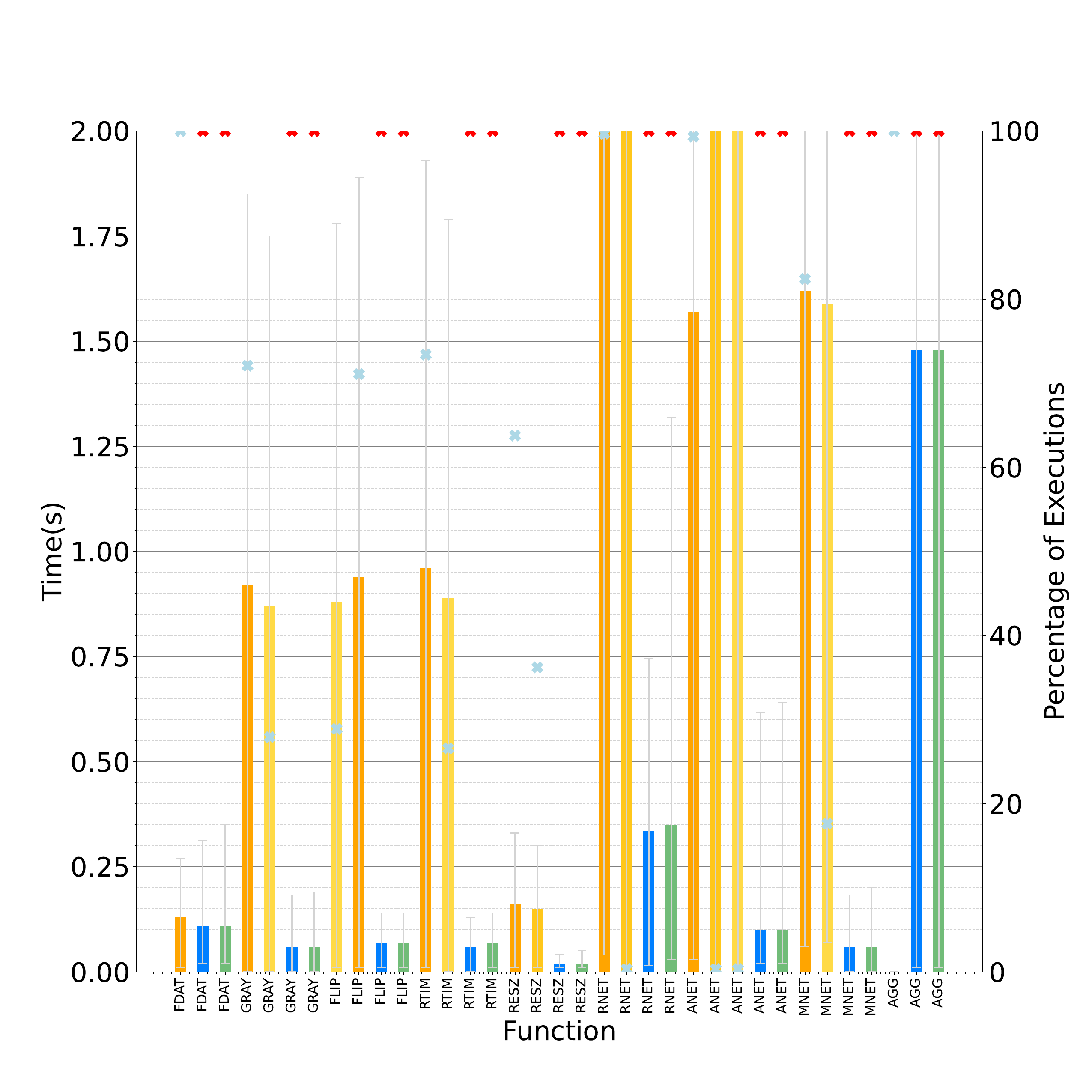}
    \label{fig:appendix-central-europe-image}
    }
    
\caption{Hardware Heterogeneity Across different Datacenters for Graph and Image workflow, medium payload , 1 RPS, 300s}\label{fig:appendix-plot-1}
\end{figure}

Our analysis reveals that $t_c \approx 290$ms for AzS and $t_c \approx 51.5$ms for AzN, accounting for approximately $40\%$ and $66.1\%$ of the total inter-function cold-start overhead, respectively. The difference in $t_c$ across environments aligns with the distinct scaling mechanisms in AzS and AzN, where AzS relies on a task hub with multiple storage services (queues, blobs, tables), whereas AzN scales through event hub partitions, effectively reducing initialization overhead. The high goodness-of-fit ($R^2 = 0.92$ for AzS and $R^2 = 0.99$ for AzN) supports the validity of our regression model in estimating $t_i(p)$ and isolating $t_c$.

\subsection{Hardware Heterogeneity Plots for East USA and Central Europe}
\label{sec:appendix-func-variability-plots}

Fig.~\ref{fig:appendix-plot-1} provides additional plots here on the hardware heterogeneity and performance variability for different global regions of Azure and AWS when running FaaS workflows. This complements Fig.~\ref{fig:graph-img-wf-exec-times} in the main article and supports \ref{ta:func:variability-2}.

In US East, AWS has 4 different CPU architectures for Lambda functions (Intel Xeon @ 2.5, 2.9 and 3GHz and AMD EPYC) while Azure has 3 Intel Xeon CPUs (E5-2673 v4 @ 2.3GHz, Platinum 8272CL @ 2.60GHz and Platinum 8171M @ 2.60GHz).
In Central Europe, AWS has 3 Intel CPUs (Intel Xeon @ 2.5, 2.9 and 3GHz) while Azure runs only only 1 hardware type (AMD EPYC 7763).

\end{document}